\documentclass[reprint, aps, pra, superscriptaddress, citeautoscript, floatfix]{revtex4-2}

\usepackage{silence}

\usepackage{amsfonts}
\usepackage{amsmath}
\usepackage{amssymb}
\usepackage{dsfont}
\usepackage{bbm}
\usepackage{multirow}
\usepackage{physics2}
\usepackage{ragged2e}  
\usepackage{siunitx}
\usepackage{tikz}
\usepackage{scalerel}

\usepackage{subcaption}

\usepackage{mathtools}
\usepackage[scr=dutchcal,scrscaled=1.03]{mathalpha}

\usepackage{xr-hyper}
\usepackage{hyperref}
\usepackage[capitalise]{cleveref}

\usepackage{graphicx}
\graphicspath{{figures}}

\usepackage{xcolor}
\usepackage{soul}
\usepackage{framed}

\catcode`\%=12\relax
\DeclareSIUnit[number-unit-product = ]\percent{
\catcode`\%=14\relax

\colorlet{shadecolor}{yellow}
\usephysicsmodule{ab}
\usephysicsmodule{ab.braket}

\captionsetup{justification=justified, singlelinecheck=false}
\DeclareCaptionJustification{justified}{\justifying}

\newcommand\numberthis{\addtocounter{equation}{1}\tag{\theequation}}

\DeclareRobustCommand\sampleline[1]{%
  \tikz\draw[#1] (0,0) (0,\the\dimexpr\fontdimen22\textfont2\relax)
  -- (10pt,\the\dimexpr\fontdimen22\textfont2\relax);%
}

\DeclareMathOperator\Tr{Tr}
\DeclareMathOperator\Span{span}
\DeclareMathOperator\sinc{sinc}
\DeclareMathOperator*{\argmax}{arg\,max}
\DeclareMathOperator*{\argmin}{arg\,min}

\newcommand\intlabel{{\mathrm{I}}}
\newcommand\timeorder{{\mathcal{T}}}
\newcommand\parity{{\mathcal{P}}}
\newcommand\Mparity{{\mathcal{M}}}
\newcommand\extlabel{{\mathrm{ext}}}
\newcommand\efflabel{{\mathrm{eff}}}
\newcommand\barelabel{{\mathrm{idle}}}
\newcommand\contlabel{{\mathrm{c}}}
\newcommand\targlabel{{\mathrm{t}}}
\newcommand\slabel{{\mathrm{s}}}
\newcommand\drivelabel{{\mathrm{d}}}
\newcommand\midlabel{{\mathrm{m}}}
\newcommand\CNOTlabel{{\mathrm{CNOT}}}
\newcommand\fourier{{\mathcal{F}}}
\newcommand\ESD{{\mathcal{E}}}
\newcommand\hilbert{{\mathcal{H}}}
\newcommand\floqham{{\mathscr{H}}}
\newcommand\floqkam{{\mathscr{K}}}
\newcommand\eincoh{\varepsilon^\mathrm{incoh}}
\newcommand\pulselabel{{\mathrm{p}}}

\newcommand\ramplabel{{\mathrm{r}}}
\newcommand\caplabel{{\mathrm{cap}}}
\newcommand\indlabel{{\mathrm{ind}}}
\newcommand\rampfunc{{\mathcal{R}}}

\newcommand\inlabel{{\mathrm{in}}}
\newcommand\outlabel{{\mathrm{out}}}

\newcommand\ilabel{{\mathrm{i}}}
\newcommand\flabel{{\mathrm{f}}}
\newcommand\reflabel{{\mathrm{ref}}}
\newcommand\aclabel{{\mathrm{AC}}}
\newcommand\LZlabel{{\mathrm{LZ}}}

\newcommand\NN{{\mathbb{N}}}
\newcommand\ZZ{{\mathbb{Z}}}
\newcommand\RR{{\mathbb{R}}}
\newcommand\CRlabel{{\mathrm{CR}}}
\newcommand\Id{{I}}

\newcommand{\upsquigarrow}{%
  \raise.2ex\rlap{\kern.3ex\rotatebox[origin=c]{90}{$\rightsquigarrow$}}%
  \phantom{\uparrow}\vphantom{\uparrow}%
}
\newcommand{\downsquigarrow}{%
  \raise.2ex\rlap{\kern.0ex\rotatebox[origin=c]{-90}{$\rightsquigarrow$}}%
  \phantom{\downarrow}\vphantom{\downarrow}%
}

\makeatletter
\newsavebox{\@brx}
\newcommand{\llangle}[1][]{\savebox{\@brx}{\(\m@th{#1\langle}\)}%
  \mathopen{\copy\@brx\kern-0.5\wd\@brx\usebox{\@brx}}}
\newcommand{\rrangle}[1][]{\savebox{\@brx}{\(\m@th{#1\rangle}\)}%
  \mathclose{\copy\@brx\kern-0.5\wd\@brx\usebox{\@brx}}}
\makeatother

\def\measurehat#1{%
   \setbox0=\vbox{$\SavedStyle\hat{#1}\hfil\break$\null\par
      \setbox0=\lastbox\unskip\unpenalty\global\setbox1=\lastbox}%
   \setbox0=\hbox{\unhbox1 \unskip\unpenalty\unskip \global\setbox2=\lastbox}%
   \setbox0=\vbox{\unvbox2 \setbox0=\lastbox}%
}
\def\hathat#1{%
  \ThisStyle{%
    \measurehat{#1}\dimen0=\wd0 \measurehat{\kern0pt#1}%
    \raise.25ex\rlap{\kern\dimexpr\dimen0-\wd0$\SavedStyle\hat{\phantom{#1}}$}{\raise-0.05ex\rlap{\kern\dimexpr\dimen0-\wd0$\SavedStyle\hat{\phantom{#1}}$}{#1}}\vphantom{#1}%
  }%
}

\AddToHook{cmd/appendix/before}{%
  \crefalias{section}{appendix}%
  \crefalias{subsection}{appendix}
}

\AtBeginDocument{
  \renewcommand\Re{\operatorname{Re}}%
  \renewcommand\Im{\operatorname{Im}}%

}

\begin{document}

\title{Exploration of Fluxonium Parameters for Capacitive Cross-Resonance Gates}
\author{Eugene Y. Huang}
\email{e.y.huang@tudelft.nl}
\affiliation{QuTech and Kavli Institute of Nanoscience, Delft University of Technology, 2628 CJ, Delft, The Netherlands}
\author{Christian Kraglund Andersen}
\email{c.k.andersen@tudelft.nl}
\affiliation{QuTech and Kavli Institute of Nanoscience, Delft University of Technology, 2628 CJ, Delft, The Netherlands}
\date{\today}

\begin{abstract}
  We study the cross-resonance effect in capacitively-coupled fluxonium qubits
  and devise a simple formula for their maximum \emph{ZX} interaction strength.
  By going beyond the perturbative regime,
  we find that a CNOT gate can generally be realized in under 200\,ns with residual \emph{ZZ} limited to 50\,kHz, for fluxonium qubits with
  frequencies below 1\,GHz.
  Our analysis relies on a semi-analytical method:
  we first numerically diagonalize the Floquet Hamiltonian of the strongly-driven control qubit
  and then perturbatively incorporate the weak qubit-qubit coupling
  to obtain an effective Hamiltonian.
  We also derive frequency collision windows around harmful control--target and control--spectator transitions.
  For large fluxonium devices, we predict a collision-free yield
  that is considerably less sensitive to junction variability
  compared to transmons in the same layout.
  These results support the viability of
  an all-fluxonium cross-resonance architecture
  with only capacitive couplings.
\end{abstract}

\maketitle

\section{Introduction}

Superconducting qubits are an established platform for quantum information processing. To date, the most common architectures are based on
the transmon qubit~%
\cite{kochChargeinsensitiveQubitDesign2007},
with the largest devices integrating over a hundred such qubits
on a single chip
\cite{
abughanemSuperconductingQuantumComputers2025,%
mckayBenchmarkingQuantumProcessor2023,%
acharyaQuantumErrorCorrection2025%
}.
However, native gate errors
necessitate the use of error-correction protocols to build fault-tolerant logical qubits
from a large number of physical qubits
\cite{%
fowlerSurfaceCodesPractical2012,%
zhaoRealizationErrorCorrectingSurface2022,%
abdurakhimovTechnologyPerformanceBenchmarks2024,%
mckayBenchmarkingQuantumProcessor2023,%
acharyaQuantumErrorCorrection2025%
}.
Given current error rates,
running useful quantum algorithms
would require millions of physical qubits
\cite{%
gidneyHowFactor20482021,%
leeEvenMoreEfficient2021,%
dalzellQuantumAlgorithmsSurvey2023%
}
and it remains unclear how this could be realized with existing technology
\cite{%
bravyiFutureQuantumComputing2022,%
bravyiHighthresholdLowoverheadFaulttolerant2024,%
mohseniHowBuildQuantum2025%
}.
Further improvements to gate fidelity and architectural scalability
could significantly reduce the overhead for quantum error correction and advance the prospects for future fault-tolerant quantum computers,
motivating the continued research into alternative superconducting qubits
\cite{
gyenisExperimentalRealizationProtected2021,%
hyyppaUnimonQubit2022,%
menciaIntegerFluxoniumQubit2024,%
ardatiUsingBifluxonTunneling2024,%
rousseauEnhancingDissipativeCat2025,%
puttermanHardwareefficientQuantumError2025,%
levineDemonstratingLongCoherenceDualRail2024,%
huangLogicalMultiqubitEntanglement2026%
}.

The fluxonium qubit is one such alternative that has seen rapid adoption,
thanks in part to its compatibility with transmon-style circuit-QED designs
\cite{%
manucharyanFluxoniumSingleCooperPair2009,%
nguyenBlueprintHighPerformanceFluxonium2022%
},
\nocite{
nguyenBlueprintHighPerformanceFluxonium2022,%
sunCharacterizationLossMechanisms2023,%
zhuangNonMarkovianRelaxationSpectroscopy2026%
}
using the same capacitively coupled resonator for dispersive readout~%
\cite{%
zhuCircuitQEDFluxonium2013,%
stefanskiImprovedFluxoniumReadout2024%
}
and a single-axis drive for qubit control~%
\cite{
zwanenburgSinglequbitGatesRotatingwave2025,%
rowerSuppressingCounterRotatingErrors2024%
}.
Single- and two-qubit gates have recently been demonstrated on fluxonium qubits with errors below \num{e-3}
\cite{%
rowerSuppressingCounterRotatingErrors2024,%
somoroffMillisecondCoherenceSuperconducting2023,%
dingHighFidelityFrequencyFlexibleTwoQubit2023,%
lin24DaysStableCNOT2025%
},
matching the performance of the best two-qubit transmon devices
\cite{%
liRealizationHighFidelityCZ2024,%
marxer999FidelitySingleQubit2025%
}.
Additionally,
the highly nonlinear fluxonium potential
confers a large anharmonicity that
can mitigate leakage and, thus, improve error-correction performance
\cite{%
acharyaQuantumErrorCorrection2025,%
miaoOvercomingLeakageQuantum2023%
}.
Further benefits of the fluxonium qubit include
reduced dielectric loss
due to its smaller charge dipole moment
\cite{
nguyenBlueprintHighPerformanceFluxonium2022,%
sunCharacterizationLossMechanisms2023,%
zhuangNonMarkovianRelaxationSpectroscopy2026%
},
and suppressed Purcell decay into its readout resonator,
owing to its typically sub-gigahertz qubit frequency
\cite{nguyenHighCoherenceFluxoniumQubit2019,reedFastResetSuppressing2010}.

Existing fluxonium processors are, however, largely confined to two-qubit systems
while experimental and theoretical work continues to explore
the most effective strategies for scaling,
leveraging tunable interactions
\cite{
moskalenkoHighFidelityTwoqubit2022,%
zhangTunableInductiveCoupler2024,%
zhaoScalableFluxoniumQubit2025,
chakrabortyTunableSuperconductingQuantum2025%
},
auxiliary coupler states
\cite{
simakovCouplerMicrowaveActivatedControlledPhase2023,%
dingHighFidelityFrequencyFlexibleTwoQubit2023,%
singhFastMicrowavedrivenTwoqubit2026,%
rosenfeldHighFidelityTwoQubitGates2024,%
dimitrovCrossResonantGatesHybrid2025,%
kugutInteractionResilientScalableFluxonium2025,%
langeCrosstalkSuperconductingQubit2025%
}
or even hybrid architectures combining both fluxoniums and transmons
\cite{cianiMicrowaveactivatedGatesFluxonium2022,dimitrovCrossResonantGatesHybrid2025}.
On the other hand,
the most straightforward two-qubit gate schemes require only fixed direct couplings,
with static interactions activated by microwave drives or baseband flux pulses
\cite{
doganTwoFluxoniumCrossResonanceGate2023,%
lin24DaysStableCNOT2025,%
baoFluxoniumAlternativeQubit2022,%
maNativeApproachControlledZ2024,%
morenoCNOTGatesInductively2025,%
ficheuxFastLogicSlow2021,%
xiongArbitraryControlledphaseGate2022,%
nesterovProposalEntanglingGates2021%
},
\nocite{
rigettiProtocolUniversalGates2005,%
sheldonProcedureSystematicallyTuning2016,%
heyaCrossCrossResonanceGate2021,%
weiNativeTwoQubitGates2024,%
nguyenProgrammableHeisenbergInteractions2024%
}
a strategy also favored in earlier generations of transmon processors
\cite{
chowImplementingStrandScalable2014,%
barendsSuperconductingQuantumCircuits2014,%
krinnerRealizingRepeatedQuantum2022,%
marquesLogicalqubitOperationsErrordetecting2022,%
acharyaIntegrationThroughSapphireSubstrate2025,%
wardEchoCrossResonance2026%
}.
We consider
flux pulsing undesirable as it
displaces the fluxonium from its half-flux bias point
(where its spectrum is symmetric),
leaving it sensitive to dephasing from flux noise
\cite{maNativeApproachControlledZ2024,ateshianTemperatureMagneticFieldDependence2025}.
Moreover,
baseband flux pulses are often severely distorted as they propagate towards the qubit,
necessitating careful characterization and precompensation
\cite{rolTimedomainCharacterizationCorrection2020,hellingsCalibratingMagneticFlux2025}.
Drive-activated interactions,
on the other hand,
are relatively easy to calibrate
and preserve the qubits at their operational bias point.

Among the possible drive-activated gates between qubits with fixed couplings,
the cross-resonance (CR) gate is, arguably, the simplest in terms of implementation
\cite{
rigettiProtocolUniversalGates2005,%
sheldonProcedureSystematicallyTuning2016,%
nesterovProposalEntanglingGates2021,%
heyaCrossCrossResonanceGate2021,%
ficheuxFastLogicSlow2021,%
xiongArbitraryControlledphaseGate2022,%
weiNativeTwoQubitGates2024,%
nguyenProgrammableHeisenbergInteractions2024%
}.
In the CR gate,
both qubits are driven at the resonance frequency of one qubit in the pair,
reusing the single-axis drive lines shared with single-qubit control.
This hardware-efficient architecture was used to realize
a record \num{1121} transmon qubits
on a single chip
\cite{castelvecchiIBMReleasesFirstever2023}.
However,
the small anharmonicity inherent to transmons lead eventually to unavoidable
frequency collisions,
impeding further scaling of this architecture
\nocite{
chamberlandTopologicalSubsystemCodes2020,%
hertzbergLaserannealingJosephsonJunctions2021,%
morvanOptimizingFrequencyAllocation2022%
}
\cite{%
hertzbergLaserannealingJosephsonJunctions2021,%
morvanOptimizingFrequencyAllocation2022%
}.
Given the fluxonium qubit's
significantly higher anharmonicity,
a natural question to ask is whether the CR architecture
is implementable with fluxoniums
and how large such devices could scale.

The CR gate can be applied between fluxonium qubits coupled either
capacitively or inductively
\cite{doganTwoFluxoniumCrossResonanceGate2023,lin24DaysStableCNOT2025}.
While inductive coupling offers a larger on-off ratio for the effective two-qubit interaction,
it is difficult to engineer in a scalable manner due to the small size of typical flux loops
\cite{%
nguyenBlueprintHighPerformanceFluxonium2022,%
lin24DaysStableCNOT2025,%
morenoCNOTGatesInductively2025%
}.
In comparison,
the capacitor pads of a fluxonium,
while smaller than those of transmons,
can easily interface with coupling elements
\cite{dingHighFidelityFrequencyFlexibleTwoQubit2023,singhFastMicrowavedrivenTwoqubit2026}.
For this reason, we focus on capacitive couplings,
despite their theoretical drawbacks,
with the aim of exploring the potential of
the \emph{simplest} multi-fluxonium design.

The rest of the paper is organized as follows.
In
\cref{sec:model},
we introduce the model Hamiltonian
that forms the basis of our analysis,
consisting of two capacitively coupled fluxonium qubits
with one qubit driven and the other idle.
In \cref{sec:parameters-for-fast-gates},
we determine the maximum speed at which two-qubit gates can be performed
while keeping residual always-on interactions below an acceptable threshold.
Our semi-analytical approach
allows us to investigate the dynamics under strong CR drives beyond the perturbative regime,
where we observe a universal saturation in the effective two-qubit interaction strength.
This observation enables us to empirically derive a simple yet accurate formula
for the minimum time required to implement a CNOT gate, valid for strong drive amplitudes.
In \cref{sec:frequency-collisions},
we analyze the primary sources of unintentional resonances,
including from spectator qubits,
that can induce coherent errors during the CR gate
and determine the minimum detuning required to suppress them.
These detuning windows are then used in Monte Carlo simulations
to calculate the probability of zero frequency collisions in a qubit lattice
as a function of system size.

\section{Model}
\label{sec:model}

\begin{figure}
  \begin{subcaptiongroup}
    \includegraphics{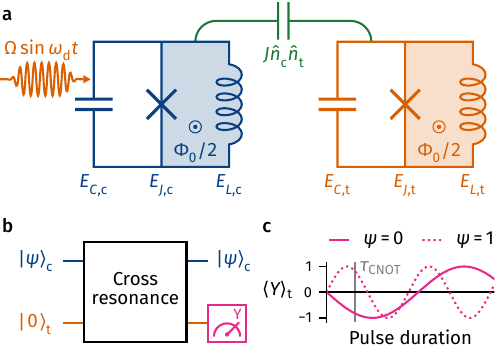}
    \phantomcaption\label{fig:cross-resonance-circuit}
    \phantomcaption\label{fig:cross-resonance-example-calibration}
    \phantomcaption\label{fig:cross-resonance-example-graph}
  \end{subcaptiongroup}
  \caption{%
    Overview of the cross-resonance effect between two capacitively-coupled fluxoniums.
    (\subref{fig:cross-resonance-circuit})
    Circuit diagram showing the control qubit
    (left, \textbf{blue})
    driven by a pulse modulated at the frequency of the target qubit
    (right, \textbf{orange})
    with peak amplitude $\Omega$.
    Both qubits are half-flux ($\Phi_0/2$) biased and have individual charging energies $E_{C,q}$, Josephson energies $E_{J,q}$ and inductive energies $E_{L,q}$,
    and are coupled via their charge operators $\hat n_q$ with coupling coefficient $J$.
    (\subref{fig:cross-resonance-example-calibration})
    A CNOT gate is calibrated by first preparing the control qubit
    in $|\psi\rangle_\contlabel=|0\rangle_\contlabel$
    or $|1\rangle_\contlabel$
    and the target qubit in $|0\rangle_\targlabel$,
    and then applying the cross-resonance pulse at a fixed amplitude for varying durations,
    followed by a measurement of the target qubit along the $Y$ axis.
    (\subref{fig:cross-resonance-example-graph})
    The control qubit ideally remains in its initial eigenstate
    while the target qubit rotates about the $X$ axis (in the $ZY$ plane)
    at an angular rate conditioned on the control-qubit eigenstate,
    $|\psi\rangle_\contlabel$.
    When the pulse duration is $\tau_\CNOTlabel$,
    a net phase difference of $\pi$ accumulates
    between the two conditional rotations, realizing the CNOT gate
    up to single-qubit phases and an additional target-qubit rotation.
  }
  \label{fig:cross-resonance-schematic}
\end{figure}

The CR effect is a drive activated
two-qubit interaction between two transversely coupled qubits.
For superconducting qubits, this transverse coupling can be either
capacitive
or
inductive.
Specifically, one of the qubits, called the \emph{control}, is driven at the frequency of its coupled neighbor, called the \emph{target}.
The resulting interaction rotates the target qubit with a Rabi rate conditional on the state of the control qubit,
naturally generating the CNOT gate up to single-qubit gates.

We are interested in the case of pure, direct capacitive coupling, which we model using the Hamiltonian
($\hbar=1$)
\begin{equation}
  \hat H(\Omega, t)
  =
  \hat H_\contlabel(\Omega, t)
  + \hat H_\targlabel
  + J\,\hat n_\contlabel \otimes \hat n_\targlabel
  \,,
  \label{eq:system-hamiltonian}
\end{equation}
where
$\hat H_\contlabel(\Omega, t)$
and
$\hat H_\targlabel$
act on, respectively, the control and target qubits,
$\Omega$ is the CR drive amplitude,
$J$ is the capacitive coupling strength
and $\hat n_q$ is the charge operator of qubit $q$.
The individual Hamiltonians are
\begin{align}
  \hat H_\contlabel(\Omega, t)
  &\coloneqq
  \hat H_\contlabel^\barelabel
  +
  \Omega
  \sin(\omega_\drivelabel t)
  \hat n_\contlabel
  \label{eq:control-hamiltonian}
  \,,\\
  \hat H_\targlabel
  &\coloneqq
  \hat H_\targlabel^\barelabel
  \,,
\end{align}
where
$
  \hat H_q^\barelabel
  =
  \sum_j
  E_q^j
  \ketbra|j><j|
$
is the diagonalized fluxonium Hamiltonian.
We refer to the states
$\{|j\rangle\}_{j\in\NN_0}$
(where $\NN_0=\{0,1,2,\ldots\}$)
as the bare basis,
with corresponding energies
$\{E_q^j\}_{j\in\NN_0}$ for qubit $q$.
We allow the drive amplitude $\Omega$ to be time-\emph{dependent},
and write it as $\Omega(t)$ to indicate this explicitly when needed.
No generality is lost by assuming the control qubit is capacitively driven.
We show in
\cref{sec:capacitive-inductive-drive-symmetry}
that,
in the context of this work,
capacitive and inductive driving are equivalent up to a rescaling of drive amplitude.

The energies
$E_q^j$
and matrix elements
$\langle i|\hat n_q|j \rangle$
follow from the fluxonium Hamiltonian
\cite{manucharyanFluxoniumSingleCooperPair2009,smithQuantizationInductivelyShunted2016}
\begin{equation}
  \hat H_q^\barelabel
  \coloneqq
  4 E_{C,q} \hat n_q^2
  + \frac{1}{2}E_{L,q} \hat\varphi_q^2
  - E_{J,q} \cos(\hat\varphi_q - \varphi_{\text{ext},q})
  \,,
  \label{eq:fluxonium-hamiltonian}
\end{equation}
where $\hat\varphi_q$ is the phase operator conjugate to $\hat n_q$.
The fluxonium qubit has three design parameters:
its charging energy $E_{C,q}$,
inductive energy $E_{L,q}$
and Josephson energy $E_{J,q}$.
Its high-coherence sweet spot
is achieved when the flux loop is threaded by half a flux quantum, i.e., $\varphi_{\text{ext}, q} = \pi$.
Hence,
the full design space of the two-qubit circuit is seven-dimensional,
comprising
$(\{E_{J,q}, E_{C,q}, E_{L,q}\}, J)$,
see \cref{fig:cross-resonance-circuit}.
The coupling strength $J$ is indirectly determined by requiring
residual $ZZ$ interactions to be equal to, or below, a designed threshold.
The drive amplitude $\Omega$ should ideally be calibrated to minimize the CNOT gate duration
so it is not considered an independent design parameter.

\subsection{Effective Hamiltonian}
\label{sec:main-effective-hamiltonian}

Let
$|\widetilde{\imath\jmath}\rangle$
for $i,j\in\NN_0$
be the dressed eigenstates of the \emph{undriven}
(but coupled)
Hamiltonian $\hat H(\Omega=0)$
with eigenenergies $\widetilde E^{ij}$.
Here, the state $|\widetilde{\imath\jmath}\rangle$ is defined to be
the dressed eigenstate closest to the state $\ket|ij>$,
found by maximizing the sum of squared overlaps
$
  \sum_{ij}
  \left|
  \langle ij|\widetilde{\imath\jmath}\rangle
  \right|^2
$.
The two-qubit computational subspace is spanned by
$|\widetilde{00}\rangle$,
$|\widetilde{01}\rangle$,
$|\widetilde{10}\rangle$ and
$|\widetilde{11}\rangle$,
and we define the dressed two-qubit Pauli operators as
\begin{equation}
  \hat{\widetilde{AB}}
  \coloneqq
  \sum_{ijk\ell}
  \ %
  |\widetilde{\imath\jmath}\rangle
  \langle ij|
  \hat A_q \hat B_q
  |k\ell\rangle
  \langle \widetilde{k\ell}|
\end{equation}
for $A,B\in\{X,Y,Z,\Id\}$,
where $\hat X_q$, $\hat Y_q$, $\hat Z_q$ and $\hat \Id_q$
are the Pauli and identity operators
acting on qubit $q$.

It is conventional to track the qubits in the rotating frame transformed by the change-of-basis unitary
\begin{equation}
  \hat R(t)
  \coloneqq
  \exp
  \ab[
    -it
    \ab(
      \frac{\bar{\omega}_\contlabel}{2}
      \hat{\widetilde{ZI}}
      + \frac{\bar{\omega}_\targlabel}{2}
      \hat{\widetilde{IZ}}
    )
  ]
  \,,
  \label{eq:rotating-frame-average-qubit-frequency}
\end{equation}
where
\begin{subequations}
\begin{align}
  \bar\omega_\contlabel
  &\coloneqq
  \frac{1}{2}
  \ab(
    \widetilde{E}^{10}
    - \widetilde{E}^{00}
    + \widetilde{E}^{11}
    - \widetilde{E}^{01}
  )
  \,,\\
  \bar\omega_\targlabel
  &\coloneqq
  \frac{1}{2}
  \ab(
    \widetilde{E}^{01}
    - \widetilde{E}^{00}
    + \widetilde{E}^{11}
    - \widetilde{E}^{10}
  )
\end{align}
\end{subequations}
are the averaged qubit frequencies.
In the rotating frame, the idle Hamiltonian
(in the computational subspace)
transforms into
\begin{align}
\hat H_\CRlabel(\Omega=0)
\coloneqq{}&
\hat R(t)
[
  \hat H(\Omega=0)
  - i\partial_t
]
\hat R^\dagger(t)
\\
={}&
\frac{1}{4}
\mu_{ZZ}(\Omega=0)
\hat{\widetilde{ZZ}}
+ \text{const.}
\end{align}
where
\begin{equation}
  \mu_{ZZ}(\Omega=0)
  \coloneqq
  (
    \widetilde{E}^{11}
    - \widetilde{E}^{10}
  )
  - (
    \widetilde{E}^{01}
    - \widetilde{E}^{00}
  )
\end{equation}
is the so-called residual $ZZ$ rate.
This is an undesired always-on interaction that entangles connected qubits when idling or performing other gates.
It is an unavoidable nuisance in a CR architecture with only direct capacitive couplings
\cite{%
kandalaDemonstrationHighFidelityCnot2021,%
weiHamiltonianEngineeringMulticolor2022,%
nguyenBlueprintHighPerformanceFluxonium2022,%
lin24DaysStableCNOT2025,%
linVerifyingAnalogyTransversely2025%
}.

For a nonzero but slowly varying $\Omega$,
and $\omega_\drivelabel = \bar\omega_t$,
the rotating frame Hamiltonian can be effectively written as
(see \cref{sec:effective-cr-hamiltonian} for the derivation)
\begin{equation}
  \begin{split}
    \hat H_\CRlabel(\Omega)
    ={}&
    \phantom{{}+{}}
    \frac{1}{2}
    \mu_{ZX}(\Omega)
    \mathmakebox[\widthof{$\hat{\widetilde{ZX}}$}]{\hat{\widetilde{
      \mathmakebox[\widthof{$ZI$}]{ZX}
    }}}
    +
    \frac{1}{2}
    \mu_{IX}(\Omega)
    \mathmakebox[\widthof{$\hat{\widetilde{IX}}$}]{\hat{\widetilde{
      \mathmakebox[\widthof{$ZI$}]{IX}
    }}}
    \\
    &+
    \frac{1}{2}
    \mathmakebox[\widthof{$\mu_{ZX}(\Omega)$}]{\mu_{ZI}(\Omega)}
    \mathmakebox[\widthof{$\hat{\widetilde{ZX}}$}]{\hat{\widetilde{ZI}}}
    +
    \frac{1}{2}
    \mathmakebox[\widthof{$\mu_{IX}(\Omega)$}]{\mu_{IZ}(\Omega)}
    \mathmakebox[\widthof{$\hat{\widetilde{IX}}$}]{\hat{\widetilde{IZ}}}
    \\
    &+
    \frac{1}{4}
    \mathmakebox[\widthof{$\mu_{ZX}(\Omega)$}]{\mu_{ZZ}(\Omega)}
    \mathmakebox[\widthof{$\hat{\widetilde{ZX}}$}]{\hat{\widetilde{
      \mathmakebox[\widthof{$ZI$}]{ZZ}
    }}}
    \,.
  \end{split}
  \label{eq:effective-hamiltonian-main}
\end{equation}
The coefficients $\mu_{AB}$ can be calculated perturbatively in $J$
in terms of matrix elements in the Floquet mode basis
of the driven control qubit,
$\{|\Phi_j(\Omega, t)\}_{j\in\NN_0}$.
These modes satisfy the eigenvalue equation
\cite{shirleySolutionSchrodingerEquation1965,grifoniDrivenQuantumTunneling1998}
\begin{gather}
  \ab(
  \hat H_\contlabel(\Omega, t)
  - i\partial_t
  )
  |\Phi_j(\Omega, t)\rangle
  =
  \epsilon_j(\Omega)|\Phi_j(\Omega, t)\rangle
  \,,
  \label{eq:floquet-eigenvalue-equation}
\end{gather}
with boundary condition
$
|\Phi_j(\Omega,0)\rangle
=
|\Phi_j(\Omega,\frac{2\pi}{\omega_\drivelabel})\rangle
$.
The eigenvalues $\{\epsilon_j(\Omega)\}_{j\in\NN_0}$ are called the Floquet quasienergies.
We assign the indices such that
$|\Phi_j(\Omega, t)\rangle$
connects to the single-qubit eigenstate
$|j\rangle$
as $\Omega\to0$.
Because Floquet modes are time-dependent and coperiodic with the drive,
the transition moments between Floquet modes have Fourier series expansions:
\begin{equation}
  \langle \Phi_j(\Omega, t)|
  \hat O_\contlabel
  |\Phi_i(\Omega, t) \rangle
  \eqqcolon
  \sum_{k\in\ZZ}
  O^{[k]ji}_\contlabel(\Omega)
  e^{
    ik\omega_\drivelabel t
  }
  \,,
  \label{eq:fourier-coefficients-wrt-floquet-modes}
\end{equation}
where $\hat O_\contlabel$ is either
the charge $\hat n_\contlabel$
or phase  operator $\hat \varphi_\contlabel$.

A CNOT gate is realized with the $ZX$ term
of \cref{eq:effective-hamiltonian-main},
which rotates the target qubit
at an angular rate of $|\mu_{ZX}|$
and with an orientation that depends on the control qubit state.
By timing the CR pulse,
one can accumulate a rotation difference of exactly $\pi$,
which is locally equivalent to the CNOT gate
(see
\cref{%
fig:cross-resonance-example-calibration,%
fig:cross-resonance-example-graph%
}).
Assuming $J$ can be treated perturbatively,
the $ZX$ strength is
\begin{equation}
  \mu_{ZX}(\Omega)
  =
  -J
  \Delta p(\Omega)
  n^{10}_\contlabel
  n^{10}_\targlabel
  + O(J^3)
  \,,
  \label{eq:zx-coefficient-main}
\end{equation}
where
$n_q^{ji}\coloneqq \langle j|\hat n_q|i\rangle$
are the charge matrix elements in the \emph{bare} eigenbasis and
\begin{equation}
  \Delta p(\Omega)
  \coloneqq
  \frac{
    n^{[-1]11}_\contlabel(\Omega)
    - n^{[-1]00}_\contlabel(\Omega)
  }{
    n_\contlabel^{10}
  }
  \label{eq:conditional-polarization-main}
\end{equation}
is the \emph{conditional polarization},
normalized by the charge dipole moment.
The numerator
involves the Fourier coefficients of
\cref{eq:fourier-coefficients-wrt-floquet-modes}
and represents the difference between the drive-induced mean charge polarizations of the
$1$ and $0$ states, oscillating at the
$-\omega_\drivelabel$
harmonic.
\Cref{eq:zx-coefficient-main} tells us that
the conditional polarization
$\Delta p$
is directly proportional to the effective $ZX$ strength.
From this perspective, the mechanism of the CR effect is intuitive.
Driving the control qubit off-resonantly forces charge oscillations with an \emph{initial-state dependent} amplitude;
and since the control and target qubits are coupled via their charge operators,
the target is resonantly driven by the oscillating control-qubit polarization.

We note that the unwanted $\mu_{ZZ}$ coefficient scales as $J^2$ whereas $\mu_{ZX}$ is first order in $J$.
Hence,
the CR architecture is practicable because $\mu_{ZZ}$ can be made small
while $\mu_{ZX}$ remains sufficient to perform a CNOT gate within a reasonable time.
The other coefficients of
\cref{eq:effective-hamiltonian-main}
do not significantly hinder gate fidelity either.
The unconditional drive $\mu_{IX}$ can take arbitrary values if we add an additional in-phase drive directly on the target qubit
\cite{
  nesterovCnotGatesFluxonium2022,%
  sundaresanReducingUnitarySpectator2020,%
  sheldonProcedureSystematicallyTuning2016%
}.
The dynamic target-qubit detuning $\mu_{IZ}$ can be mitigated with either
a constant drive detuning,
resonant target driving
\cite{sundaresanReducingUnitarySpectator2020},
or an echo sequence
\cite{sheldonProcedureSystematicallyTuning2016}.
The control-qubit detuning $\mu_{ZI}$
commutes with the other interactions
and can therefore be corrected with a virtual-$Z$ before or after the CNOT gate
\cite{mckayEfficientGatesQuantum2017}.

\subsection{Drive-induced decoherence}
\label{sec:model-master-equation}

Maximizing the $ZX$ interaction during the CR gate
requires strongly driving the control qubit,
which can significantly alter its coherence properties from its undriven behavior
\cite{%
  mundadaFloquetEngineeredEnhancementCoherence2020,%
  huangEngineeringDynamicalSweet2021,%
  nguyenProgrammableHeisenbergInteractions2024,%
  wangQuantumControlNoise2024%
}.
Let $\rho$ be the density matrix of the two-fluxonium system.
When weakly coupled to an environment, the density matrix evolves according to the generic master equation
\cite{huangEngineeringDynamicalSweet2021}
\begin{equation}
  \frac{d\rho}{dt}
  =
  \sum_n
  \Gamma_n
  \mathbb{D}
  [\hat L_n]
  \rho
\end{equation}
where
$
  \mathbb{D}
  [\hat L]
  \rho
  =
  \hat L\rho \hat L^\dagger
  -
  (
  \hat L^\dagger \hat L \rho
  +
  \rho\hat L^\dagger \hat L
  )/2
$
is the Lindblad superoperator and $\hat L_n$ are the jump operators through which the system couples with the environment, with corresponding rates $\Gamma_n$.

In the uncoupled case, $J=0$, we consider decay, excitation and dephasing of the target qubit,
described by the jump operators
\begin{align}
  \hat L_\targlabel^{1\to0}
  &=
  \ketbra|0><1|
  \,,\\
  \hat L_\targlabel^{0\to1}
  &=
  \ketbra|1><0|
  \,,\\
  \hat L_\targlabel^\phi
  &=
  \frac{
    \ketbra|0><0|
    - \ketbra|1><1|
  }{2}
  \,,
\end{align}
with associated rates
$\Gamma_\targlabel^{1\to0}$,
$\Gamma_\targlabel^{0\to1}$
and $\Gamma_\targlabel^\phi$ respectively.
Heating into higher excited states is negligible at millikelvin temperatures.
For the control qubit, its jump operators are expressed in terms of the Floquet modes (evaluated at $t=0$) as
\begin{align}
  \hat L_\contlabel^{i\to j}(\Omega)
  &=
  \ketbra|\Phi_j(\Omega)><\Phi_i(\Omega)|
  \label{eq:control-fluxonium-jump-operator}
  \,,\\
  \hat L_\contlabel^\phi(\Omega)
  &=
  \frac{
    \ketbra|\Phi_0(\Omega)><\Phi_0(\Omega)|
    - \ketbra|\Phi_1(\Omega)><\Phi_1(\Omega)|
  }{2}
  \,,
  \label{eq:control-fluxonium-dephasing-operator}
\end{align}
with associated rates
$\Gamma_\contlabel^{i\to j}(\Omega)$
and $\Gamma_\contlabel^\phi(\Omega)$,
respectively.
Because the control qubit is strongly driven, heating into higher excited states
cannot be ignored even in a zero-temperature environment.

A CNOT gate implemented using a square pulse with amplitude
$\Omega$
has the duration
$t_g(\Omega)=\tfrac12 \pi/|\mu_{ZX}(\Omega)|$;
therefore,
its average incoherent gate infidelity
$\eincoh(\Omega)$
can be approximated by
\cite{abadImpactDecoherenceFidelity2025}
\begin{align}
  \eincoh
  &=
  \eincoh_\contlabel
  + \eincoh_\targlabel
  \,,\\
  \eincoh_\contlabel
  &=
  \frac{2}{5}
  \bigl(
    \Gamma_\contlabel^1
    + \Gamma_\contlabel^\phi
  \bigr)
  t_g
  + \frac{1}{2}
  \Gamma_\contlabel^\ell
  t_g
  \,,\\
  \eincoh_\targlabel
  &=
  \frac{2}{5}
  \bigl(
    \Gamma_\targlabel^1
    + \Gamma_\targlabel^\phi
  \bigr)
  t_g
  \,,
\end{align}
where
$\Gamma_q^1 \coloneqq \Gamma_q^{0\to1} + \Gamma_q^{1\to0}$
is qubit $q$'s relaxation rate
and
$
\Gamma_\contlabel^\ell
\coloneqq
\sum_{j=2}^\infty
\bigl(
  \Gamma_\contlabel^{0\to j}
  + \Gamma_\contlabel^{1\to j}
\bigr)
$
is the control qubit's leakage rate.
In a very simple model where the control qubit is susceptible only to dielectric loss
into a zero-temperature bath,
its dissipation rates are given by ($\Omega$-dependence omitted)
\cite{mundadaFloquetEngineeredEnhancementCoherence2020,huangEngineeringDynamicalSweet2021}
\begin{align}
  \Gamma_\contlabel^{i\to j}
  &=
  \sum_{k\in\ZZ}
  \bigl|
    n_\contlabel^{[k]ji}
  \bigr|^2
  S(k\omega_\drivelabel + \epsilon_j - \epsilon_i)
  \label{eq:floquet-relaxation-rate}
  \,,\\
  \Gamma_\contlabel^\phi
  &=
  \sum_{k\in\ZZ}
  \frac{1}{2}
  \bigl|
    n_\contlabel^{[k]11}
    -
    n_\contlabel^{[k]00}
  \bigr|^2
  S(k\omega_\drivelabel)
  \label{eq:floquet-dephasing-rate}
  \,,
\end{align}
where
\begin{equation}
  S(\omega)
  =
  \begin{cases}
    16 E_{C,\contlabel} \tan\delta
    &
    \text{for } \omega > 0
    \,,
    \\
    0
    &
    \text{for } \omega \leq 0
  \end{cases}
  \label{eq:noise-spectral-density}
\end{equation}
is the noise spectral density
representing the lossy capacitor
with loss tangent $\tan\delta$ at \emph{zero temperature}.
We use this idealized noise model to regularize the problem of finding an optimal drive amplitude
since $|\mu_{ZX}(\Omega)|$ can become very large when the Floquet qubit modes hybridize with the plasmon modes.
These modes are highly susceptible to dielectric losses
and, hence, should be avoided in a real implementation.
Minimizing $\eincoh_\contlabel(\Omega)$
rather than maximizing $|\mu_{ZX}(\Omega)|$
is thus the better suited objective.
Nevertheless, both goals are equivalent for small drive amplitudes.
We emphasize that the errors estimated with this simple model
should \emph{not} be interpreted as a realistic incoherent error budget.

\section{Parameters for fast gates}
\label{sec:parameters-for-fast-gates}

The objective of this section is to find design parameters that maximize the $ZX$ interaction given a fixed $ZZ$ budget.
In Subsection~\ref{sec:control-qubit-conditional-polarizability},
we study the maximum conditional polarization $\Delta p$ for the control qubit only.
In Subsection~\ref{sec:cnot-duration-landscape},
we explore the minimum CNOT gate time landscape
as a function of the control and target fluxonium parameters,
at a fixed $ZZ$ budget.
Finally, in Subsection~\ref{sec:time-domain-simulation},
we present time-domain simulations that numerically verify the predicted CNOT speed.

We primarily consider only a fixed
$E_{L,q}/E_{J,q}$ ratio of $0.25$
to reduce the dimensionality of the optimization problem.
There is an \emph{approximate} one-dimensional symmetry in fluxonium parameters
where the eigenstates
for other 
$E_{L,q}/E_{J,q}$ ratios between $0.1$ and $0.5$
are approximately related by a rescaling of coordinates,
see \cref{sec:double-well-renormalization}.
In essence,
we expect the $ZX$ and $ZZ$ rates for two different parameter sets to be roughly equal
provided the control and target qubits across the two configurations have
similar frequency and anharmonicity.

\subsection{Control-qubit conditional polarizability}
\label{sec:control-qubit-conditional-polarizability}

\begin{figure*}[t]
  \begin{subcaptiongroup}
    \includegraphics{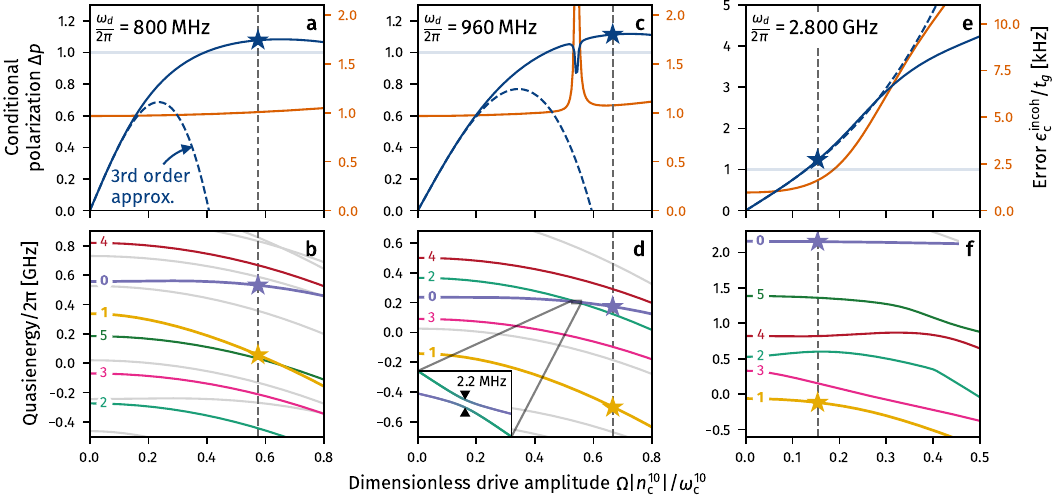}
    \phantomcaption\label{fig:optimal-drive-amplitude-delta-p-typical}
    \phantomcaption\label{fig:optimal-drive-amplitude-spectrum-typical}
    \phantomcaption\label{fig:optimal-drive-amplitude-delta-p-small-ac}
    \phantomcaption\label{fig:optimal-drive-amplitude-spectrum-small-ac}
    \phantomcaption\label{fig:optimal-drive-amplitude-delta-p-large-ac}
    \phantomcaption\label{fig:optimal-drive-amplitude-spectrum-large-ac}
  \end{subcaptiongroup}
  \caption{%
    (%
    \subref{fig:optimal-drive-amplitude-delta-p-typical},
    \subref{fig:optimal-drive-amplitude-delta-p-small-ac},
    \subref{fig:optimal-drive-amplitude-delta-p-large-ac}%
    )
    Conditional polarization
    $\Delta p$ (\textbf{blue}) and
    incoherent error rate
    $\eincoh_\contlabel/t_g$
    (\textbf{orange})
    as a function of the dimensionless drive amplitude
    $\Omega|n_\contlabel^{10}|/\omega_\contlabel^{10}$
    for representative drive frequencies (top left label).
    The numerically obtained $\Delta p$ (\textbf{solid blue})
    is compared with the
    third-order perturbative result (\textbf{dashed blue}).
    (%
    \subref{fig:optimal-drive-amplitude-spectrum-typical},~%
    \subref{fig:optimal-drive-amplitude-spectrum-small-ac},~%
    \subref{fig:optimal-drive-amplitude-spectrum-large-ac}%
    )~%
    Floquet quasienergy spectrum of the off-resonantly-driven control fluxonium,
    also as a function of
    $\Omega|n_\contlabel^{10}|/\omega_\contlabel^{10}$
    with the same drive frequency as the top panel.
    Levels are labeled adiabatically, ignoring ``small'' avoided crossings, see main text.
    The \textbf{inset} of
    (\subref{fig:optimal-drive-amplitude-spectrum-small-ac})
    shows the avoided crossing between the 0 and 2 modes,
    where the color of each level is based on its cycle-averaged energy.
    The optimal amplitude
    (\textbf{star} $\bigstar$)
    minimizes
    $\eincoh_\contlabel \propto \eincoh_\contlabel/(t_g \Delta p)$.
    The control fluxonium has parameters
    $E_{J,\contlabel}/2\pi = \qty{4}{\GHz}$,
    $E_{C,\contlabel}/2\pi = \qty{1}{\GHz}$
    and $E_{L,\contlabel}/2\pi = \qty{1}{\GHz}$.
  }
  \label{fig:optimal-drive-amplitude}
\end{figure*}

We first focus on a typical fluxonium with parameters:
$E_{J,\contlabel}/2\pi = \qty{4}{\GHz}$,
$E_{C,\contlabel}/2\pi = \qty{1}{\GHz}$ and
$E_{L,\contlabel}/2\pi = \qty{1}{\GHz}$,
which yield a qubit frequency of
$\omega_\contlabel^{10}/{2\pi} \approx \qty{582}{\MHz}$.
As shown in \cref{fig:optimal-drive-amplitude-delta-p-typical},
we calculate the conditional polarization $\Delta p$ at a fixed drive frequency $\omega_\drivelabel$. Here, the drive amplitude $\Omega$ is normalized by
$\omega_\contlabel^{10}/|n_\contlabel^{10}|$
to make a clear comparison to typical amplitudes used in single-qubit gates.
For example,
$\Omega|n_\contlabel^{10}|/\omega_\contlabel^{10} = 0.5$
generates a $\pi$ rotation in one Larmor period for a resonant capacitive drive.
Therefore,
typically
$\Omega|n_\contlabel^{10}|/\omega_\contlabel^{10} \sim 0.1$
when performing single-qubit gates
\cite{%
rowerSuppressingCounterRotatingErrors2024,%
zwanenburgSinglequbitGatesRotatingwave2025,%
zhangUniversalFastFluxControl2021%
}.
Given a drive frequency $\omega_\drivelabel$,
we determine the optimal drive amplitude $\Omega^\star$
(marked by the star in
\cref{fig:optimal-drive-amplitude-delta-p-typical})
such that the incoherent CNOT error contribution
$\eincoh_\contlabel(\Omega^\star)$
is minimized.
Moving on, we define $\Delta p^\star \coloneqq \Delta p(\Omega^\star)$.
Note the dimensionless optimal amplitude
$\Omega^\star|n_\contlabel^{10}|/\omega_\contlabel^{10}$
and conditional polarization $\Delta p^\star$ are constant under uniform rescaling of
$E_{C,\contlabel}$,
$E_{L,\contlabel}$,
$E_{J,\contlabel}$
and $\omega_\drivelabel$.

\Cref{fig:optimal-drive-amplitude-delta-p-typical}
exemplifies the simplest case, for a drive frequency of
$\omega_\drivelabel/2\pi = \qty{800}{\MHz}$, where
$\Delta p$ increases with $\Omega$ at first,
but eventually reaches a local maximum.
In other words,
increasing $\Omega$ beyond this point
will \emph{increase} the duration necessary to generate the CNOT gate.
This saturation characteristic is also present in transmons and C-shunted flux qubits
\cite{tripathiOperationIntrinsicError2019,malekakhlaghFirstprinciplesAnalysisCrossresonance2020,wareCrossresonanceInteractionsSuperconducting2019,chowSimpleAllMicrowaveEntangling2011}.
The dashed line in \cref{fig:optimal-drive-amplitude-delta-p-typical}
shows the value of $\Delta p$
calculated from perturbatively obtained Floquet modes, up to $O(\Omega^3)$.
Although initially agreeing with the numerical result,
the perturbative model fails to accurately predict both the maximum $\Delta p$ or the value of $\Omega$ at which it occurs.
The solid orange line in
\cref{fig:optimal-drive-amplitude-delta-p-typical}
is the incoherent error contribution from the control qubit due to dielectric loss
into a zero-temperature bath,
as defined in \cref{sec:model-master-equation},
assuming a dielectric loss tangent of $\tan\delta = \num{e-6}$
\nocite{
  pappasAlternatingbiasAssistedAnnealing2024,%
  hertzbergLaserannealingJosephsonJunctions2021,%
  zhangHighperformanceSuperconductingQuantum2022,%
  muthusubramanianWaferscaleUniformityDolanbridge2024,%
  wangHighcoherenceFluxoniumQubits2025,%
  pishchimovaImprovingJosephsonJunction2023,%
  osmanSimplifiedJosephsonjunctionFabrication2021%
}
\cite{nguyenHighCoherenceFluxoniumQubit2019,somoroffMillisecondCoherenceSuperconducting2023,wangHighcoherenceFluxoniumQubits2025}.
It is divided by the total gate duration and expressed as a rate in \unit{\kHz}
since $t_g$ depends not only on $\Delta p$
but also linearly on $J$ and $n_\targlabel^{10}$.
To be more precise, the optimal amplitude is found by
\begin{equation}
  \Omega^\star
  \coloneqq
  \argmin_\Omega
  \frac{
    \eincoh_\contlabel(\Omega)
  }{
    t_g(\Omega)\Delta p(\Omega)
  }
  \,,
\end{equation}
although this is equivalent to minimizing
$\eincoh_\contlabel(\Omega)$
since $t_g(\Omega) \propto 1/\Delta p(\Omega)$.

As previously mentioned,
we evaluate the incoherent errors in the basis of the Floquet modes,
thus, it is insightful to also consider the quasienergy spectrum,
see \cref{fig:optimal-drive-amplitude-spectrum-typical}.
We label the quasienergy levels, similar to
Ref.~\cite{nesterovMeasurementinducedStateTransitions2024},
whereby
$\Omega$
is increased in small steps
and the newly calculated Floquet modes are labeled according to
their maximum cycle-averaged overlap with the previous modes.
This labeling is not exactly adiabatic,
as apparent from the crossing between modes $1$ and $5$.
Here, the gap size is numerically computed to be very small (less than one hertz).
Thus, for all intents and purposes, this is a level crossing rather than an anti-crossing.
In general, our state identification algorithm ignores quasienergy gaps less than
\qty{2\pi\times 10}{\MHz}
by lifting the near degeneracy using
the cycle-averaged energy quantum number
\cite{leDefiningWellorderedFloquet2020,leMissingQuantumNumber2022}.
For a realistic ramp rate, the qubit will traverse diabatically with high probability
so these collisions do not contribute meaningfully to the coherent error budget.
Exact level crossings occur between the colored levels and their gray replicas in another Brillouin zone
since they belong to distinct two-fold symmetry groups,
as explained in
\cref{sec:time-glide-reflection-symmetry}.

While
\cref{%
  fig:optimal-drive-amplitude-delta-p-typical,%
  fig:optimal-drive-amplitude-spectrum-typical%
}
show the typical situation where there is a clear optimal $\Delta p^\star$,
the behavior in other instances can be more complicated.
See, for example, \cref{%
  fig:optimal-drive-amplitude-delta-p-small-ac,%
  fig:optimal-drive-amplitude-spectrum-small-ac%
},
where
$\omega_\drivelabel/2\pi=\qty{960}{\MHz}$.
Here, the small avoided crossing between modes~$0$ and $2$,
with a gap of \qty{2\pi\times 2.2}{\MHz},
causes a small drop in $\Delta p$ while significantly increasing the incoherent errors due to hybridization with the leakage modes.
Additionally, in \cref{%
  fig:optimal-drive-amplitude-delta-p-large-ac,%
  fig:optimal-drive-amplitude-spectrum-large-ac%
},
where
$\omega_\drivelabel/2\pi=\qty{2.8}{\GHz}$,
there is a large avoided crossing between modes~$1$ and $3$
around $\Omega|n_\contlabel^{10}|/\omega_\contlabel^{10} \approx 0.2$.
The optimal drive strength $\Omega^\star$ occurs at a relatively modest value
before $\Delta p$ has plateaued
due to the large monotonic increase in the incoherent error at stronger drive.
This last case is an explicit example where the optimal drive amplitude does \emph{not} maximize $\Delta p$.

\begin{figure}
  \begin{subcaptiongroup}
    \includegraphics{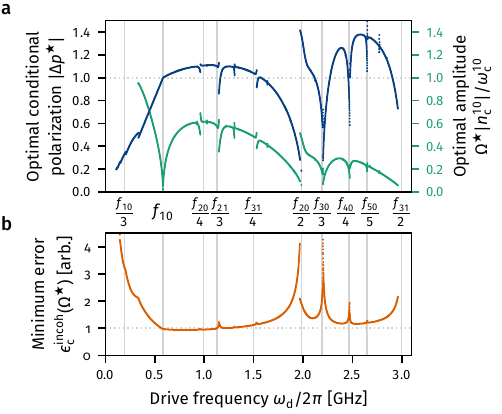}
    \phantomcaption\label{fig:polarization-spectrum-delta-p}
    \phantomcaption\label{fig:polarization-spectrum-error}
  \end{subcaptiongroup}
  \caption{%
    (\subref{fig:polarization-spectrum-delta-p})
    Optimal conditional polarization magnitude
    $|\Delta p^\star|$
    (\textbf{blue})
    and corresponding
    optimal drive amplitude $\Omega^\star$
    (\textbf{green}),
    as a function of the drive frequency $\omega_\drivelabel$.
    The absolute value of $\Delta p^\star$ is plotted
    since its sign flips at $\omega_\drivelabel = \omega^{10}_\contlabel$.
    The amplitude is optimized to minimize
    the error contribution $\eincoh_\contlabel(\Omega^\star)$,
    shown in
    (\subref{fig:polarization-spectrum-error}).
    Resonances with bare transition frequencies
    (\textbf{vertical lines})
    are annotated with
    $f_{ij} \coloneqq (E_\contlabel^i - E_\contlabel^j)/2\pi$.
    The control fluxonium has the same parameters as in
    \cref{fig:optimal-drive-amplitude}.
  }
  \label{fig:polarization-spectrum}
\end{figure}

More generally, we notice that the optimal conditional polarization decreases significantly when $\omega_\drivelabel$ is below $\omega^{10}_\contlabel$, see \cref{fig:polarization-spectrum-delta-p},
which in turn leads to slow gates and therefore large incoherent errors, see \cref{fig:polarization-spectrum-error}.
This establishes a clear preferred direction for the CR gate:
the control qubit should be the lower frequency qubit of the pair.
When
$\omega_\drivelabel$
is just above
$\omega^{10}_\contlabel$,
we have $|\Delta p^\star| \approx 1$.
The conditional polarization remains above unity
until $\omega_\drivelabel$ approaches
half of the control fluxonium's
$0\!\to\!2$ frequency, due to the hybridization caused by two-photon resonances.
At even higher frequencies, around $\omega_\drivelabel/2\pi \approx \qty{2.6}{\GHz}$,
$|\Delta p^\star|$ peaks at roughly~$1.4$.
The incoherent error is larger for higher drive frequencies,
but it remains low when we drive away from any resonances.
The target qubit can therefore be highly detuned without compromising gate speed.
This frequency flexibility stands in contrast to transmons,
for which the target qubit should be detuned by no more than the control qubit's anharmonicity
to facilitate fast gates
\cite{tripathiOperationIntrinsicError2019,malekakhlaghFirstprinciplesAnalysisCrossresonance2020,wareCrossresonanceInteractionsSuperconducting2019}.

\begin{figure}
  \begin{subcaptiongroup}
    \includegraphics{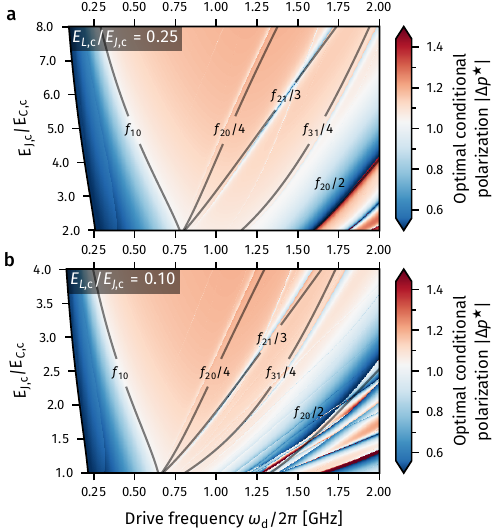}
    \phantomcaption\label{fig:polarization-spectrum2d-0.25}
    \phantomcaption\label{fig:polarization-spectrum2d-0.10}
  \end{subcaptiongroup}
  \caption{%
    Optimal conditional polarization
    $|\Delta p^\star|$
    as a function of the drive frequency $\omega_\drivelabel$
    and the control fluxonium Josephson-to-charging energy ratio
    $E_{J,\contlabel}/E_{C,\contlabel}$,
    while fixing
    $E_{L,\contlabel}/E_{J,\contlabel} = (0.25, 0.10)$
    and $E_{C,\contlabel}/2\pi = (1.0, 1.3)\,\unit{\GHz}$ in panels
    (\subref{fig:polarization-spectrum2d-0.25}, \subref{fig:polarization-spectrum2d-0.10}),
    respectively.
    The lower frequency cutoff is equal to one third of the control-qubit frequency.
    Resonances with bare transition
    (\textbf{black curves})
    are annotated with
    $f_{ij} \coloneqq (E_\contlabel^i - E_\contlabel^j)/2\pi$.
  }
  \label{fig:polarization-spectrum2d}
\end{figure}

Since fluxonium qubits typically have frequencies below \qty{1}{\GHz}
to mitigate dielectric loss
\cite{nguyenHighCoherenceFluxoniumQubit2019,wangHighcoherenceFluxoniumQubits2025}, 
it is auspicious to design the nominal target-qubit frequencies to sit between
$f_{10}$ and
$f_{20}/4$ in \cref{fig:polarization-spectrum}.
This frequency interval can be broadened by raising the control qubit's anharmonicity,
for example, by increasing $E_{J,\contlabel}/E_{C,\contlabel}$.
With this insight in mind,
we see in \cref{fig:polarization-spectrum2d-0.25}
that a larger $E_{J,\contlabel}/E_{C,\contlabel}$ ratio
leads to a wider frequency interval
where $|\Delta p^\star| > 1$.
This arises from two complementary causes.
First, the qubit frequency is lower for a larger Josephson energy.
Second,
the energy levels of the higher excited states increase as a function of the Josephson energy,
widening
the wedge between
$f_{10}$ and $f_{20}/4$.
Consequently, a large $E_{J,\contlabel}/E_{C,\contlabel}$ ratio reduces the probability of frequency collisions in the sub-gigahertz band.

We further notice that changing
$E_{L,\contlabel}/E_{J,\contlabel}$
does not qualitatively affect our observations,
see
\cref{fig:polarization-spectrum2d-0.10} for
$E_{L,\contlabel}/E_{J,\contlabel} = 0.10$
and a renormalized
$E_{C,\contlabel}/2\pi = \qty{1.3}{\GHz}$.
The
$E_{J,\contlabel}/E_{C,\contlabel}$
range is rescaled to span from \numrange{1}{4}
so that the control-qubit frequencies stay comparable to
those in
\cref{fig:polarization-spectrum2d-0.25}.
When the drive frequency is below $f_{20}/2$, the heatmap in both
\cref{fig:polarization-spectrum2d-0.25,fig:polarization-spectrum2d-0.10}
looks very similar.
The resonances above $f_{20}/2$ differ noticeably
because smaller Josephson and inductive energies
compress the higher energy levels, producing a more crowded spectrum.
Although the high frequency behavior of $|\Delta p^\star|$ is less relevant for the CR effect between sub-gigahertz fluxoniums, this detail is relevant for CR gates between a control fluxonium and a target transmon
\cite{cianiMicrowaveactivatedGatesFluxonium2022,dimitrovCrossResonantGatesHybrid2025},
or a microwave oscillator
\cite{zhengCrossresonanceControlOscillator2025}.

Conveniently, within the frequency range of interest,
the approximation $|\Delta p^\star| \approx 1$ holds reasonably well.
Combining this with \cref{eq:zx-coefficient-main},
we arrive at a simple formula for the minimum CNOT gate time
between any two capacitively-coupled fluxoniums
in terms of the coupling strength and their charge dipole moments:
\begin{equation}
  \tau_\CNOTlabel^\star
  =
  \frac{\pi/2}{
    J
    |
    n_\contlabel^{10}
    n_\targlabel^{10}
    |
  }
  \label{eq:simple-cnot-speed-limit}
  \,.
\end{equation}
This speed limit is easy to evaluate, valid beyond simple perturbative approximations and, thus, different to the limit given in Ref.~\cite{nesterovCnotGatesFluxonium2022}.
In particular,
Ref.~\cite{nesterovCnotGatesFluxonium2022}
assumes $\Omega/(\omega^{10}_\targlabel - \omega^{10}_\contlabel)$ is small
and that gate speed is primarily limited by off-resonant driving of the control qubit.

\subsection{CNOT duration landscape}
\label{sec:cnot-duration-landscape}

Increasing the coupling strength $J$
decreases the duration of the CNOT gate,
as per the speed-limit formula in
\cref{eq:simple-cnot-speed-limit}.
In practice, however,
we also need to keep the residual $ZZ$ rate below a certain threshold,
which indirectly determines the value of $J$.
By fixing
$E_{L,q}/E_{J,q} = 0.25$ for both control and target fluxoniums,
and their qubit frequencies to
$\omega^{10}_\contlabel/2\pi=\qty{500}{\MHz}$
and $\omega^{10}_\targlabel/2\pi=\qty{800}{\MHz}$,
we reduce the number of free system parameters to just two.
Thus, we can directly extract the expected CNOT time
$\tau_\CNOTlabel^\star$ as a function of the control and target
$E_{J,q}/E_{C,q}$ ratios.
We confirm in \cref{sec:extra-cnot-duration-landscapes}
that the CNOT time is largely independent of the precise
control and target qubit frequencies we have chosen.

\begin{figure}
  \begin{subcaptiongroup}
    \includegraphics{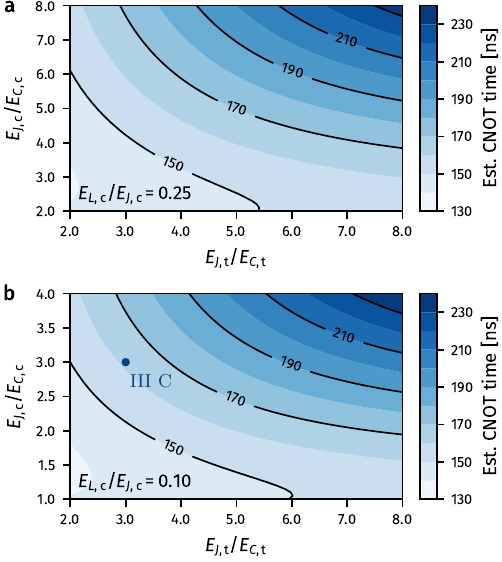}
    \phantomcaption\label{fig:cnot-duration-equal}
    \phantomcaption\label{fig:cnot-duration-small-inductance}
  \end{subcaptiongroup}
  \caption{%
    Estimated CNOT time for two capacitively-coupled fluxonium qubits with frequencies
    $
    \omega^{10}_{(\contlabel, \targlabel)}/2\pi
    =
    (500, 800)\,\unit{\MHz}
    $
    and a coupling strength $J$ determined by
    $|\mu_{ZZ}|/{2\pi} = \qty{50}{\kHz}$,
    for varying Josephson-to-charging energy ratios
    $E_{J,q}/E_{C,q}$.
    The inductive-to-Josephson energy ratio of the target fluxonium is
    $E_{L,\targlabel}/E_{J,\targlabel}=0.25$ in both panels
    and that of the control is
    (\subref{fig:cnot-duration-equal})
    $E_{L,\contlabel}/E_{J,\contlabel}=0.25$
    and
    (\subref{fig:cnot-duration-small-inductance})
    $0.10$.
    The round \textbf{marker} in
    (\subref{fig:cnot-duration-small-inductance})
    corresponds to the system parameters used
    in \cref{sec:time-domain-simulation}.%
  }
  \label{fig:cnot-duration}
\end{figure}

With our fixed parameters,
and $|\mu_{ZZ}|/2\pi=\qty{50}{\kHz}$,
\cref{fig:cnot-duration-equal}
shows
$\tau_\CNOTlabel^\star$
lies in the range of \qtyrange{130}{230}{\ns}.
In particular,
for
$E_{J,q}/E_{C,q} \lesssim 4$,
we see
$\tau_\CNOTlabel^\star$
is lower than \qty{150}{\ns}.
In general, smaller
$E_{J,q}/E_{C,q}$
ratios lead to faster CNOT gates;
but a small
control-qubit $E_{J,\contlabel}/E_{C,\contlabel}$
decreases its anharmonicity,
thereby increasing the likelihood that the CR drive
accidentally excites leakage transitions.
A relatively fast CNOT under \qty{170}{\ns}
is still achievable with
$E_{J,\contlabel}/E_{C,\contlabel} = 8$,
provided that
$E_{J,\targlabel}/E_{C,\targlabel} < 3$.
In \cref{fig:cnot-duration-small-inductance},
we again notice that
for $E_{L,\contlabel}/E_{C,\contlabel} = 0.10$,
we find similar CNOT times
compared to the $E_{L,\contlabel}/E_{C,\contlabel} = 0.25$ case
in
\cref{fig:cnot-duration-equal},
after appropriately rescaling the control qubit $E_{J,\contlabel}$.

The key takeaway from this subsection and
Subsection~\ref{sec:control-qubit-conditional-polarizability} is as follows.
To minimize frequency collisions and maximize CNOT speed,
the control qubit should have a \emph{low frequency} and a \emph{high anharmonicity}
(large
$E_{J,\contlabel}/E_{C,\contlabel}$),
whereas
the target qubit should have a \emph{higher frequency} and
a \emph{lower anharmonicity}
(smaller
$E_{J,\targlabel}/E_{C,\targlabel}$).

\subsection{Time-domain simulation}
\label{sec:time-domain-simulation}

To complement the semi-analytical analysis,
we now verify the gate performance
with numerical time-domain simulations.
For simplicity, we focus only on a single example system given by the Hamiltonian parameters in
\cref{table:time-domain-parameters}.
The Hilbert space is truncated down to 32 dimensions,
consisting of the dressed product states with 8 control fluxonium and 4 target fluxonium levels.
In this system,
the bare control-qubit frequency is \qty{502}{\MHz}
and the bare target-qubit frequency is \qty{805}{\MHz}.
The residual $ZZ$ rate is
$\mu_{ZZ}/2\pi \approx \qty{-50}{\kHz}$.
The inductive-to-Josephson energy ratios are
$E_{L,\contlabel}/E_{J,\contlabel} = 0.10$
and
$E_{L,\targlabel}/E_{J,\targlabel} = 0.25$
and the Josephson-to-charging energy ratios for both fluxoniums are
$E_{J,q}/E_{C,q} \approx 3$.
These system parameters correspond to the marker in
\cref{fig:cnot-duration-small-inductance}.

\begin{figure}[t]
  \includegraphics{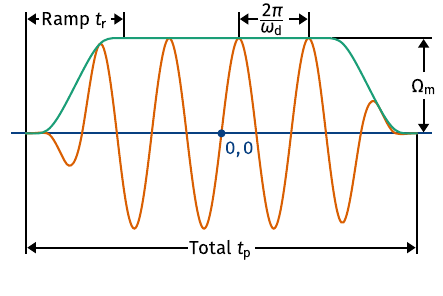}
  \caption{%
    Soft square CR pulse used to drive the control qubit in time-domain simulations.
    The non-zero pulse duration is $t_\pulselabel$
    which includes two Planck-taper ramps of duration $t_\ramplabel$ each from zero
    to a midpoint plateau amplitude of $\Omega_\midlabel$.
    The envelope (\textbf{green}) is modulated by $\sin(\omega_\drivelabel t)$
    to synthesize the microwave signal (\textbf{orange}).
    The equation for the envelope is given in
    \cref{eq:cnot-envelope-formula}.
  }
  \label{fig:cnot-envelope}
\end{figure}

\begin{table}
\centering
  \begin{tabular}{lwc{5em}wc{5em}wc{5em}wc{5em}}
    \hline\hline\\[-2.3ex]
    & $E_{J,q}/2\pi$ & $E_{C,q}/2\pi$ & $E_{L,q}/2\pi$ & $J/2\pi$ \\[0.5ex]
    & [GHz] & [GHz] & [GHz] & [MHz]
    \\[0.5ex]
    \hline\\[-2ex]
    Control & 5.60 & 1.87 & 0.56 & \multirow{2}{*}{97} \\[0.5ex]
    Target & 3.52 & 1.18 & 0.88 & \\[0.5ex]
    \hline
    \hline
  \end{tabular}
\caption{
  Hamiltonian parameters used in the time-domain simulations of
  \cref{sec:time-domain-simulation}.
  The bare control-qubit frequency is \qty{502}{\MHz}
  and the bare target-qubit frequency is \qty{805}{\MHz}.
  The $ZZ$ rate is
  $\mu_{ZZ}/2\pi \approx \qty{-50}{\kHz}$.
}
\label{table:time-domain-parameters}
\end{table}

The CR drive
is modulated by a soft square envelope,
shown pictorially in
\cref{fig:cnot-envelope},
or in equation form,
\begin{equation}
  \Omega(t)
  =
  \Omega_\midlabel
  \times
  \begin{cases}
    \hfil 1
    &
    |t| \leq
    \tfrac{t_\pulselabel}{2} - t_\ramplabel
    \,,\\[1ex]
    \hfil \mathcal{R}\ab(|t| - \tfrac{t_\pulselabel}{2})
    &
    \tfrac{t_\pulselabel}{2} - t_\ramplabel
    <
    |t|
    <
    \tfrac{t_\pulselabel}{2}
    \,,\\[1ex]
    \hfil 0
    &
    \text{otherwise,}
  \end{cases}
  \label{eq:cnot-envelope-formula}
\end{equation}
where
\begin{equation}
  \mathcal{R}(t)
  \coloneqq
  \ab[
    1
    +
    \exp\left(
      \frac{
        t_\ramplabel
      }{
        t_\ramplabel
        - t
      }
      -
      \frac{
        t_\ramplabel
      }{
        t
      }
    \right)
  ]^{-1}
  \label{eq:planck-taper}
\end{equation}
is a Planck-tapered ramp function varying from $1$ to $0$ over $0<t<t_\ramplabel$.
The gate is therefore active over the interval
$[-t_\pulselabel/2, +t_\pulselabel/2]$.
The smooth ramps aim to ensure the 0 and 1 states of the control qubit are adiabatically transported to their corresponding Floquet modes.

After computing the full gate unitary,
we truncate it down to a $4\times 4$ matrix acting on the computational subspace.
This sub-matrix is further transformed into a rotating frame where the target-qubit frequency is reduced by the \emph{drive frequency}
$\omega_\drivelabel$
(as opposed to the average target-qubit frequency).
From this, we calculate the average infidelity to the CNOT gate
after applying compensating $Z$ and $X$ rotations on the control and target qubits, respectively;
see \cref{sec:cnot-fidelity-computation} for details.

\begin{figure}
  \begin{subcaptiongroup}
  \includegraphics{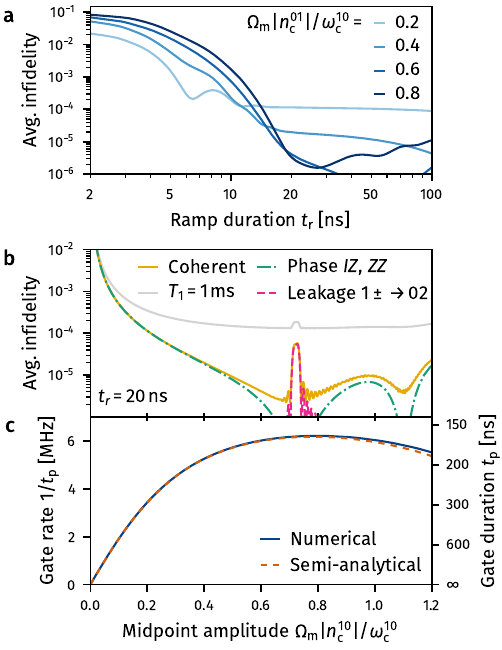}
    \phantomcaption\label{fig:time-domain-ramp-sweep-error}
    \phantomcaption\label{fig:time-domain-amplitude-sweep-error}
    \phantomcaption\label{fig:time-domain-amplitude-sweep-duration}
  \end{subcaptiongroup}
  \caption{%
    (\subref{fig:time-domain-ramp-sweep-error})
    Optimized average CNOT infidelity as a function of both the ramp duration $t_\ramplabel$ and midpoint amplitude $\Omega_\midlabel$.
    (\subref{fig:time-domain-amplitude-sweep-error})~%
    Optimized infidelity
    and
    (\subref{fig:time-domain-amplitude-sweep-duration})
    corresponding gate rate (inverse gate duration $1/t_\pulselabel$)
    as a function of $\Omega_\midlabel$
    for fixed $t_\ramplabel=\qty{20}{\ns}$.
    The coherent error (\textbf{solid yellow})
    in (\subref{fig:time-domain-amplitude-sweep-error})
    is decomposed into phase error
    (from $IZ$ and $ZZ$, \textbf{dash-dotted green})
    and leakage error (\textbf{dashed pink}).
    Incoherent error (\textbf{solid gray})
    is added assuming $T_1=\qty{1}{\ms}$.
    The numerically optimized gate rate (\textbf{solid blue})
    in (\subref{fig:time-domain-amplitude-sweep-duration})
    is compared to the semi-analytically derived,
    \cref{eq:zx-coefficient-main},
    gate rate (\textbf{dashed orange}).
  }
  \label{fig:time-domain}
\end{figure}

The simulations show
significant control-qubit errors for ramps shorter than \qty{20}{\ns},
see \cref{fig:time-domain-ramp-sweep-error}.
Here, the total pulse duration $t_\pulselabel$
and drive frequency $\omega_\drivelabel$
are optimized at each point to minimize infidelity.
For longer ramps, we generally observe coherent errors below \num{e-4} for a range of drive amplitudes.
Moreover, we see that smaller amplitude gates possess larger infidelity because the ratio of the
$ZZ$ to $ZX$ strength is larger,
which tilts the target qubit's rotation axis away from the equator.

For a fixed ramp duration
$t_\ramplabel = \qty{20}{\ns}$,
\cref{fig:time-domain-amplitude-sweep-error}
shows that
coherent gate errors below \num{e-4}
are consistently achieved across a large range of drive amplitudes.
Here, the total coherent error is plotted as the solid yellow line, together with an estimate of the incoherent error shown as the solid gray line.
The incoherent error estimate uses the simple formula
$\tfrac{4}{5}t_\pulselabel/T_1$
where $T_1=\qty{1}{\ms}$
\cite{abadUniversalFidelityReduction2022}, which yields a minimum error just above \num{e-4}.
More specifically, the total coherent error can be decomposed into a phase error
(arising from nonzero $IZ$ and $ZZ$ during the gate),
plotted in dash-dotted green,
or leakage out of the computational subspace,
plotted in dashed pink.
Leakage in this case is the result of
the energies
$\epsilon_1(\Omega) + E_\targlabel^1 + 4\omega_\drivelabel$
and
$\epsilon_0(\Omega) + E_\targlabel^2$
coming onto resonance
at
$\Omega|n^{10}_\contlabel|/\omega^{10}_\contlabel \approx 0.7$,
see also the discussion in
\cref{sec:frequency-collisions}
and further simulations in
\cref{sec:comparison-with-numerical-simulations-esd}.
Residual ripples remain in the coherent error for larger amplitudes
because the avoided crossing is traversed once at both the start and end ramps,
producing Landau--Zener--St\"uckelberg interference
\cite{shevchenkoLandauZenerStuckelberg2010}.
The coherent error is generally dominated by phase error,
except near
$\Omega_\midlabel|n^{10}_\contlabel|/\omega^{10}_\contlabel = 0.7$
and $1.1$,
where the phase error drops below \num{e-9}.
We hypothesize this occurs because the unconditional drive induced on the target qubit at certain amplitudes
cancels the net $IZ$ and $ZZ$ components,
akin to the rotary pulses of
Ref.~\cite{sundaresanReducingUnitarySpectator2020}.

Having numerically calculated the CNOT duration,
we can verify that the gate rate (inverse of the gate duration) matches with the prediction from \cref{eq:zx-coefficient-main},
see \cref{fig:time-domain-amplitude-sweep-duration}.
The predicted gate rate
for a time-varying pulse
is determined by averaging the semi-analytic expression for the $ZX$ rate in
\cref{eq:zx-coefficient-main}
over the pulse shape.
Both the numerical and semi-analytical results agree very well at low amplitudes,
but deviate up to \qty{3}{\percent} at the largest amplitudes considered.
The fastest gate is achieved in \qty{161}{\ns},
in accordance with the estimates of
\cref{fig:cnot-duration-small-inductance}.
Although $\Delta p$ often saturates above unity,
the crude approximation of
\cref{eq:simple-cnot-speed-limit},
which takes $\Delta p^\star = 1$,
is accurate in this instance because
the underestimation can be viewed
as a correction factor to account for a finite ramp duration.

\section{Frequency collisions}
\label{sec:frequency-collisions}

Driving the control qubit at the amplitude needed to saturate the $ZX$ rate
can induce undesired multiphoton transitions,
as anticipated in
\cref{fig:time-domain-amplitude-sweep-error}.
These transitions stem from resonances between Floquet quasienergies and should be avoided.
In this section,
we establish general guidelines for choosing fluxonium parameters
that circumvent unwanted resonances
and ensure that fast CNOT gates between qubit pairs remain feasible,
even in the presence of random fabrication disorder.

Using the insights from
\cref{sec:parameters-for-fast-gates},
we fix the control-qubit parameters to
$E_{J,\contlabel}/2\pi = \qty{4.0}{\GHz}$,
$E_{C,\contlabel}/2\pi = \qty{1.2}{\GHz}$
and $E_{L,\contlabel}/2\pi = \qty{0.4}{\GHz}$.
These parameters produce a low-frequency control qubit with a high anharmonicity,
providing a wide range for positioning the target-qubit frequencies.
Moreover, a small $E_{L,\contlabel}/E_{J,\contlabel}$ ratio minimizes relaxation via inductive losses,
see \cref{sec:double-well-renormalization}.
To analyze potential frequency collisions, we use an effective energy spectral density description of the driven control qubit's charge operator.
Throughout this section,
we consider a frequency collision to be an unwanted resonant interaction
inducing a coherent error exceeding \num{e-4}.
This threshold reflects the approximate incoherent error
expected from  a \qty{100}{\ns} gate
applied between qubits with a \qty{1}{\ms} lifetime.

\subsection{Control-qubit leakage}
\label{sec:control-fluxonium-transitions}

\begin{figure}
  \begin{subcaptiongroup}
    \includegraphics{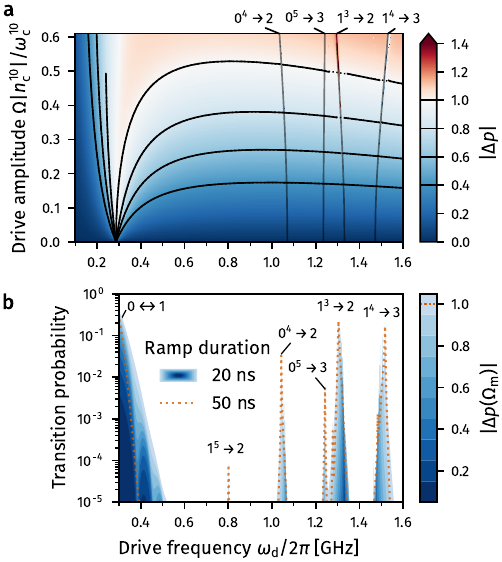}
    \phantomcaption\label{fig:control-qubit-collisions-heatmap}
    \phantomcaption\label{fig:control-qubit-collisions-peaks}
  \end{subcaptiongroup}
  \caption{%
    (\subref{fig:control-qubit-collisions-heatmap})
    Control-qubit conditional polarization $|\Delta p|$
    as a function of drive frequency $\omega_\drivelabel$ and amplitude $\Omega$.
    The contour lines of $|\Delta p|$
    (\textbf{solid black}) for
    $0.8$ and $1.0$
    do not extend to the lower drive frequencies
    because these values are not attained there.
    Avoided crossings in the Floquet quasienergy spectrum occur along the
    near-vertical \textbf{translucent} lines.
    The energy levels participating in the avoided crossing is annotated on the top spine with the superscript indicating the transition photon order.
    The contour lines are intentionally undefined near avoided crossings because they would otherwise introduce large discontinuities.
    Spline fits to the solid lines (\textbf{dotted}) are used to interpolate across the gaps.
    (\subref{fig:control-qubit-collisions-peaks})~%
    Control-qubit transition probability averaged over initial $0$  and $1$ states,
    after applying a Planck-tapered square pulse
    with ramp duration \qty{20}{\ns}
    (\textbf{filled blue})
    or \qty{50}{\ns}
    (\textbf{dotted orange})
    with drive frequency
    $\omega_\drivelabel$
    and midpoint amplitude $\Omega_\midlabel$,
    such that the midpoint conditional polarization $|\Delta p(\Omega_\midlabel)|$
    is constant.
    The \textbf{blue shading} indicates the value of
    $|\Delta p(\Omega_\midlabel)|$
    for the \qty{20}{\ns} ramp,
    whereas $|\Delta p(\Omega_\midlabel)| = 1.0$
    for the \qty{50}{\ns} ramp.
    The transitions associated with each peak are annotated in the same format as
    (\subref{fig:control-qubit-collisions-heatmap}).
    Control-qubit parameters are
    $E_J/2\pi = \qty{4.0}{\GHz}$,
    $E_C/2\pi = \qty{1.2}{\GHz}$
    and $E_L/2\pi = \qty{0.4}{\GHz}$.
  }
  \label{fig:control-qubit-collisions}
\end{figure}

To simulate the control-qubit transition probabilities,
we use the envelope from
\cref{eq:cnot-envelope-formula}.
In this numerical simulation,
the fluxonium is truncated to its lowest six eigenstates.
Because the transition probabilities are susceptible to interference effects
that depend on the pulse duration $t_\pulselabel$,
we wish to somehow average $t_\pulselabel$ away.
This effective phase averaging is emulated by evolving unitarily during the ramp portions of the envelope,
and applying a completely dephasing channel in the basis of Floquet modes during the plateau.

To determine what midpoint amplitude $\Omega_\midlabel$ to use,
we first solve the inverse problem of finding the amplitude required to achieve a given conditional polarization $|\Delta p|$ for a range of drive frequencies,
see \cref{fig:control-qubit-collisions-heatmap}.
Here, the solid contour lines are overlaid to show where
$|\Delta p|$ is equal to steps between $0.4$ and $1.0$.
Notice that
$|\Delta p|$ is not well defined
near certain resonances
(avoided crossings in the quasienergy spectrum).
These resonances are annotated with the notation
$i^k\!\to\!j$ meaning a $k$-photon transition from $i$ to $j$.
The resonances bend with drive amplitude because the energies of the dressed initial and final states are Stark shifted.
Given a desired $|\Delta p|$,
we use an arbitrary spline fit to the contour lines to map
a drive frequency
$\omega_\drivelabel$
to a drive amplitude
$\Omega_\midlabel$.
As already discussed in
\cref{sec:control-qubit-conditional-polarizability},
driving below the control-qubit frequency results in slow gates.
Thus, we do not consider this case any further.

Having constructed a map from
$\omega_\drivelabel$
to
$\Omega_\midlabel$,
we can now calculate the probability for the control qubit to transition out of its initial state
as a function of $\omega_\drivelabel$
and $|\Delta p(\Omega_\midlabel)|$,
this is shown in \cref{fig:control-qubit-collisions-peaks}
as an average over initial $0$ and $1$ states.
The filled shaded blue curves correspond to fixing the ramp duration to
\qty{20}{\ns},
with $|\Delta p(\Omega_\midlabel)|$ indicated by its fill color.
The widest peak is, as expected, the $0\leftrightarrow1$ transition when the control qubit is driven near resonance.
Other major leakage transitions only occur when
$\omega_\drivelabel/2\pi > \qty{1}{\GHz}$.
We also consider a slightly longer ramp of \qty{50}{\ns}
(shown in dotted orange).
We see that the width of its $0\leftrightarrow1$ peak roughly halves compared to the
\qty{20}{\ns} case.
This is intuitive since, for a first-order transition,
the peak width is expected to be inversely proportional to the ramp duration.
By contrast, the peak widths of the higher-order leakage transitions
show little difference between the two ramp durations.

We summarize the observed collision bounds in
\cref{table:control-qubit-collision-windows}
for the longer ramp duration of \qty{50}{\ns}
and for $\Delta p = 0.8$ and $1.0$.
The longer ramp is chosen here
to minimize the bounds on the $0\leftrightarrow1$ collision.
The bounds on the multiphoton processes are asymmetric around the bare transition frequency
because the energies of the involved states are Stark shifted by the strong drive.
A lower bound for the $0\leftrightarrow1$ transition is not provided
since we do not consider CR drives below the control-qubit frequency.
For the target qubit, which is ideally sub-gigahertz,
the only relevant unwanted transition is the $0\leftrightarrow1$ bitflip.
Thus, based on suppressing control-qubit transitions alone,
\cref{table:control-qubit-collision-windows}
suggests that the target-qubit frequency can sit between
\qty{362}{\MHz} and \qty{1}{\GHz},
for a drive amplitude enabling $\Delta p = 0.8$.

\begin{table}[htbp]
  \centering
  \begin{tabular}{wc{5em}wc{8em}wc{5em}wc{5em}}
    \hline\hline\\[-2.3ex]
    \multirow{2}{*}{Transition}
    &
    Bare frequency
    &
    \multicolumn{2}{c}{Bounds [MHz]}
    \\[0.5ex]
    & [MHz]
    & $\Delta p = 0.8$
    & $\Delta p = 1.0$
    \\[0.5ex]
    \hline\\[-2.3ex]
    \hline\\[-2ex]
    $\mathmakebox[\widthof{$0^0$}]{0}\leftrightarrow1$
    &
    $\phantom0286$
    &
    $
    \phantom{+00,\,}
    {+76}
    $
    &
    $
    \phantom{+00,\,}
    {+86}
    $
    \\
    $1^3\rightarrow2$
    &
    $1332$
    &
    $-32, \phantom0{+5}$
    &
    $-48, \phantom0{+6}$
    \\
    $0^4\rightarrow2$
    &
    $1070$
    &
    $-19, \phantom0{-8}$
    &
    $-36, \phantom0{-9}$
    \\
    $1^4\rightarrow3$
    &
    $1472$
    &
    $
    \phantom0{+6},
    +37
    $
    &
    $
    \phantom0{+6},
    +61
    $
    \\
    $0^5\rightarrow3$
    &
    $1235$
    &
    $
    \phantom0{+2},
    \phantom0{+6}
    $
    &
    $
    \phantom0{+0},
    +12
    $
    \\
    $1^5\rightarrow2$
    &
    $\phantom0799$
    &

    &
    $
    \phantom0{+1},
    \phantom0{+3}
    $
    \\[0.5ex]
    \hline
    \hline
  \end{tabular}
  \caption{
    Frequency collision bounds on control-qubit transitions extracted from
    \cref{fig:control-qubit-collisions-peaks}
    with a probability threshold of \num{e-4}.
    The bounds are expressed as a pair of lower, upper detunings relative to the bare frequency.
  }
  \label{table:control-qubit-collision-windows}
\end{table}

\subsection{Effective energy spectral density}
\label{sec:effective-energy-spectral-density}

In this section, we briefly introduce the \emph{effective energy spectral density} (effective ESD) 
of the control fluxonium's charge operator,
which will allow us to visualize
the frequency bands that may drive undesired two-body transitions.
The effective ESD arises naturally in the interaction picture
of the individual Hamiltonians, whose time-evolution unitaries are
\begin{align}
  \hat U_\contlabel(t_1, t_0)
  &=
  \mathcal{T}
  \exp
  \ab(
  -i\!
  \int_{t_0}^{t_1}
  \hat H_\contlabel(\Omega(t), t)
  \,
  dt
  )
  \,,\\
  \hat U_\targlabel(t_1, t_0)
  &=
  \exp\bigl(
    {-i}\hat H_\targlabel
    \times
    (t_1 - t_0)
  \bigr)
  \,.
\end{align}
Recall, $\hat H_\contlabel$ and $\hat H_\targlabel$
are the individual Hamiltonians of the control and target qubits, respectively.
If we assume $\Omega(t)$ drives the CR gate from the time
$t_\ilabel$ to the time $t_\flabel$,
the change-of-basis unitary into the interaction picture is
$
\hat A(t)
=
\hat A_\contlabel(t)
\otimes
\hat A_\targlabel(t)
$
where
$
  \hat A_q(t)
  =
  \hat U_q^\dagger(t, t_\ilabel)
$.
The \emph{coupled} Hamiltonian in the interaction picture then becomes
\begin{equation}
  \hat H_\intlabel(t)
  =
  J\,
  \hat n_{\contlabel,\intlabel}(t)
  \hat n_{\targlabel,\intlabel}(t)
\end{equation}
where
$
  \hat n_{q,\intlabel}(t)
  \coloneqq
  \hat A_q(t)
  \hat n_q
  \hat A_q^\dagger(t)
$
are the interaction-picture charge operators.

From the interaction-picture Hamiltonian,
it is clear that transitions occur due to the coupling
$\hat n_{\contlabel,\intlabel}(t) \hat n_{\targlabel,\intlabel}(t)$.
Therefore, we can calculate the transition probability from the product state
$\ket|ki>$ to $\ket|\ell j>$
within perturbation theory as
\begin{equation}
  P^{ki\to \ell j}
  =
  J^2
  \bigl|
    n_{\targlabel}^{ji}
  \bigr|^2
  \biggl|
  \int_{t_\ilabel}^{t_\flabel}
  \!\!
  n_{\contlabel,\intlabel}^{\ell k}(t)
  e^{i\omega^{ji}_\targlabel t}
  \,
  dt
  \biggr|^2
  ,
  \label{eq:unregularized-scattering-fourier-transform}
\end{equation}
where
$\omega_q^{ji} \coloneqq E^j_q - E^i_q$.
Given that $\Omega(t)$ is explicitly defined to be zero outside the integration domain,
we ``can'' extend the integration limits and simply obtain
\begin{equation}
  P^{ki\to \ell j}
  =
  J^2
  \bigl|
  n_{\targlabel}^{ji}
  \bigr|^2
  \bigl|
  \fourier\{
    n_{\contlabel,\intlabel}^{\ell k}
  \}
  (-\omega_\targlabel^{ji})
  \bigr|^{\mathrlap{2}}
  \,,
  \label{eq:transition-probability-unequal-control}
\end{equation}
using the Fourier transform convention
$
  \fourier\{f\}(\omega)
  =
  \int_{-\infty}^\infty
  f(t)
  e^{-i\omega t}
  \,dt
$.
We can simplify
\cref{eq:transition-probability-unequal-control}
further by averaging (summing) over initial (final) control states:
\begin{align}
  P^{i\to j}_\targlabel
  ={}&
  J^2
  \bigl|n_{\targlabel}^{ji}\bigr|^2
  \ESD(-\omega_\targlabel^{ji})
  \,,
  \label{eq:transition-probability-target-only}
\end{align}
where
\begin{equation}
  \ESD(\omega)
  =
  \frac{1}{2}
  \sum_{\ell=0}^\infty
  \sum_{k=0}^1
  \bigl|
    \fourier\{
      n_{\contlabel,\intlabel}^{\ell k}
    \}
  (\omega)
  \bigr|^2
  \label{eq:esd-formula-main}
\end{equation}
is the effective energy spectral density of the control fluxonium's charge operator.

\Cref{eq:transition-probability-target-only}
is useful since we have fixed the control-fluxonium parameters,
so the effective ESD needs to be computed only once,
but the allocation of target-fluxonium parameters remain to be determined.
The extension of the integration limits in going from
\cref{eq:unregularized-scattering-fourier-transform}
to
\cref{eq:transition-probability-unequal-control}
requires careful justification since the integrand does not always decay at long times.
For a more rigorous derivation of
\cref{eq:transition-probability-target-only}
using Floquet theory,
see \cref{sec:effective-energy-spectral-density-derivation}.
To be precise, the ESD we calculate
is averaged over all carrier phases
as opposed to a fixed sinusoidal carrier.

\subsection{Control-fluxonium effective ESD}
\label{sec:control-fluxonium-effective-esd}

\begin{figure*}[t]
  \begin{subcaptiongroup}
    \includegraphics{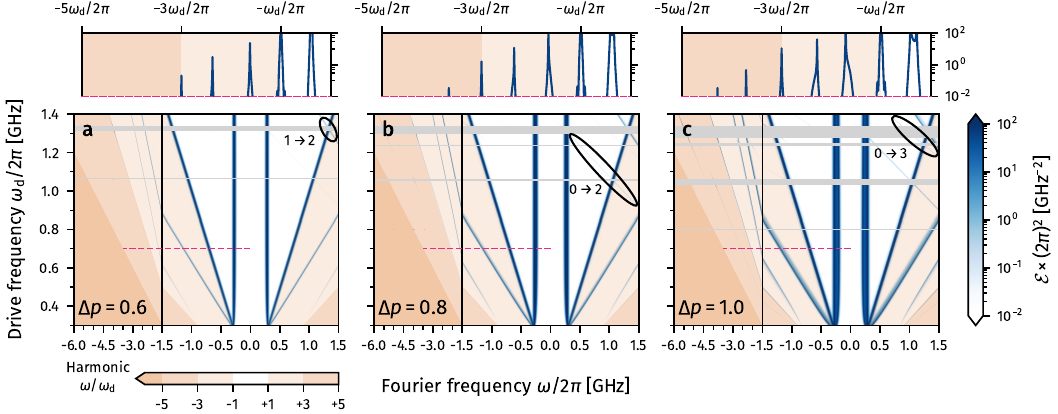}
    \phantomcaption\label{fig:charge-operator-esd-0p6}
    \phantomcaption\label{fig:charge-operator-esd-0p8}
    \phantomcaption\label{fig:charge-operator-esd-1p0}
  \end{subcaptiongroup}
  \caption{%
    Effective energy spectral density of the control fluxonium's charge operator
    $\ESD_\contlabel^{nn}$
    for conditional polarizations
    (%
      \subref{fig:charge-operator-esd-0p6}%
    )
    $\Delta p = 0.6$,
    (%
      \subref{fig:charge-operator-esd-0p8}%
    )
    $\Delta p = 0.8$,
    (%
    \subref{fig:charge-operator-esd-1p0}%
    )
    $\Delta p = 1.0$.
    Line cuts at $\omega_\drivelabel = \qty{700}{\MHz}$ are shown along the top row.
    The horizontal axis is compressed between
    \qty{-6.0}{\GHz}
    to \qty{-1.5}{\GHz}
    to conserve space.
    The gray horizontal bands mask drive frequencies
    which scatter the control qubit
    and invalidates the ESD,
    see \cref{table:control-qubit-collision-windows}.
    The background gradient partitions the viewport into a range of
    $\omega/\omega_\drivelabel$
    ratios,
    as indicated by the color bar in the bottom left.
    The gradient steps occur on odd harmonics
    $|\omega|/\omega_\drivelabel \in 2\NN_0 + 1$.
    Leakage sideband peaks are circled
    and annotated with the control-fluxonium transition induced.
    Control-fluxonium parameters are the same as in
    \cref{fig:control-qubit-collisions}.
  }
  \label{fig:charge-operator-esd}
\end{figure*}

Having established the control-fluxonium ESD as a central quantity of interest, we now distill its main features.
We calculate the ESD for an envelope with the Planck-taper ramp function,
\cref{eq:planck-taper},
with a ramp duration of $t_\ramplabel=\qty{50}{\ns}$
and visualize it in
\cref{fig:charge-operator-esd}
as a function of the drive frequency
(limiting the calculations above the control-qubit frequency).
The plateau duration is idealized to be infinitely long,
see
\cref{sec:ft-infinitely-long-square-pulse}.
Since the charge matrix elements connecting the computational states of a fluxonium
typically have less than unit magnitudes,
only the visible peaks of
\cref{fig:charge-operator-esd}
can induce transitions with probability greater than or approximately \num{e-4}.
The gray horizontal bands mask the drive frequencies
which scatter the control qubit
and represent collision windows already documented in
\cref{table:control-qubit-collision-windows}.
The pair of vertical peaks near
$\omega/2\pi \approx \qty{\pm300}{\MHz}$
are the drive-photon unassisted
$0\leftrightarrow 1$ bitflip transitions of the control fluxonium.
Because the qubit frequency is Stark shifted downwards
when driven with a positive detuning,
these peaks broaden with increasing $\Delta p$.
The slanted peaks correspond to drive-photon assisted transitions.
The background gradient divides the plot window into intervals of the
$\omega/\omega_\drivelabel$
ratios,
as mapped by the color bar in the bottom left.
The gradient steps occur on odd harmonics
$|\omega|/\omega_\drivelabel \in 2\NN_0 + 1$.
The first harmonic
$|\omega| = \omega_\drivelabel$
is the desired response that resonantly drives the target qubit.
However, the third harmonic
$|\omega| = 3\omega_\drivelabel$,
for instance,
is undesired as it may overlap with a neighboring fluxonium's leakage transition frequency.
Recall that our Fourier sign convention
implies that negative frequencies of the ESD excite the target fluxonium,
whereas positive frequencies de-excite.
As such, we see
in the top panels of \cref{fig:charge-operator-esd}
that multi-photon excitations increase in likelihood
with increasing drive strength.

Only exact odd-harmonic peaks are allowed when the fluxonium is biased at half flux
due to parity symmetry.
Therefore, a peak with an even number of drive photons must also involve a control-fluxonium transition.
For instance,
in \cref{fig:charge-operator-esd-0p6}
where $\Delta p = 0.6$,
there are four peaks between
$\omega_\drivelabel < |\omega| < 3\omega_\drivelabel$.
These are the $\pm2\omega_\drivelabel$ bitflip sideband transitions
where the control fluxonium undergoes either
$0\!\to\!1$ or $1\!\to\!0$.
Additionally,
in the top right corner of
\cref{fig:charge-operator-esd-0p6},
we annotate
the $-2\omega_\drivelabel + \tilde\omega^{21}_\contlabel$ leakage sideband.
(The tilde signifies the Stark-shifted transition frequency.)
As the drive amplitude increases to induce
$\Delta p = 0.8$
in
\cref{fig:charge-operator-esd-0p8},
another leakage sideband
is discernible:
$-3\omega_\drivelabel + \tilde\omega^{20}_\contlabel$.
In
\cref{fig:charge-operator-esd-1p0}
where $\Delta p = 1.0$,
the $-4\omega_\drivelabel + \tilde\omega^{30}_\contlabel$ sideband
is visible.
As expected,
a stronger drive
enhances the higher-harmonic responses.

When the control fluxonium has a large anharmonicity,
like in this case,
the (annotated) leakage sidebands are only relevant
when the drive frequency is over one gigahertz.
Therefore, the main source of potential frequency collisions are
the third harmonics and second-harmonic bitflip sidebands.
The third harmonic can be avoided if the target fluxoniums'
$1\to2$ transition is beyond
three times the maximum target-qubit frequency
(e.g., beyond \qty{3}{\GHz}).
Unfortunately, the bitflip sideband for a low frequency drive
easily collides with a higher qubit-frequency spectator.
Because the sideband frequency
is Stark shifted as the drive amplitude is ramped up or down,
its collision with a spectator transition can be interpreted
as Landau--Zener--St\"uckelberg interference
\cite{shevchenkoLandauZenerStuckelberg2010}.
Hence,
optimal control may nullify this collision
by engineering destructive interference.
However, investigating this
or other pulse optimization techniques
lies beyond the scope of this work.

\subsection{Control--target resonances}
\label{sec:control-target-transitions}

\begin{figure}
  \begin{subcaptiongroup}
    \includegraphics{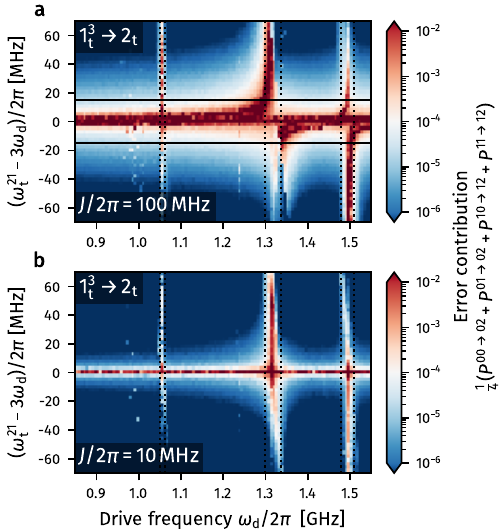}
    \phantomcaption\label{fig:control-target-third-harmonic-strong}
    \phantomcaption\label{fig:control-target-third-harmonic-weak}
  \end{subcaptiongroup}
  \caption{%
    Target-fluxonium leakage error
    induced by driving near the third subharmonic of its $1\!\to\!2$ frequency.
    The CR drive is resonant with the target qubit,
    $\omega_\drivelabel = \omega_\targlabel^{10}$.
    The vertical axis is the detuning between the transition frequency
    and the third drive harmonic.
    The default coupling strength
    is used in (\subref{fig:control-target-third-harmonic-strong})
    while a weaker strength
    is used in (\subref{fig:control-target-third-harmonic-weak}),
    see lower-left label.
    \textbf{Solid} black lines
    in (\subref{fig:control-target-third-harmonic-strong})
    define a \qty{\pm15}{\MHz} detuning window,
    outside of which the error is (generally) below \num{e-4}.
    \textbf{Dotted} black lines reproduce the collision windows from
    \cref{table:control-qubit-collision-windows}
    for $\Delta p = 0.8$.
  }
  \label{fig:control-target-third-harmonic}
\end{figure}

\begin{figure}
  \begin{subcaptiongroup}
    \includegraphics{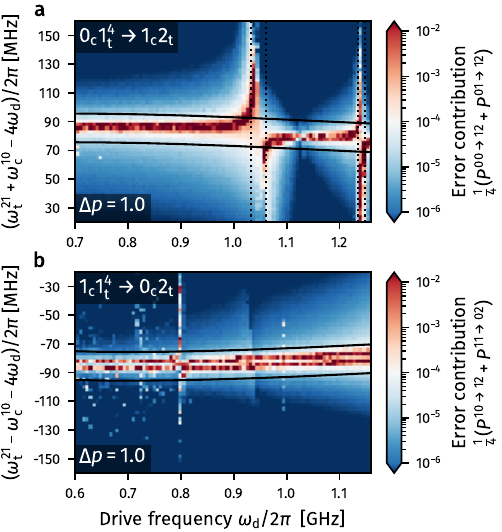}
    \phantomcaption\label{fig:control-target-fourth-harmonic-plus}
    \phantomcaption\label{fig:control-target-fourth-harmonic-minus}
  \end{subcaptiongroup}
  \caption{%
    Target-fluxonium leakage error
    assisted by
    (\subref{fig:control-target-fourth-harmonic-plus})~%
    a~control-qubit excitation
    and
    (\subref{fig:control-target-fourth-harmonic-minus})~%
    control-qubit decay,
    induced by driving near the
    fourth subharmonic of its $1\!\to\!2$ frequency,
    at a larger drive amplitude inducing $\Delta p = 1.0$.
    The vertical axis is the detuning between the
    \emph{bare} transition frequency
    and the fourth drive harmonic.
    \textbf{Solid} black lines
    define a
    \qty{\pm10}{\MHz} window
    centered on the \emph{Stark-shifted} transition frequency.
    \textbf{Dotted} black lines reproduce the collision windows from
    \cref{table:control-qubit-collision-windows} for $\Delta p = 1.0$.
  }
  \label{fig:control-target-fourth-harmonic}
\end{figure}

Although we can use the ESD along with
\cref{eq:transition-probability-target-only}
to accurately predict leakage probabilities,
we choose, for simplicity, to only use the ESD as a guide for deciding
which leakage transitions are important.
The size of the collision windows we present here are based on
transition probabilities obtained by numerically solving the Schr\"odinger equation
as in \cref{sec:control-fluxonium-transitions}.
We use the envelope from
\cref{eq:cnot-envelope-formula}
with a ramp duration of \qty{50}{\ns}
and plateau amplitude inducing
$\Delta p = 0.8$,
unless otherwise stated.
Again, we ignore interference effects
by completely dephasing the system
in the basis of Floquet modes,
see
\cref{sec:control-fluxonium-transitions}.

We start by simulating the collision probabilities between
our control fluxonium, whose parameters we have fixed,
and a target fluxonium whose parameters will be randomly sampled with:
$E_{J,\targlabel}/2\pi$
between (2 and 10)~\unit{\GHz},
$E_{C,\targlabel}/2\pi$
between (1.0 and 1.5)~\unit{\GHz},
and $E_{L,\targlabel}/2\pi$
between (0.1 and 2.0)~\unit{\GHz}.
Further,
we retain only instances where
$E_{L,\targlabel}/E_{J,\targlabel}$
is between 0.1 and 0.5.
We also fix $J/2\pi = \qty{100}{\MHz}$
since this leads to a $\mu_{ZZ}/2\pi$
between (10 and 100) \unit{\kHz}
in \qty{90}{\percent} of instances
when uniformly sampling from this parameter space. The $0\to3$ transition frequency
$\omega^{30}_\targlabel/2\pi$
exceeds \qty{5}{\GHz}
in \qty{99}{\percent} of instances, when uniformly sampling.
From
\cref{fig:charge-operator-esd},
we conclude that
exciting the target fluxonium to the $3$ state
is only possible in rare cases.
For this reason, the only leakage channel we consider is the $1\to2$ transition.

Let us first consider the contribution to the average gate infidelity
due to the target fluxonium leaking into the $2$ state
when driving near
$\omega_\drivelabel \approx \tfrac13\omega^{21}_\targlabel$, see Fig.~\ref{fig:control-target-third-harmonic}.
Because the $0$ and $1$ states are hybridized by the CR drive
$\omega_\drivelabel = \omega^{10}_\targlabel$,
both initial states are equally likely to be scattered into the $2$ state and the error is calculated as the weighted sum of transition probabilities
\cite{nesterovCnotGatesFluxonium2022},
\begin{equation}
  \frac14(
    P^{00\to02}
    + P^{01\to02}
    + P^{10\to12}
    + P^{11\to12}
  )
  \,.
  \label{eq:target-fluxonium-leakage-error-contribution}
\end{equation}
In \cref{fig:control-target-third-harmonic},
the data points are projected down onto the span of the two axes
and binned into a grid
with each pixel taking the maximum value (error contribution) of its bin. 
The dotted vertical lines in
\cref{fig:control-target-third-harmonic}
depicts the collision bounds from
\cref{table:control-qubit-collision-windows}
for $\Delta p = 0.8$.
Outside of the dotted bounds,
the target-fluxonium leakage is mostly a narrow horizontal peak,
as expected from
\cref{eq:transition-probability-target-only}.
We notice, in particular, the cases when
$\omega_\drivelabel/2\pi$
is near
\qty{1.070}{\GHz},
\qty{1.332}{\GHz}
or
\qty{1.472}{\GHz},
which are subharmonics of the
$0\!\to\!2$,
$1\!\to\!2$
and $1\!\to\!3$
transitions of the control qubit, respectively.
Significant control-qubit leakage is expected,
according to
\cref{fig:control-qubit-collisions-peaks},
when the drive is insufficiently detuned from these subharmonics.

When the detuning between
$3 \omega_\drivelabel$ and  $\omega^{21}_\targlabel$
is at least
\qty{\pm15}{\MHz}
(solid horizontal lines in
\cref{fig:control-target-third-harmonic-strong})
the leakage error is mostly limited below \num{e-4},
except near
$\omega_\drivelabel/2\pi\approx\tfrac13\omega^{21}_\contlabel/2\pi=\qty{1.332}{\GHz}$
where a level-repulsion-like feature exists,
``pushing'' leakage out of the simple boundaries.
This behavior is not predicted by
\cref{eq:transition-probability-target-only}.
If a very small coupling strength $J/2\pi=\qty{10}{\MHz}$ is chosen instead,
see \cref{fig:control-target-third-harmonic-weak},
the ``level repulsion'' is no longer significant.
This suggests the feature arises from higher-order $J$ corrections.
A single collision bound uniform in $\omega_\drivelabel$
is insufficient to fully suppress leakage errors below \num{e-4} for all parameters.
However, for simplicity, we assume that excess leakage outside of our chosen bounds
in \cref{fig:charge-operator-esd-1p0}
is improbable
so that we can keep using univariate bounds.
We have, nonetheless, included additional padding around the strict \num{e-4} threshold
to mitigate underestimating the likelihood of this collision.

We similarly consider the leakage via the
$01\!\to\!12$
and $11\!\to\!02$
transitions,
see \cref{%
fig:control-target-fourth-harmonic-plus,%
fig:control-target-fourth-harmonic-minus%
}, respectively.
These likewise show errors
resulting from driving near their fourth subharmonic frequency.
The error contributions are, respectively,
$\frac14(P^{10\to12} + P^{11\to12})$
and
$\frac14(P^{00\to12} + P^{01\to12})$.
The data displayed here is for a higher drive amplitude, inducing
$\Delta p = 1.0$, to intentionally overestimate this particular collision bound.
The vertical axis is the detuning of the fourth drive harmonic
$4\omega_\drivelabel$ from the bare transition frequency
$\omega_\targlabel^{21} \pm \omega_\contlabel^{10}$.
Because these transitions involve a control bitflip,
the resonance frequency is Stark shifted from the bare frequency,
which is why the vertical limits are not symmetric around zero.
The collision bounds (solid lines) are curved
because they are offset (\qty{\pm10}{\MHz})
from the \emph{Stark-shifted} transition frequency,
which is both drive frequency and amplitude dependent.

The Stark shift on the control-qubit frequency is monotonic with drive amplitude (to an extent),
or equivalently $\Delta p$.
When $\Delta p \geq 0.8$,
the qubit frequency shifts by approximately
\qty{10}{\MHz}
per \num{0.05} change in $\Delta p$.
Therefore,
the leakage frequency can be detuned from resonance
by varying the plateau amplitude.
Further, if the collision bounds were derived using $\Delta p = 0.8$
(see \cref{fig:control-spectator-leakage}
in \cref{sec:additional-frequency-collision-analysis}),
they would be even narrower than \qty{\pm10}{\MHz}.
A change in $\Delta p$ of $\pm0.05$ is therefore an overestimate.
Hence, calibrating the amplitude to avoid these resonances
(e.g., avoiding the leakage spike in
\cref{fig:time-domain-amplitude-sweep-error})
should be straightforward
and we do not include them in our final frequency collision model.

Lastly, the transition
$11\!\to\!02$
can also be driven at its second subharmonic frequency,
but only when $\omega_\drivelabel/2\pi \gtrsim \qty{1.1}{\GHz}$.
Since we aim to limit qubit frequencies below one gigahertz,
it is unlikely this resonance condition will be met.
Nevertheless,
the detuning
$
(
  \omega^{21}_\targlabel
  - \tilde{\omega}^{10}_\contlabel
  - 2\omega_\drivelabel
)/2\pi
$
should be outside of the interval
(\num{-20} to \num{40})~\unit{\MHz}.

\subsection{Control--spectator resonances}
\label{sec:control-spectator-transitions}

\begin{figure}
  \begin{subcaptiongroup}
    \includegraphics{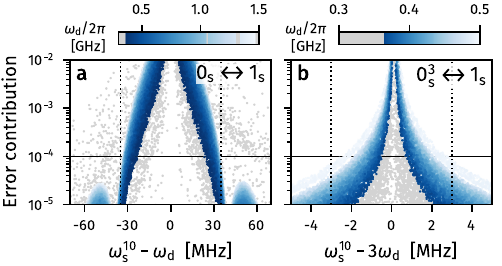}
    \phantomcaption\label{fig:spectator-bitflip-first-harmonic}
    \phantomcaption\label{fig:spectator-bitflip-third-harmonic}
  \end{subcaptiongroup}
  \caption{%
    Spectator bitflip error induced by driving
    nearly on resonance with
    (\subref{fig:spectator-bitflip-first-harmonic})~%
    the spectator frequency and
    (\subref{fig:spectator-bitflip-third-harmonic})~%
    its third subharmonic.
    The vertical axis is the error contribution
    $
    \frac15(
      P^{00\to01}
      + P^{01\to00}
      + P^{10\to11}
      + P^{11\to10}
    )
    $.
    The \textbf{blue shading} of the scatter points corresponds to the drive frequency,
    as indicated by the color bar above.
    The extracted collision bounds
    (\textbf{dotted vertical})
    are:
    (\subref{fig:spectator-bitflip-first-harmonic})~%
    \qty{\pm35}{\MHz}
    and
    (\subref{fig:spectator-bitflip-third-harmonic})~%
    \qty{\pm3}{\MHz}.
  }
  \label{fig:spectator-bitflip}
\end{figure}

\begin{figure}
  \begin{subcaptiongroup}
    \includegraphics{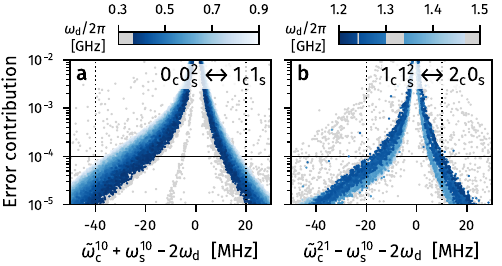}
    \phantomcaption\label{fig:control-and-spectator-bswap}
    \phantomcaption\label{fig:control-and-spectator-control-leakage}
  \end{subcaptiongroup}
  \caption{%
    (\subref{fig:control-and-spectator-bswap})
    bSWAP error
    and
    (\subref{fig:control-and-spectator-control-leakage})
    spectator-assisted control-leakage
    error
    induced when the detuning
    (horizontal axis)
    between the
    Stark-shifted transition frequency
    and the second drive harmonic
    is small.
    The vertical axis
    in
    (\subref{fig:control-and-spectator-bswap})
    is
    $
    \frac15(
      P^{00\to11}
      + P^{11\to00}
    )
    $
    and in
    (\subref{fig:control-and-spectator-control-leakage})~%
    $
    \frac14P^{11\to20}
    $.
    The \textbf{blue shading} of the scatter points corresponds to the color bar.
    The extracted collision bounds
    (\textbf{dotted vertical})
    are:
    (\subref{fig:control-and-spectator-bswap})~%
    ($-40$ to $+20$)~\unit{\MHz}
    and
    (\subref{fig:control-and-spectator-control-leakage})~%
    ($-20$ to $+10$)~\unit{\MHz}.
  }
  \label{fig:control-and-spectator-primary}
\end{figure}

The neighbors of a control qubit must have distinct frequencies
so that a CR interaction can be enabled with one target
while the other spectators ideally undergo an effective identity gate.
When the control qubit is driven near the frequency of a spectator,
bitflips on the spectator can contribute to the average infidelity of
what should be the two-qubit identity.
Considering the basis of the control and spectator qubits,
we can quantify the infidelity contribution as
\cite{nesterovCnotGatesFluxonium2022}
\begin{equation}
  \frac15(
    P^{00\to01}
    + P^{01\to00}
    + P^{10\to11}
    + P^{11\to10}
  )
  \label{eq:spectator-fluxonium-bitflip-error-contribution}
  \,,
\end{equation}
shown in \cref{fig:spectator-bitflip-first-harmonic} as a function of drive detuning.
Here, each scatter point represents a random spectator whose parameters are drawn
from the same domain as the target fluxoniums previously,
with a randomly assigned drive frequency.
The drive amplitude is again chosen to induce $\Delta p = 0.8$.
Points with a drive frequency within the collision bounds of
\cref{table:control-qubit-collision-windows}
are colored gray.
The collision bounds (vertical dotted lines)
ignore the gray points and are determined to be
\qty{\pm35}{\MHz}.

As shown in \cref{fig:spectator-bitflip-third-harmonic},
when a spectator's qubit frequency is
far detuned from the CR drive,
it can still be resonantly excited by its third harmonic.
However, the collision bound,
\qty{\pm3}{\MHz},
is very small
and this collision is unlikely to transpire
since three times the lowest target-qubit frequency
will often be over a gigahertz,
where no qubits are nominally situated.
Therefore, we exclude this resonance in our final collision model.

In addition to errors involving only spectator transitions,
there is also the possibility joint control--spectator transitions.
For instance,
the second drive harmonic can unintentionally generate
a bSWAP gate
between the control and spectator qubits
\cite{nesterovProposalEntanglingGates2021},
the error incurred by this is
$
\frac15(
  P^{00\to11}
  + P^{11\to00}
)
$,
see \cref{fig:control-and-spectator-bswap}.
The extracted collision bounds for this error channel are
($-40$ to $+20$)~\unit{\MHz}.
Because this interaction can be activated
with a sub-gigahertz spectator by drive frequencies
$\omega_\drivelabel/2\pi$ below \qty{\sim600}{\MHz},
we unfortunately expect it to be a likely source of collisions.

A decay in the spectator qubit can induce leakage in the control qubit
via $11\!\to\!20$,
with collision bounds of
($-20$ to $+10$)~\unit{\MHz}
between the second drive harmonic and the transition frequency,
see \cref{fig:control-and-spectator-control-leakage}.
This collision is unlikely for a control qubit with high anharmonicity
since,
with our chosen control-qubit parameters,
the drive frequency $\omega_\drivelabel/2\pi$ needs to be at least
\qty{1.2}{\GHz}.

Finally,
spectator leakage driven by the third drive harmonic,
or by the fourth harmonic
and assisted by a control bitflip,
has similar behavior to the equivalent channel for target fluxoniums, see
\cref{%
  fig:control-target-third-harmonic,%
  fig:control-target-fourth-harmonic%
}.
The error heatmaps are included in
\cref{sec:additional-frequency-collision-analysis}
as \cref{fig:control-spectator-leakage}
for reference.
We extract identical collision bounds of \qty{\pm15}{\MHz} 
for the detuning
$(\omega_\slabel^{21} - 3\omega_\drivelabel)/2\pi$
and \qty{\pm10}{\MHz}
for
$(\omega_\slabel^{21} \pm \tilde\omega_\contlabel^{10} - 4\omega_\drivelabel)/2\pi$.
We do not consider the fourth-harmonic resonance to be a
collision since,
as previously argued,
the resonance can be detuned by adjusting the plateau amplitude.
This mitigation strategy should work in the presence of multiple spectators as long as their
$1\!\to\!2$
frequencies are spaced apart by tens of megahertz,
otherwise tuning the amplitude can bring the drive onto resonance
with another spectator's leakage transition.

Further spectator errors
are excluded from the final collision model
because their collision bounds are either too small or
involve improbably high qubit frequencies,
see \cref{sec:additional-frequency-collision-analysis}.

\subsection{Zero-collision yield}
\label{sec:zero-collision-yield}

An important consideration in fixed-frequency architectures with microwave activated gates
is the onset of frequency collisions in larger systems.
The case of transmons in a CR architecture has previously been studied using
Monte Carlo simulations, which showed that the standard deviation in transmon frequencies
must be less than \qty{10}{\MHz}
(\qty{0.4}{\percent} relative standard deviation in the critical current density)
to effectively scale beyond a thousand qubits,
even when targeting a gate error of just \num{e-2}
\cite{hertzbergLaserannealingJosephsonJunctions2021,chamberlandTopologicalSubsystemCodes2020,morvanOptimizingFrequencyAllocation2022}.
Such stringent fabrication precision is only achievable using junction annealing methods
\cite{%
  hertzbergLaserannealingJosephsonJunctions2021,%
  zhangHighperformanceSuperconductingQuantum2022,%
  pappasAlternatingbiasAssistedAnnealing2024%
}.

\begin{figure}
  \begin{subcaptiongroup}
    \includegraphics{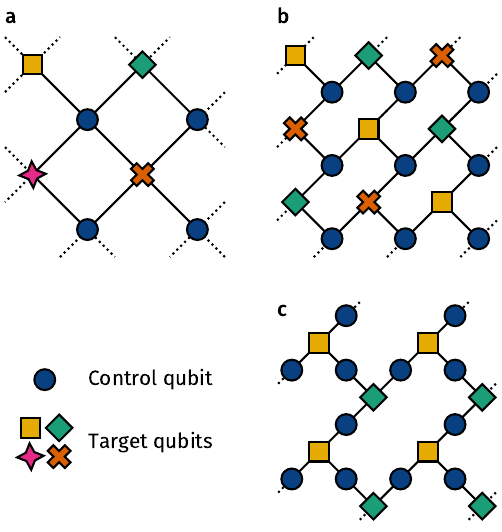}
    \phantomcaption\label{fig:connectivity-square}
    \phantomcaption\label{fig:connectivity-hexagon}
    \phantomcaption\label{fig:connectivity-heavyhex}
  \end{subcaptiongroup}
  \caption{%
    Qubit frequency allocation on a unit cell of the
    (%
      \subref{fig:connectivity-square}%
    )
    square lattice,
    (%
      \subref{fig:connectivity-hexagon}%
    )
    hexagonal lattice
    and
    (%
      \subref{fig:connectivity-heavyhex}%
    )
    heavy-hex lattice.
    All control qubits have nominally the same parameters
    while there are up to four species of target qubits.
    Edges represent always-on capacitive connections
    and dotted lines are connections to the adjacent tiles.
  }
  \label{fig:connectivity}
\end{figure}

Here, we show the results of similar Monte Carlo simulations to assess whether frequency collisions
ultimately limit the scalability of the fluxonium CR architecture, when targeting a gate error of \num{e-4}.
The probability of having zero frequency collisions (the zero-collision yield)
is calculated for three different qubit connectivities:
square, hexagonal and heavy-hexagonal.
Their geometric arrangements and qubit frequency allocation patterns are shown in
\cref{fig:connectivity}.
The qubits alternate between control and target roles,
with only a single low frequency control-qubit species.
This ensures that no two adjacent qubits are strongly driven by a CR tone
during parallel operation of CNOT gates,
which could otherwise lead to additional multi-photon resonances
not analyzed in this work.
The collision windows
from the previous sections
let us quickly determine whether a high-fidelity CNOT gate is possible between
a control qubit and any of its neighbors by comparing only their
(Stark-shifted)
transition spectra.
Given a graph of connected fluxoniums,
we can randomly perturb their nominal parameters
and then check for frequency collisions in each control-qubit vertex neighborhood.
\Cref{table:frequency-collision-model}
lists the final simplified collision windows used in the Monte Carlo simulations.
Note that the dispersive shift of the $2$ state by neighboring qubits
is below one megahertz,
which is far smaller than the collision bounds.

All three lattices support quantum error correction codes
of arbitrarily large distances.
However,
the control qubit has 4 neighbors in the square lattice,
3 in the hexagonal lattice
and 2 in the heavy-hex lattice.
The reduced neighbor count is beneficial
not only for decreasing the likelihood of frequency collisions,
but also for reducing the sum of residual $ZZ$ interactions
and the capacitive loading on each fluxonium.
To compare the yield more fairly across the different lattice types,
we use the code distance as the common measure of lattice size.
For defining the code distance,
we use the rotated surface code for the square lattice
\cite{%
bombinOptimalResourcesTopological2007,%
horsmanSurfaceCodeQuantum2012,%
tomitaLowdistanceSurfaceCodes2014%
}
and a time-dynamic embedding of the surface code for the hexagonal lattice
\cite{mcewenRelaxingHardwareRequirements2023,eickbuschDemonstrationDynamicSurface2025}.
Following
Ref.~\cite{hertzbergLaserannealingJosephsonJunctions2021},
we use the heavy-hexagon code
for the heavy-hex lattice,
which has a similar (pseudo-)threshold as the surface code
\cite{chamberlandTopologicalSubsystemCodes2020,benitoComparativeStudyQuantum2025}.

Based on the findings of this work,
we pick the nominal target-qubit parameters shown in \cref{table:monte-carlo-parameters}.
We choose a uniform inductive energy of
$E_{L,\targlabel}/2\pi = \qty{1.0}{\GHz}$
and vary the Josephson energy to adjust the qubit frequency.
The charging energy for the target qubit is also fixed at
$E_{C,\targlabel}/2\pi = \qty{1.0}{\GHz}$.
The control-qubit parameters remain as
$E_{J,\contlabel}/2\pi = \qty{4.0}{\GHz}$,
$E_{C,\contlabel}/2\pi = \qty{1.2}{\GHz}$
and $E_{L,\contlabel}/2\pi = \qty{0.4}{\GHz}$.
The reasoning behind the parameters in \cref{table:monte-carlo-parameters} is detailed in
\cref{sec:target-qubit-frequency-allocation}.

\begin{table}[htbp]
  \centering
  \def\arraystretch{1.1}
  \begin{tabular}{wc{2.4em}wc{7.5em}wc{7.5em}wc{7.5em}}
    \hline\hline\\[-2.3ex]
    Index
    &
    Transition
    &
    Detuning
    &
    Bounds$/2\pi$ [MHz]
    \\[0.5ex]
    \hline\hline\\[-2.3ex]
    1
    &
    $0_i \leftrightarrow 1_i$
    &
    $\omega_i^{10} - \omega_\drivelabel\phantom0$
    &
    $-80,+\infty$
    \\
    2
    &
    $1_i^3 \to 2_i$
    &
    $\tilde{\omega}_i^{21} - 3\omega_\drivelabel$
    &
    $\pm60$
    \\
    3
    &
    $0_i^4 \to 2_i$
    &
    $\tilde{\omega}_i^{20} - 4\omega_\drivelabel$
    &
    $\pm80$
    \\
    4
    &
    $0_i^5 \to 3_i$
    &
    $\tilde{\omega}_i^{30} - 5\omega_\drivelabel$
    &
    $\pm25$
    \\[0.5ex]
    \hline\\[-2.3ex]
    5
    &
    $1_j^3 \to 2_j$
    &
    $\omega_j^{21} - 3\omega_\drivelabel$
    &
    $\pm15$
    \\
    6
    &
    $1_i1_j^2 \to 0_i2_j$
    &
    $\omega_j^{21} - \tilde{\omega}_i^{10} - 2\omega_\drivelabel$
    &
    $-20, +40$
    \\
    7
    &
    $1_i1_j^2 \to 2_i0_j$
    &
    $\tilde\omega_i^{21} - \omega_j^{10} - 2\omega_\drivelabel$
    &
    $-20, +10$
    \\[0.5ex]
    \hline\\[-2.3ex]
    8
    &
    $0_k \leftrightarrow 1_k$
    &
    $\omega_k^{10} - \omega_\drivelabel$
    &
    $\pm35$
    \\
    9
    &
    $0_i0_k^2 \leftrightarrow 1_i1_k$
    &
    $
    \tilde\omega_i^{10}
    + \omega_k^{10}
    - 2\omega_\drivelabel$
    &
    $-40, +20$
    \\[0.5ex]
    \hline\hline
  \end{tabular}
  \caption{
    Collision bounds of likely resonances.
    The label $i$ refers to any control qubit in the lattice and
    $j$ refers to any of its neighboring qubits.
    Another neighbor unequal to $j$ is labeled $k$.
    A cross-resonance drive on $i$ has frequency
    $\omega_\drivelabel$ which could be the qubit frequency of any of its neighbors.
    Collisions 1--4 involve only the control-qubit~$i$,
    5--7 involve also a neighbor $j$
    and 8 and 9 involve a spectator $k\neq j$ with $\omega_\drivelabel = \omega_j^{10}$.
    All bounds, except 1, 5 and 8, are expressed as detunings relative to the
    Stark-shifted resonance frequency.
  }
  \label{table:frequency-collision-model}
\end{table}

\begin{figure}
  \includegraphics{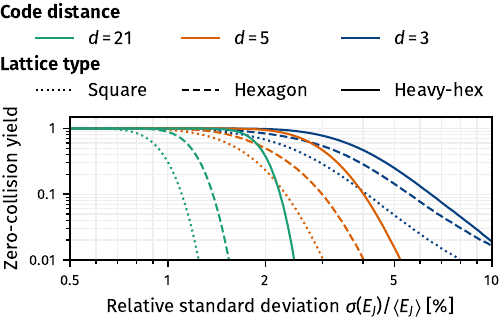}
  \caption{%
    Monte Carlo simulated zero-collision yield
    as a function of the relative standard deviation in the Josephson energies
    $\sigma(E_J)/\langle E_J\rangle$.
    The yield is calculated for code distances
    $d=21$ (\textbf{green}),
    $d=5$ (\textbf{orange})
    and $d=3$ (\textbf{blue})
    for the square lattice (\textbf{dotted}),
    hexagonal lattice (\textbf{dashed})
    and heavy-hex lattice (\textbf{solid}).
    The data points represent the collision-free fraction of \num{e6} samples.
  }
  \label{fig:zero-collision-yield}
\end{figure}

\begin{table}[htbp]
\centering
  \begin{tabular}{wl{5.5em}wc{2.2em}wc{2.2em}wc{2.2em}wc{2.2em}wc{4em}wc{4em}}
    \hline\hline\\[-2.3ex]
    \multirow{2}{*}{Lattice} &
    \multicolumn{4}{c}{$E_{J,\mathrm{t}}/2\pi$}
    & $E_{C,\mathrm{t}}/2\pi$
    & $E_{L,\mathrm{t}}/2\pi$
    \\[0.5ex]
    &
    \multicolumn{4}{c}{[GHz]}
    & [GHz]
    & [GHz]
    \\[0.5ex]
    \hline\\[-2ex]
    Square & 4.6 & 3.7 & 3.4 & 3.1 & 1.0 & 1.0\\[0.5ex]
    Hexagon & 4.5 & 3.5 & 3.1 & & 1.0 & 1.0\\[0.5ex]
    Heavy-hex & 4.4 & 3.2 & & & 1.0 & 1.0\\[0.5ex]
    \hline
    \hline
  \end{tabular}
\caption{
  Nominal target-qubit parameters used in the Monte Carlo simulation.
  The target-qubit frequencies are distributed between
  \qty{422}{\MHz}
  to
  \qty{920}{\MHz}.
  The control-qubit parameters are:
  $E_{J,\contlabel}/2\pi = \qty{4.0}{\GHz}$,
  $E_{C,\contlabel}/2\pi = \qty{1.2}{\GHz}$
  and $E_{L,\contlabel}/2\pi = \qty{0.4}{\GHz}$,
  corresponding to a qubit frequency of
  \qty{286}{\MHz}.
}
\label{table:monte-carlo-parameters}
\end{table}

As seen in \cref{fig:zero-collision-yield},
our proposed parameters can generally produce high-yield devices up to a distance of \num{21} for all three lattices considered,
provided the relative standard deviation (RSD) in the Josephson energy,
$\sigma(E_J)/\langle E_J\rangle$,
is reasonably low.
We assume the qubits' $E_J$ are normally distributed
and that the inductive energies $E_L$ vary independently and normally with
$\sigma(E_L)/\langle E_L\rangle$
equal to one-tenth of $\sigma(E_J)/\langle E_J\rangle$
\cite{cianiMicrowaveactivatedGatesFluxonium2022}.
The capacitive energies $E_C$ are treated as static.
For $d=3$, $5$ and $21$, both the square and hexagonal lattices have
\num{17}, \num{49} and \num{881} qubits, respectively, whereas the heavy-hex lattice has \num{23}, \num{65} and \num{1121}. Specifically, for a distance of \num{21} and a yield exceeding one-half,
the RSD must be less than \qty{1.0}{\percent} for the square lattice,
\qty{1.2}{\percent} for the hexagonal lattice
and \qty{2.0}{\percent} for the heavy-hex lattice.
According to Ref.~\cite{hertzbergLaserannealingJosephsonJunctions2021},
one thousand transmons in a heavy-hex arrangement requires an RSD below \qty{0.3}{\percent}.
Again, we emphasize that our collision windows are derived using a significantly stricter error threshold of \num{e-4},
compared to \num{e-2} in Ref.~\cite{hertzbergLaserannealingJosephsonJunctions2021}.

Laser annealed transmon junctions
have achieved RSD values below \qty{0.5}{\percent}
\cite{hertzbergLaserannealingJosephsonJunctions2021,zhangHighperformanceSuperconductingQuantum2022},
which would comfortably lead to high-yield devices with our architecture, even for distance \num{21} layouts.
Without post-fabrication tuning,
the best room temperature junction resistance RSD is around
\qtyrange{2}{6}{\percent}
\cite{%
  hertzbergLaserannealingJosephsonJunctions2021,%
  osmanSimplifiedJosephsonjunctionFabrication2021,%
  zhangHighperformanceSuperconductingQuantum2022,%
  pishchimovaImprovingJosephsonJunction2023,%
  muthusubramanianWaferscaleUniformityDolanbridge2024,%
  wangHighcoherenceFluxoniumQubits2025%
},
which is sufficient to reliably produce a distance \num{3} device.
Note, however, that fluxoniums also have smaller junction areas than transmons,
and as such,
there may be a larger spread in the junction resistance
by a factor of two to three
\cite{%
  osmanSimplifiedJosephsonjunctionFabrication2021,%
  pishchimovaImprovingJosephsonJunction2023,%
  wangHighcoherenceFluxoniumQubits2025%
}.

Finally, we note that precise control over the fluxoniums' inductive energies
is not necessary.
Only collision types 1, 8 and 9 of
\cref{table:frequency-collision-model}
contribute significantly to frequency crowding,
and these collisions involve only the qubit frequencies.
As long as the qubit frequency is tunable via single-junction annealing,
scaling the fluxonium CR architecture up to a few thousand qubits
will not be hindered by frequency collisions.

\section{Conclusion}

In this work,
we have shown that an all-fluxonium cross-resonance (CR) architecture
leveraging only direct capacitive coupling provides a practical and easy to implement path toward larger multi-fluxonium processors.
Despite the characteristically small charge dipole moment
associated with low-frequency fluxonium qubits,
our exploration indicates that a CR-based CNOT gate is generally achievable
in \qty{200}{\ns}
with a residual $ZZ$ rate of \qty{50}{\kHz}.
Meanwhile,
the large anharmonicity of the fluxonium complements the microwave-activated CR architecture by mitigating frequency collisions.
Assuming a Josephson energy disorder of \qty{1}{\percent}
(achievable with junction annealing
\cite{%
  hertzbergLaserannealingJosephsonJunctions2021,%
  zhangHighperformanceSuperconductingQuantum2022,%
  pappasAlternatingbiasAssistedAnnealing2024%
}),
we find that zero-collision yields remain above \num{0.5}
for square lattices supporting up to a distance\nobreakdash-21 rotated surface code.
This represents a substantial improvement over
an all-transmon square lattice,
where, for the same variation in Josephson energies,
Ref.~\cite{hertzbergLaserannealingJosephsonJunctions2021}
calculated a yield below \num{0.01}
already for the much smaller distance-3 layout.
Beyond the scalability of the all-fluxonium CR architecture,
calibrating the CR-based CNOT gate is also straightforward
\cite{sheldonProcedureSystematicallyTuning2016,pattersonCalibrationCrossResonanceTwoQubit2019};
our time-domain simulations of a two-fluxonium system
show coherent errors comfortably below \num{e-4}
when using only soft square pulses.

Our study of the CR effect in the strong driving regime further
revealed a universal saturation characteristic in the effective $ZX$ strength,
slightly exceeding the bare qubit-qubit exchange coupling rate
$J|n_\contlabel^{10}n_\targlabel^{10}|$.
This holds across a wide range of qubit-qubit detunings,
provided the lower-frequency qubit acts as the control,
in sharp contrast to the narrow straddling regime
typically needed for fast CR gates between transmons
\cite{tripathiOperationIntrinsicError2019,malekakhlaghFirstprinciplesAnalysisCrossresonance2020}.
However,
the drive amplitude that saturates the $ZX$ interaction
can be up to five times larger than
what is used in resonant single-qubit driving.
This can lead to non-negligible heating and a degradation in qubit coherence
\cite{
zwanenburgSinglequbitGatesRotatingwave2025,%
mundadaFloquetEngineeredEnhancementCoherence2020%
}.
Such strong driving can also enable multiphoton resonances,
which may induce leakage despite the large anharmonicity of the fluxonium.
Nevertheless, we still find that the majority of frequency collisions arise from
first-order resonances between closely spaced qubit frequencies,
which could be effectively minimized by using spectral pulse shaping techniques
\cite{%
motzoiSimplePulsesElimination2009,%
liExperimentalErrorSuppression2024,%
hyyppaReducingLeakageSingleQubit2024%
}.

As a blueprint for future experiments,
we propose a lattice of alternating control and target qubits,
where the higher-frequency targets serve as auxiliary qubits for parity measurements in a surface code or similar error-correction framework.
The measurement fidelity of these higher-frequency auxiliary qubits can then be enhanced
by using flux-pulse-assisted readout
while avoiding errors from swapping with temporarily resonant neighbors
\cite{stefanskiImprovedFluxoniumReadout2024}.

\begin{acknowledgments}
The authors acknowledge support from the Dutch Research Council (NWO).
Additionally, E.Y.H. acknowledges support from Holland High Tech (TKI).
\end{acknowledgments}

\section*{Data Availability}
All numerical data are archived at 4TU.ResearchData
\cite{huangDataUnderlyingPublication2026}
and the
code used for generating and visualizing the data is available on GitHub
\cite{huangCodeExplorationFluxonium}.

\appendix

\section{Capacitive--inductive drive symmetry}
\label{sec:capacitive-inductive-drive-symmetry}

To argue that the results of the main text
are independent of whether the control fluxonium is capacitively or inductively driven,
we show the charge dipole moments
between Floquet modes
are \emph{mostly} identical in both cases.
The calculations here are carried out in the extended Hilbert space,
where the time-periodic Hamiltonian,
\cref{eq:control-hamiltonian},
is made time independent, see also
\cref{eq:control-floquet-hamiltonian},
by promoting the time parameter $t$
to an operator $\hat\tau$.
Its conjugate pair is the frequency operator
$-i\hat\partial_\tau$.
Together, they satisfy the canonical commutation relations (CCR) just like
$\hat\varphi$ and $\hat n$:
$
  [\hat \tau, -i\hat\partial_\tau]
  = [\hat \varphi, \hat n]
  = i.
$
Only the control fluxonium \emph{in isolation} is analyzed
in this section,
so the `$\contlabel$' subscripts are dropped for simplicity.

The drive term of
\cref{eq:control-floquet-hamiltonian}
can be absorbed into the charge operator by redefining it as
\begin{equation}
  \hat n'
  \coloneqq
  \hat n
  +
  \frac{\Omega_\caplabel}{8E_C}
  \sin(\omega_\drivelabel \hat\tau)
  \,,
  \label{eq:redefined-charge-operator}
\end{equation}
where we add the subscript `$\caplabel$'
to distinguish this amplitude from the inductive drive amplitude introduced later.
The Floquet Hamiltonian in terms of $\hat n'$ is
\begin{equation}
  \hat\floqham
  =
  \hat H_0'
  - i\hat\partial_\tau
  + \frac{
    \Omega_\caplabel^2
  }{
    32E_C
  }
  \cos(2\omega_\drivelabel\hat\tau)
  - \frac{
    \Omega_\caplabel^2
  }{
    32E_C
  }
  \,,
  \label{eq:intermediate-primed-inductive-driving-hamiltonian}
\end{equation}
where $\hat H_0'$
is the fluxonium Hamiltonian
from
\cref{eq:fluxonium-hamiltonian}
but with $\hat n$ replaced by
$\hat n'$.
Because the term absorbed into $\hat n'$
commutes with $\hat\varphi$,
we preserve their CCR:
$[\hat\varphi, \hat n']=i$.
However,
the frequency operator should be redefined to
\begin{equation}
  -i\hat\partial_\tau'
  \coloneqq
  -i\hat\partial_\tau
  +
  \frac{\omega_\drivelabel\Omega_\caplabel}{8E_C}
  \cos(\omega_\drivelabel\hat\tau)
  \hat\varphi
  +
  \frac{\Omega_\caplabel^2}{32E_C}
  \cos(2\omega_\drivelabel\hat\tau)
  \label{eq:redefined-frequency-operator}
\end{equation}
such that it commutes with
$\hat n'$.
The last term
is included to eliminate
the $\cos(2\omega_\drivelabel\hat\tau)$ term in
\cref{eq:intermediate-primed-inductive-driving-hamiltonian},
which 
is an unphysical
coupling of different harmonics of the same Floquet mode.
Its inclusion does not affect the commutativity between
$\hat n'$
and
$-i\hat\partial_\tau'$.
The Floquet Hamiltonian in terms of
the redefined operators becomes
\begin{equation}
  \hat\floqham
  =
  \hat H_0'
  - i\hat\partial_\tau'
  + \Omega_\indlabel
    \cos(\omega_\drivelabel\hat \tau)
    \hat\varphi
  - 2E_C
  \left(
    \frac{\Omega_\indlabel}{\omega_\drivelabel}
  \right)^2
  \!,
  \label{eq:primed-inductive-driving-hamiltonian}
\end{equation}
where
$
\Omega_\indlabel
\coloneqq
-\omega_\drivelabel\Omega_\caplabel/(8E_C)
$
is a rescaled amplitude.

Given the operator pairs
$(
\hat\tau,
-i\hat\partial_\tau'
)$
and
$(
\hat\varphi,
\hat n'
)$
satisfy the CCR,
it follows from
the Stone--von Neumann theorem
that there exists
a unitary $\hat U$ such that
\cite{hallStoneNeumannTheorem2013}
\begin{subequations}
  \label{eq:svn-change-of-basis-action}
  \begin{align}
    \hat U \hat\varphi\,
    \hat U^\dagger
    &=
    \hat \varphi
    \,,
    &
    \hat U \hat\tau\,
    \hat U^\dagger
    &=
    \hat \tau
    \,,\\
    \hat U
    \hat n'
    \hat U^\dagger
    &=
    \hat n
    \,,
    &
    \hat U
    (
      -i
      \hat \partial_\tau'
    )
    \hat U^\dagger
    &=
    -i
    \hat \partial_\tau
    \,.
  \end{align}
\end{subequations}
In this case,
the unitary can be formulated explicitly as
\begin{equation}
  \hat U
  =
  \exp\!\biggl(
    \frac{i\Omega_\caplabel}{8E_C}
    \sin(\omega_\drivelabel\hat\tau)
    \hat\varphi
    +
    \frac{i\Omega_\caplabel^2}{64E_C\omega_\drivelabel}
    \sin(2\omega_\drivelabel\hat\tau)
    \!
  \biggr)
  \,\mathrlap{.}
\end{equation}
By
\cref{eq:svn-change-of-basis-action},
the transformed Hamiltonian
$
\hat\floqkam
\coloneqq
\hat U
\hat\floqham
\hat U^\dagger
$
has the same form as
\cref{eq:primed-inductive-driving-hamiltonian}
but with the primed operators replaced with their unprimed counterparts.
Physically,
this corresponds to inductive driving at a rescaled amplitude
$\Omega_\indlabel$,
up to a global energy shift
$\propto (\Omega_\indlabel/\omega_\drivelabel)^2$.

Let
$\epsilon_j + k\omega_\drivelabel$
and
$|\Phi^{j,k}_\caplabel\rrangle$
for
$j\in\NN_0, k\in\ZZ$
be
eigenpairs of
$\hat\floqham$.
Since $\hat\floqkam$
is unitarily similar
to $\hat\floqham$,
it shares the same eigenspectrum but with transformed eigenstates,
$
|\Phi^{j,k}_\indlabel\rrangle
=
\hat U
|\Phi^{j,k}_\caplabel\rrangle
$.
Let
$
n^{[k]ji}_\beta
=
\llangle
\Phi^{j,k}_\beta
| \hat n |
\Phi^{i,0}_\beta
\rrangle
$
where $\beta$ stands for
`$\caplabel$' or `$\indlabel$'.
The matrix elements
in the two bases can be related by sandwiching
\cref{eq:redefined-charge-operator}
between bras and kets of the `$\caplabel$' basis
to get
\begin{equation}
  n^{[k]ji}_\indlabel
  =
  n^{[k]ji}_\caplabel
  -
  \frac{
    i\Omega_\caplabel
  }{
    16E_C
  }
  \delta_{i,j}
  (
    \delta_{k,+1}
    - \delta_{k,-1}
  )
  \,.
\end{equation}
The elements only differ when $i=j$
and $k=\pm1$;
therefore, the relaxation rate, 
\cref{eq:floquet-relaxation-rate},
is the same when calculated using either basis.
Further,
$
n^{[\pm1]jj}_\beta
- n^{[\pm1]ii}_\beta
$
is the same in either basis so the dephasing rate,
\cref{eq:floquet-dephasing-rate},
and conditional polarization,
\cref{eq:conditional-polarization-main},
are also unaffected.
On the contrary,
the collision bounds of
\cref{table:frequency-collision-model}
do depend on the style of driving.
The bounds for collision 8
depend on the magnitude of
$n^{[\pm1]jj}_\beta$;
and the bounds for collisions 1--4,
involving only the control fluxonium,
depend on the matrix elements of
the drive operator (charge or phase)
and the drive amplitude.

The equivalence between capacitive and inductive driving
provides the freedom to choose either
$\hat\floqham$
or
$\hat\floqkam$
when calculating the Floquet eigenstates.
We find that, for the drive frequencies and amplitudes of interest,
using the Hamiltonian with inductive driving
requires a smaller Hilbert space to accurately compute the first few eigenstates
\cite{debernardisBreakdownGaugeInvariance2018,rothOptimalGaugeMultimode2019}.
One rationale is that $\hat\varphi$
couples the ground and first-excited states
less strongly than $\hat n$ does to the higher-excited states.
Another justification
comes from considering the energy offset
proportional to $(\Omega_\indlabel/\omega_\drivelabel)^2$
in \cref{eq:primed-inductive-driving-hamiltonian}.
For small $\omega_\drivelabel$,
the eigenenergies
of $\hat\floqham$
bend sharply as the drive amplitude increases,
signaling increased mixing of basis states.

Finally,
the apparent linear vanishing of $\Delta p$
as $\omega_\drivelabel$
falls below the control-qubit frequency $\omega_\contlabel$
can be elucidated in this section with a little further analysis.
The relationship between the charge and phase matrix elements
in the `$\indlabel$' eigenbasis can be derived from the commutation relation
\begin{equation}
  [
    \hat\varphi,
    \hat\floqkam
  ]
  =
  i(8E_C)\hat n
  \,.
\end{equation}
Evaluating the matrix elements in the `$\indlabel$' basis gives
\begin{equation}
  \frac{
    i(
      \epsilon_j
      - \epsilon_i
      + k\omega_\drivelabel
    )
  }{
    8E_C
  }
  \varphi^{[k]ji}_\indlabel
  =
  n^{[k]ji}_\indlabel
  \label{eq:phase-charge-relation-inductive}
  \,,
\end{equation}
which is a generalization of
$
i\omega_\contlabel
\varphi^{10}
/(8E_C)
=
n^{10}
$.
Define
$
\Delta m
\coloneqq
(
\varphi^{[1]11}_\indlabel
- \varphi^{[1]00}_\indlabel
)/\varphi^{10}
$
as the magnetic analog to $\Delta p$.
\Cref{eq:phase-charge-relation-inductive}
implies that
$|\Delta p/\Delta m| = \omega_\drivelabel/\omega_\contlabel$.
For $\omega_\drivelabel \ll \omega_\contlabel$,
the control qubit is magnetized adiabatically and $\Delta m \approx 1$.
Hence,
in the low drive frequency limit,
we have
$|\Delta p| \approx \omega_\drivelabel/\omega_\contlabel$.

\section{Effective Hamiltonian derivation}
\label{sec:effective-cr-hamiltonian}

In this section,
we make the $J$-dependence of the system Hamiltonian
in \cref{eq:system-hamiltonian}
explicit.
For a fixed $J\in\RR$ and a time-dependent envelope $\Omega(t)$,
the Schr\"odinger equation for the coupled system is
\begin{equation}
  i\partial_t
  \ket|\psi(t)>
  =
  \hat H(J,\Omega(t),t)
  \ket|\psi(t)>
  \,.
  \label{eq:schrodinger-equation}
\end{equation}
Its solution is a path
$\ket|\psi(t)>$
in the combined Hilbert space
$\hilbert = \hilbert_\contlabel\otimes\hilbert_\targlabel$,
where $\hilbert_q$ is the Hilbert space of qubit $q$ in isolation.
For a slowly varying envelope $\Omega(t)$,
we might expect $\ket|\psi(t)>$ to follow an adiabatic-like trajectory.
However,
standard adiabatic theory fails 
because the drive carrier
$\sin(\omega_\drivelabel t)$ varies rapidly.
Crucially though, since the carrier is periodic,
its time-dependence can be separated from the envelope
using Floquet theory
\cite{grifoniDrivenQuantumTunneling1998}.

By numerically solving the Floquet eigenproblem,
constructed later,
our approach enables a non-perturbative treatment of $\Omega$.
Moreover,
the derivation presented in this section provides a formal justification for the
accuracy of the semi-analytical heuristic in
Ref.~\cite{tripathiOperationIntrinsicError2019}.
Alternatively,
a full perturbative treatment of both $\Omega$ and $J$ can be carried out
using time-dependent Schrieffer--Wolff perturbation theory,
as in Ref.~\cite{malekakhlaghFirstprinciplesAnalysisCrossresonance2020}.

\subsection{Floquet theory in an extended Hilbert space}
\label{sec:extended-hilbert-space}

Consider a two-parameter Hamiltonian
$
  \hat H(J,\Omega(s),\tau)
$,
where the time-dependence of the envelope and carrier has been separated into
$s$ and $\tau$, respectively.
We can define the partial differential equation,
\begin{equation}
  i(\partial_s + \partial_\tau)
  \ket|\Psi(s,\tau)>
  =
  \hat H(J,\Omega(s),\tau)
  \ket|\Psi(s,\tau)>
  \,,
  \label{eq:two-time-schrodinger-equation}
\end{equation}
whose solutions
have the property that
$\ket|\psi(t)>=\ket|\Psi(t,t)>$
satisfies the original Schr\"odinger equation \eqref{eq:schrodinger-equation}
\cite{grifoniDrivenQuantumTunneling1998}.
In addition, if
$\ket|\Psi(s,\tau)>$
is $\tau$-coperiodic with the Hamiltonian
at some $s$,
$
|\Psi(s,\tau)\rangle
=
|\Psi(s,\tau+\frac{2\pi}{\omega_\drivelabel})\rangle
$,
then this property must extend to all $s$.
This allows us to restrict the function space of
\cref{eq:two-time-schrodinger-equation}
to study only Fourier series solutions of the form
\begin{equation}
  \ket|\Psi(s, \tau)>
  =
  \sum_{i,j\in\NN_0}
  \sum_{k\in\ZZ}
  c_{ijk}(s)
  e^{ik\omega_d \tau}
  |ij\rangle
  \label{eq:fourier-decomposition}
\end{equation}
where $c_{ijk}(s)$ are complex coefficients
and $\{|ij\rangle\}_{i,j\in\NN_0}$ is the bare basis for $\hilbert$.

One can view the complex exponentials
$
  \{e^{ik\omega_\drivelabel \tau}\}_{k\in\ZZ}
$
as a basis for
an \emph{auxiliary space},
$
\mathcal{T}
\coloneqq
L^2\bigl(\bigl[0,\frac{2\pi}{\omega_\drivelabel}\bigr]\bigr)
$,
the space of $(2\pi/\omega_\drivelabel)$-periodic square-integrable functions.
In this context,
$|\Psi(s,\tau)\rangle$
is an $s$-parameterized path in the \emph{extended} Hilbert space
$
\hilbert^\extlabel
\coloneqq
\mathcal{T}
\otimes
\hilbert
$
\cite{sambeSteadyStatesQuasienergies1973}.
We adopt the double-ket notation to emphasize this point of view:
\begin{equation}
  \hilbert^\extlabel
  \ni
  |\Psi(s)\rrangle
  =
  \sum_{i,j\in\NN_0}
  \sum_{k\in\ZZ}
  c_{ijk}(s)
  |e_k\rangle
  \otimes
  |ij\rangle
  \,,
\end{equation}
where
$\ket|e_k>$
is simply the function
$(\tau \mapsto e^{ik\omega_\drivelabel \tau}) \in \mathcal{T}$
in ket notation.
The overlap of two states
$|\alpha\rrangle, |\beta\rrangle$
in
$\hilbert^\extlabel$
is inherited from its tensor product factors, so it is naturally
\begin{equation}
  \llangle\alpha|\beta\rrangle
  =
  \frac{\omega_\drivelabel}{2\pi}
  \int_0^{\frac{2\pi}{\omega_\drivelabel}}
  \langle
  \alpha(\tau)|\beta(\tau)
  \rangle
  \,d\tau
  \,.
\end{equation}

In the extended Hilbert space,
the two-time equation,
\cref{eq:two-time-schrodinger-equation},
is promoted to a genuine Schr\"odinger equation,
\begin{equation}
  i\partial_s
  |\Psi(s)\rrangle
  =
  \hat\floqham(J, \Omega(s))
  |\Psi(s)\rrangle
  \,,
\end{equation}
by defining the Floquet Hamiltonian
$\hat\floqham \coloneqq \hat H - i\hat\partial_\tau$
where
\begin{equation}
  \hat\partial_\tau
  \coloneqq
  \sum_{k\in\ZZ}
  ik\omega_\drivelabel
  |e_k\rangle
  \langle e_k|
\end{equation}
acts on $\mathcal{T}$.
The advantage of the extended space representation is that
time-\emph{dependent} Floquet modes
become time-\emph{independent} eigenstates
of the Floquet Hamiltonian.
Further,
because
$\hat{\mathscr{H}}(J,\Omega)$
has no explicit time-dependence,
the evolution is manifestly adiabatic
if $\Omega$ is slowly varying.
As only the control qubit is driven,
we can instead decompose the extended space as
$
\hilbert^\extlabel
=
\hilbert_\contlabel^\extlabel
\otimes
\hilbert_\targlabel
$
where
$
\hilbert_\contlabel^\extlabel
\coloneqq
\mathcal{T}
\otimes
\hilbert_\contlabel
$.
The control qubit is then endowed with its own Floquet Hamiltonian,
\begin{equation}
  \hat \floqham_\contlabel(\Omega)
  \coloneqq
  \hat H_{\contlabel}^\barelabel
  - i\hat\partial_{\tau}
  +
  \Omega
  \sin(\omega_\drivelabel\hat\tau)
  \hat n_\contlabel
  \,,
  \label{eq:control-floquet-hamiltonian}
\end{equation}
where
\begin{equation}
  \sin(
  \omega_\drivelabel
  \hat\tau
  )
  =
  \frac{1}{2i}
  \sum_{k\in\ZZ}
  \bigl(
  |e_{k+1}\rangle
  \langle e_k|
  - |e_{k}\rangle
  \langle e_{k+1}|
  \bigr)
\end{equation}
also acts on $\mathcal{T}$,
and the total Floquet Hamiltonian can be reorganized as
\begin{equation}
  \hat\floqham
  =
  \hat\floqham_\contlabel
  + \hat H_{\targlabel}
  + J\hat n_\contlabel \hat n_\targlabel
  \,.
  \label{eq:floquet-hamiltonian}
\end{equation}
This form lends itself to the direct application of time-independent perturbation theory.
Once the eigensystems of
$\hat\floqham_\contlabel$
and
$\hat H_{\targlabel}$
are known,
we can perturbatively incorporate the weak $J$-coupling
to block-diagonalize $\hat\floqham$.

We adopt the conventional basis where
$\langle i|\hat n_q|j\rangle$ are purely imaginary.
Consequently, the sinusoid in
\cref{eq:control-floquet-hamiltonian}
makes the Floquet Hamiltonian
possess only real matrix elements
in the \emph{bare basis},
\begin{equation}
  \{
    |e_k\rangle
    \otimes |i\rangle
    \otimes |j\rangle
    \,|\,
    k\in\ZZ
    \text{ and }
    i,j\in\NN_0
  \}
  \,,
  \label{eq:bare-extended-basis}
\end{equation}
i.e., it is manifestly time-reversal symmetric.

\subsection{Semi-analytical block diagonalization}
\label{sec:semi-analytical-block-diagonalization}

Let
$
\{|\Phi_{j,k}(\Omega)\rrangle\}_{j\in\NN_0,k\in\ZZ}
$
be instantaneous eigenstates of
$\hat\floqham_\contlabel(\Omega)$
with eigenvalues $\epsilon_j(\Omega) + k\omega_\drivelabel$.
The index $j$ is defined such that
the modes with $k=0$ are adiabatically connected to the undriven eigenstates:
\begin{gather}
  \ket|\Phi_{j,k=0}(\Omega=0,\tau)>
  \coloneqq
  \ket|j>
  \label{eq:floquet-eigenstate-adiabatic-to-undriven}
  \,,\\
  \epsilon_j(\Omega=0)
  =
  E_\contlabel^j
  \,.
\end{gather}
The phases of the $\Omega \neq 0$ modes
are chosen so that their Fourier coefficients
in the bare basis remain real.
Further,
the $k\neq0$ modes are all defined to be in phase
with the $k=0$ mode:
\begin{equation}
  \ket|\Phi_{j,k}(\Omega, \tau)>
  \coloneqq
  e^{ik\omega_\drivelabel\tau}
  \ket|\Phi_{j,0}(\Omega, \tau)>
  \label{eq:floquet-eigenstate-in-phase}
  \,.
\end{equation}
\Cref{%
eq:floquet-eigenstate-adiabatic-to-undriven,%
eq:floquet-eigenstate-in-phase%
}
uniquely fixes the Floquet eigenbasis,
with the states automatically satisfying parallel transport,
$\llangle\Phi_{j,k}(\Omega)|\partial_\Omega|\Phi_{j,k}(\Omega)\rrangle=0$,
due to our phase choice \cite{berryQuantumPhaseFive1989}.

The Floquet Hamiltonian in
\cref{eq:floquet-hamiltonian}
must be perturbatively block diagonalized
because the target qubit is resonantly driven by the CR pulse,
$\omega_\drivelabel \approx E_\targlabel^1 - E_\targlabel^0$.
The eigensystem of
$\hat\floqham(J=0,\Omega)$
therefore possesses a ladder of nearly-degenerate doublets:
\begin{subequations}
  \label{eq:bare-doublets-definition}
  \begin{align}
    |{\uparrow}_{j,k}(\Omega)\rrangle
    &\coloneqq
    \mathrlap{|\Phi_{j,k}(\Omega)\rrangle}
    \phantom{|\Phi_{j,k-1}(\Omega)\rrangle}
    \otimes
    \ket|0>
    \,,\\
    |{\downarrow}_{j,k}(\Omega)\rrangle
    &\coloneqq
    |\Phi_{j,k-1}(\Omega)\rrangle
    \otimes
    \ket|1>
    \,.
  \end{align}
\end{subequations}
These doublets also satisfy the parallel transport criteria within the subspace they span, i.e., 
$
  \llangle{\sigma}_{j,k}
  | \partial_\Omega
  | {\varsigma}_{j,k}\rrangle
  =
  0
  $
for $\sigma,\varsigma\in\{{\uparrow},{\downarrow}\}$.
This is because the diagonal entries,
$\sigma=\varsigma$,
are always zero by our phase convention and
the off-diagonal entries are also zero because
$|0\rangle$
and $|1\rangle$
are orthogonal.

A finite $J$ coupling splits the near-degeneracies, mixing the bare doublets.
Let
$
  \hat S(J, \Omega)
  =
  \sum_{n=1}^\infty
  J^n\hat S^{(n)}(\Omega)
$
be the Schrieffer--Wolff (SW) generator that rotates the bare doublets into the dressed doublets:
\begin{equation}
  |\upsquigarrow/\downsquigarrow_{j,k}(J,\Omega)\rrangle
  \coloneqq
  e^{-\hat S(J,\Omega)}
  |{\uparrow}/{\downarrow}_{j,k}(\Omega)\rrangle
  \,,
\end{equation}
such that
\begin{equation}
  \mathcal{V}_{j,k}(J,\Omega)
  =
  \Span\{
    |{\upsquigarrow}_{j,k}(J,\Omega)\rrangle
    ,
    |{\downsquigarrow}_{j,k}(J,\Omega)\rrangle
  \}
\end{equation}
are invariant subspaces of $\hat\floqham(J, \Omega)$.
Each $(j,k)$-block of the Hamiltonian is generally of the form
\begin{equation}
  \hat \floqham_{j,k}
  =
  (
    \epsilon_j
    + \delta\epsilon_j
    + k\omega_\drivelabel
  )
  + g_j
  \hat X_{j,k}
  +
  \frac{
    \Delta_\drivelabel
    + \zeta_j
  }{2}
  \hat Z_{j,k}
  \,,
  \label{eq:bd-hamiltonian}
\end{equation}
where
\begin{subequations}
  \label{eq:pauli-operators-dressed-doublets}
  \begin{align}
    \mathmakebox[\widthof{$\hat X_{j,k}$}]{\hat X_{j,k}}
    &\coloneqq
    \phantom{i}
    |{\downsquigarrow}_{j,k}\rrangle\llangle{\upsquigarrow}_{j,k}|
    \mathmakebox[\widthof{$i-i$}]{+}
    |{\upsquigarrow}_{j,k}\rrangle\llangle{\downsquigarrow}_{j,k}|
    \,,\\
    \mathmakebox[\widthof{$\hat X_{j,k}$}]{\hat Y_{j,k}}
    &\coloneqq
    i|{\downsquigarrow}_{j,k}\rrangle\llangle{\upsquigarrow}_{j,k}|
    \mathmakebox[\widthof{$i-i$}]{\phantom{i}-i}
    |{\upsquigarrow}_{j,k}\rrangle\llangle{\downsquigarrow}_{j,k}|
    \,,\\
    \mathmakebox[\widthof{$\hat X_{j,k}$}]{\hat Z_{j,k}}
    &\coloneqq
    \phantom{i}
    |{\upsquigarrow}_{j,k}\rrangle\llangle{\upsquigarrow}_{j,k}|
    \mathmakebox[\widthof{$i-i$}]{-}
    |{\downsquigarrow}_{j,k}\rrangle\llangle{\downsquigarrow}_{j,k}|
  \end{align}
\end{subequations}
are the Pauli matrices in the dressed doublet basis
(their dependence on $J$ and $\Omega$ have been omitted).
Recall that $\hat \floqham$ has only real matrix elements,
which explains why $\hat Y_{j,k}$ does not appear
in the general form of each block,
\cref{eq:bd-hamiltonian}.
The interpretation of the coefficients
are as follows.
The parenthesized constant
is the energy of the $(j,k)$-Floquet mode
with a correction $\delta\epsilon_j = O(J^2)$.
The induced Rabi rate
on the target qubit is $2g_j=O(J)$,
which may be positive or negative.
The drive detuning from the bare target-qubit frequency is
$\Delta_\drivelabel \coloneqq \omega_\drivelabel - (E_\targlabel^1 - E_\targlabel^0)$.
The control-dependent shift to the target-qubit frequency
is $\zeta_j=O(J^2)$.
Only $g_j$ represents a resonant interaction
and thus appears at first order.
The non-resonant corrections
$\delta\epsilon_j$ and $\zeta_j$
arise at second order.

The leading order contribution to $g_j(J, \Omega)$ is simply
\begin{align}
  g_j(J, \Omega)
  &=
  \llangle
  {\downsquigarrow}_{j,0}(J, \Omega)
  | \hat \floqham(J, \Omega) |
  {\upsquigarrow}_{j,0}(J, \Omega)
  \rrangle
  \\
  &=
  J
  n_\contlabel^{10}
  n_\targlabel^{10}
  p_j(\Omega) + O(J^3)
  \,,
\end{align}
where
$n_q^{10} \coloneqq \braket<1|\hat n_q|0>$
and
\begin{align}
  p_{j}(\Omega)
  \coloneqq{}&
  \frac{1}{n^{10}_\contlabel}
  \llangle
    \Phi_{j,-1}(\Omega)|
    \hat n_\contlabel
    |\Phi_{j,0}(\Omega)
  \rrangle
  \label{eq:control-polarization}
  \\
  ={}&
  \frac{1}{n^{10}_\contlabel}
  \frac{\omega_\drivelabel}{2\pi}
  \int_0^{\frac{2\pi}{\omega_\drivelabel}}
  \bra<\Phi_{j,0}(\Omega,\tau)|
  \hat n_\contlabel
  \ket|\Phi_{j,0}(\Omega,\tau)>
  e^{i\omega_\drivelabel \tau}
  \,
  d\tau
  \nonumber
\end{align}
is the charge expectation of the $j$th Floquet mode,
oscillating at the frequency $-\omega_\drivelabel$
and normalized by $n^{10}_\contlabel$.
Parity selection rules in
$\hat n_\targlabel$
forbid even powers of $J$ from entering the perturbative expansion of $g_j$.

\subsection{Parallel-transported basis}
\label{sec:parallel-transported-basis}
\begin{figure}
  \includegraphics{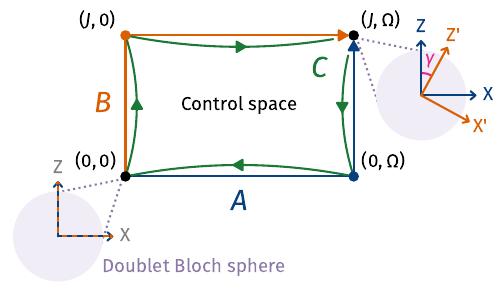}
  \caption{%
    Parallel transport of the nearly-degenerate doublets
    along different paths in control space.
    Transport along $A$,
    where $\Omega$ is first increased followed by $J$,
    yields the \textbf{blue} axes in the Bloch sphere on the right.
    Transport along $B$,
    where $J$ is first increased followed by $\Omega$,
    yields the primed \textbf{orange} axes.
    The primed axes are tilted relative to the unprimed axes
    by an angle $\gamma$
    (dependent on the control-qubit state).
    Parallel transport along the closed loop $C$ rotates the
    unprimed axes into the primed axes.
  }
  \label{fig:parallel-transport}
\end{figure}

The dressed doublets satisfy parallel transport to first order in the $J$-direction.
That is,
for
$\sigma, \varsigma \in \{\upsquigarrow, \downsquigarrow\}$,
\begin{align}
  &
  \llangle\sigma_{j,k}(J,\Omega)|
  \partial_J
  |\varsigma_{j,k}(J,\Omega)\rrangle
  \label{eq:parallel-transport-J}
  \\
  &\quad=
  \llangle\sigma_{j,k}(0,\Omega)|
  e^{\hat S(J,\Omega)}
  \partial_J
  e^{-\hat S(J,\Omega)}
  |\varsigma_{j,k}(0,\Omega)\rrangle
  \\
  &\quad=
  \underbrace{
    \llangle\sigma_{j,k}(0,\Omega)|
    [
    - \hat S^{(1)}(J,\Omega)
    - 2J\hat S^{(2)}(J,\Omega)
    ]
    |\varsigma_{j,k}(0,\Omega)\rrangle
  }_{=\,0}
  \nonumber
  \\[-1.4ex]
  &\phantom{\quad={}}{}+ O(J^2)
  \,.
\end{align}
The matrix element vanishes
by virtue of the SW transformation being a direct rotation between degenerate subspaces,
i.e., the bare doublets are in the kernel of $\hat S^{(n)}$ for all orders
\cite{bravyiSchriefferWolffTransformationQuantum2011}.
Hence, the dressed doublets can alternatively be characterized as
\begin{equation}
  |\upsquigarrow/\downsquigarrow_{j,k}(J,\Omega)\rrangle
  =
  \hat \Gamma^A_{j,k}(J,\Omega)|\upsquigarrow/\downsquigarrow_{j,k}(0, 0)\rrangle
  + O(J^3)
  \,,
  \label{eq:parallel-transport-doublets}
\end{equation}
where
$\hat\Gamma^A_{j,k}(J,\Omega): \mathcal{V}_{j,k}(0,0) \to \mathcal{V}_{j,k}(J,\Omega)$
is an isomorphism of subspaces which parallel transports its input vector
along a curve
in the control space $\{(J,\Omega)\in\RR^2\}$,
defined as the polyline from $(0, 0)$ to $(0, \Omega)$ to $(J, \Omega)$.
This is shown pictorially as the curve $A$ in \cref{fig:parallel-transport}.
Since \cref{eq:parallel-transport-J} vanishes up to $O(J^2)$,
the discrepancy becomes $O(J^3)$
in \cref{eq:parallel-transport-doublets}
after integration over a length of $J$.

Another gauge choice performs the parallel transport
$\hat\Gamma^B_{j,k}$
along
the other sides of the rectangle:
$(0,0)$ to $(J,0)$ and then $(J,\Omega)$,
see curve $B$ in \cref{fig:parallel-transport}.
We distinguish these states as the \emph{primed} doublets,
\begin{equation}
  |\upsquigarrow'/\downsquigarrow_{j,k}'(J,\Omega)\rrangle
  \coloneqq
  \hat \Gamma^B_{j,k}(J,\Omega)|\upsquigarrow/\downsquigarrow_{j,k}(0, 0)\rrangle
  \,.
\end{equation}
They satisfy, by construction, parallel transport along $\Omega$:
\begin{equation}
  \llangle\sigma'_{j,k}(J,\Omega)|
  \partial_\Omega
  |\varsigma'_{j,k}(J,\Omega)\rrangle
  = 0
  \,.
  \label{eq:primed-parallel-transport}
\end{equation}
The naturalness of the primed doublets is apparent
when considering the change of basis
$\hat P$,
whose $(j, k)$-blocks are
\begin{equation}
  \hat P_{j,k}(J,\Omega)
  =
  \ %
  \sum_{\mathclap{
    \sigma\in\text{\scriptsize$\{\upsquigarrow,\downsquigarrow\}$}
  }}
  \ %
  |\sigma_{j,k}(J, \Omega = 0)\rrangle
  \llangle\sigma'_{j,k}(J,\Omega)|
  \,.
  \label{eq:primed-doublets-change-of-basis}
\end{equation}
The inertial terms,
$
i\dot\Omega
\hat P_{j_1,k_1}
\partial_\Omega
\hat P^\dagger\vphantom{P}_{j_2,k_2}$,
introduced into the transformed Hamiltonian
when $\Omega$ is time-dependent
vanish when $(j_1, k_1)=(j_2, k_2)$ by virtue of
\cref{eq:primed-parallel-transport}.
The block-off-diagonal terms
($j_1\neq j_2$ or $k_1\neq k_2$)
can be neglected by the adiabatic theorem
provided they are much smaller than the inter-block energy gaps.
The upshot is, in the adiabatic limit,
$\hat P$
may be regarded as a time-independent transformation.

Before transforming $\hat\floqham$
by $\hat P$,
we first establish the relationship between the primed and unprimed doublets.
Let $\hat \Gamma^C_{j,k}$ transform the
unprimed to primed doublets:
\begin{equation}
  |\upsquigarrow'/\downsquigarrow_{j,k}'(J,\Omega)\rrangle
  \eqqcolon
  \hat\Gamma^C_{j,k}(J,\Omega)
  |\upsquigarrow/\downsquigarrow_{j,k}(J,\Omega)\rrangle
  \,.
  \label{eq:primed-from-unprimed-doublets}
\end{equation}
Parallel transport, \cref{eq:primed-parallel-transport}, is satisfied if and only if,
for $\sigma, \varsigma \in \{\upsquigarrow, \downsquigarrow\}$,
\begin{equation}
  \llangle\sigma_{j,k}|
  \hat\Gamma^{C\dagger}_{j,k}
  \bigl(
  \partial_\Omega
  \hat\Gamma^C_{j,k}
  \bigr)
  |\varsigma_{j,k}\rrangle
  =
  -
  \llangle\sigma_{j,k}|
  \partial_\Omega
  |\varsigma_{j,k}\rrangle
  \,.
  \label{eq:wz-phase-equality}
\end{equation}
This is essentially a two-by-two matrix differential equation and
can be easily solved for $\hat \Gamma^C_{j,k}$ up to an error of $O(J^3)$.
First,
note that the diagonal of the right-hand side
is again zero due to normalization and the fact that the matrix elements are real.
The off-diagonal term is
\begin{align}
  &\llangle{\upsquigarrow}_{j,k}(J,\Omega)|
  \partial_\Omega
  |{\downsquigarrow}_{j,k}(J,\Omega)\rrangle
  \\
  &\qquad=
  -J\llangle{\uparrow}_{j,k}(\Omega)|
  \bigl[
    \partial_\Omega
    S^{(1)}(J,\Omega)
  \bigr]
  |{\downarrow}_{j,k}(\Omega)\rrangle
  + O(J^3)
  \,.
  \nonumber
\end{align}
Again, parity selection rules in $\hat n_\targlabel$ forbid even powers of $J$.
Define the integral over $\Omega$ of the right-hand side as
\begin{equation}
  \frac{\gamma_j(J,\Omega)}{2}
  \coloneqq
  J
  \!
  \int_0^{\Omega}
  \!
  d\Omega'
  \llangle{\uparrow}_{j,k}(\Omega')|
  \bigl[
    \partial_{\Omega'}
    S^{(1)}(J,\Omega')
  \bigr]
  |{\downarrow}_{j,k}(\Omega')\rrangle
  \,.
\end{equation}
\Cref{eq:wz-phase-equality} 
can now be reformulated in a basis-independent way:
\begin{equation}
  \bigl(
  \partial_\Omega
  \hat\Gamma^C_{j,k}
  \bigr)
  \big|_{\mathcal{V}_{j,k}}
  \!
  =
  -i\hat\Gamma^{C}_{j,k}
  \frac{
    \partial_\Omega \gamma_j
  }{2}
  \hat Y_{j,k}
  + O(J^3)
  \,.
  \label{eq:wz-phase-equality-operator-form}
\end{equation}
Its naive solution is
\begin{equation}
  \hat \Gamma^C_{j,k}(J,\Omega)
  =
  \exp\Bigl(
    -
    \frac{i\gamma_j}{2}
    \hat Y_{j,k}
  \Bigr)
  + O(J^3)
  \,,
  \label{eq:wz-phase-solution}
\end{equation}
which represents
a rotation of $\gamma_j$ around the $Y$ axis of the Bloch sphere.
Geometrically,
$\hat\Gamma^C_{j,k}(J,\Omega)$
is the holonomy acting on 
$\mathcal{V}_{j,k}(J,\Omega)$
obtained by parallel transport around the rectangular loop,
curve $C$ in \cref{fig:parallel-transport},
with vertices, and in order:
$(J,\Omega)$,
$(0,\Omega)$,
$(0,0)$,
$(J,0)$.
This is also known as the Wilczek--Zee phase
\cite{wilczekAppearanceGaugeStructure1984}.
The loop collapses
when $\Omega=0$
and we have, accordingly, $\gamma_j=0$
and $\hat\Gamma^C_{j,k}$ acts trivially as expected.

The naive guess turns out to be correct
despite
$\hat Y_{j,k}$ varying with $\Omega$.
This is because
$\partial_\Omega\hat Y_{j,k}$
vanishes on
$\mathcal{V}_{j,k}$
for the following reasons.
The operator $i\hat Y_{j,k}$ is anti-Hermitian and has real matrix elements
so
$i\partial_\Omega\hat Y_{j,k}$
must inherit this property.
Since the dressed doublets
are purely real
and anti-Hermitian operators can only be imaginary along the diagonal,
the diagonal of
$i\partial_\Omega\hat Y_{j,k}$
in the dressed doublet basis
must be identically zero.
A quick computation shows the off-diagonal also vanishes:
\begin{align}
  &\llangle{\upsquigarrow}_{j,k}|
  \bigl(
    i\partial_\Omega
    \hat Y_{j,k}
  \bigr)
  |{\downsquigarrow}_{j,k}\rrangle
  \nonumber
  \\
  &\ \quad=
  \llangle{\upsquigarrow}_{j,k}|
    \partial_\Omega
  \bigl(
    i\hat Y_{j,k}
  |{\downsquigarrow}_{j,k}\rrangle
  \bigr)
  -
  \llangle{\upsquigarrow}_{j,k}|
  i\hat Y_{j,k}
  \partial_\Omega
  |{\downsquigarrow}_{j,k}\rrangle
  \\
  &\ \quad=
  \llangle{\upsquigarrow}_{j,k}|
  \partial_\Omega
  |{\upsquigarrow}_{j,k}\rrangle
  -
  \llangle{\downsquigarrow}_{j,k}|
  \partial_\Omega
  |{\downsquigarrow}_{j,k}\rrangle
  \,,
\end{align}
where the last line is zero by our phase convention.

By analogy with the primed doublets,
we can define the corresponding primed Pauli operators
of \cref{eq:primed-from-unprimed-doublets}
as
$
  \hat A'_{j,k}
  \coloneqq
  \hat\Gamma^C_{j,k}
  \hat A_{j,k}
  \hat\Gamma^{C\dagger}_{j,k}
$
for
$A\in\{X,Y,Z\}$.
Because
$\hat P_{j,k}(J,\Omega)$
maps
$\hat A'_{j,k}(J, \Omega)$
to
$\hat A_{j,k}(J, 0)$,
it is helpful to first decompose the unprimed Pauli operators
in terms of their primed counterparts as an intermediate step to transforming
the Hamiltonian blocks:
\begin{subequations}
  \label{eq:primed-axis-tilt}
  \begin{align}
    \hat X_{j,k}
    &=
    (1 - \tfrac12\gamma_j^2)
    \hat X'_{j,k}
    + \gamma_j\hat Z'_{j,k}
    + O(J^3)
    \,,\\
    \hat Y_{j,k}
    &=
    \hat Y'_{j,k}
    + O(J^3)
    \,,\\
    \hat Z_{j,k}
    &=
    (1 - \tfrac12\gamma_j^2)
    \hat Z'_{j,k}
    - \gamma_j\hat X'_{j,k}
    + O(J^3)
    \,.
  \end{align}
\end{subequations}
It then follows that
each block
$\hat\floqham_{j,k}$
transforms as
\begin{align}
  \hat\floqham_{j,k}^\efflabel(J, \Omega)
  \coloneqq{}&
  \hat P_{j,k}(J, \Omega)
  \hat\floqham_{j,k}(J, \Omega)
  \hat P_{j,k}^\dagger(J, \Omega)
  \\
  ={}&
  (
    \epsilon_j(\Omega)
    + \delta\epsilon_j(J, \Omega)
    + k\omega_\drivelabel
  )
  \label{eq:floquet-effective-hamiltonian}
  \\
  &+
  g_j'(J, \Omega)
  \hat X_{j,k}(J, 0)
  \nonumber
  \\
  &+
  \frac{
    \Delta_\drivelabel
    + \zeta_j'(J, \Omega)
  }{2}
  \hat Z_{j,k}(J, 0)
  + O(J^3)
  \,,
  \nonumber
\end{align}
where
\begin{align}
  g_j'(J, \Omega)
  &\coloneqq
  g_j(J, \Omega)
  - \tfrac12\gamma_j(J, \Omega)\Delta_\drivelabel
  \,,\\
  \zeta_j'(J,\Omega)
  &\coloneqq
  \zeta_j(J, \Omega)
  + \gamma_j(J, \Omega)g_j(J, \Omega)
\end{align}
are modified rates
due to the tilt of the unprimed
$X$~axis
towards the
primed $Z$~axis,
\cref{eq:primed-axis-tilt}.
Importantly,
$\zeta_j'$ remains second order in $J$,
so the axis correction does not change the perturbative order of the Hamiltonian coefficients.
Although the drive detuning $\Delta_\drivelabel$ is non-perturbative,
it is desirable to have it partially cancel $\zeta_j'$
when implementing a CNOT gate.
The induced Rabi rate $g_j'$ is therefore unaffected to first order
if we treat $\Delta_\drivelabel$ as a second-order term.
We do not explicitly write down
semi-analytic expressions for the second-order frequency shifts,
$\delta\epsilon_j$
and
$\zeta_j'$,
since they are not analyzed in this manuscript.

\subsection{Leaving the extended Hilbert space}
\label{sec:leaving-extended-hilbert-space}

The effective extended space Hamiltonian of
\cref{eq:floquet-effective-hamiltonian}
is only physically relevant once
we define how the states of the original space embed into the extended space.
Consider an initial state
$|\Psi(0)\rrangle$
in the subspace
$
\mathcal{V}_{j=0,k=0}(J,0)
\oplus
\mathcal{V}_{j=1,k=0}(J,0)
$.
We can parameterize its evolution in the basis of the undriven dressed doublets as
\begin{equation}
  |\Psi(s)\rrangle
  =
  \ %
  \sum_{\mathclap{j\in\{0,1\}}}
  \ %
  c_{j,0}(s)
  |{\upsquigarrow}_{j,0}(J,0)\rrangle
  + c_{j,1}(s)
  |{\downsquigarrow}_{j,0}(J,0)\rrangle
  \,,
  \label{eq:floquet-solution}
\end{equation}
for complex coefficients $c_{j,i}(s)$.
The diagonal trajectory
$\ket|\psi(t)>=\ket|\Psi(t,t)>$
in the original Hilbert space is then
\begin{equation}
  |\psi(t)\rangle
  =
  \ %
  \sum_{\mathclap{j\in\{0,1\}}}
  \ %
  c_{j,0}(t)
  |\widetilde{j0}\rangle
  + c_{j,1}(t)
  e^{-i\omega_\drivelabel t}
  |\widetilde{j1}\rangle
  \,,
  \label{eq:diagonal-solution}
\end{equation}
where the basis states are the dressed eigenstates of the coupled Hamiltonian,
introduced in
\cref{sec:main-effective-hamiltonian}.
(For consistency with the main text,
the $J$-dependence of these states is omitted.)
Define the time-dependent change of basis into the
\emph{cross-resonance frame}:
\begin{equation}
  \hat R(t)
  \coloneqq
  \exp
  \Bigl(
  {-\frac{i\omega_\drivelabel t}{2}}
  \,
  \hat{\widetilde{IZ}}
  +
  \frac{i\omega_\drivelabel t}{2}
  \Bigr)
  \,,
\end{equation}
where
$
  \hat H_\CRlabel
  \coloneqq
  \hat R
  (\hat H - i\partial_t)
  \hat R^\dagger
$
is the transformed Hamiltonian.
A stationary target qubit in the CR frame
precesses
with frequency $\omega_\drivelabel$
in the lab frame,
hence the state of
\cref{eq:diagonal-solution}
transforms into
\begin{equation}
  |\psi_\CRlabel(t)\rangle
  =
  \sum_{\mathclap{j\in\{0,1\}}}
  \ %
  c_{j,0}(t)
  |\widetilde{j0}\rangle
  + c_{j,1}(t)
  |\widetilde{j1}\rangle
  \,.
\end{equation}
If we identify
$
  |\widetilde{j0}\rangle
  \simeq
  |{\upsquigarrow}_{j,0}(J, 0)\rrangle
$ and $
  |\widetilde{j1}\rangle
  \simeq
  |{\downsquigarrow}_{j,0}(J, 0)\rrangle
$,
the CR frame solution is identical to \cref{eq:floquet-solution}.
Therefore,
we can construct an effective Hamiltonian in the CR frame
by directly taking the coefficients
of \cref{eq:floquet-effective-hamiltonian} for the $k=0$ band
and projecting them onto the two-qubit Pauli basis,
after which we obtain the Hamiltonian
\begin{gather}
  \hat H_\CRlabel^\efflabel
  \coloneqq
  \biggl(
    \frac{g_0' - g_1'}{2}
  \biggr)
  \hat{\widetilde{ZX}}
  +
  \biggl(
    \frac{g_0' + g_1'}{2}
  \biggr)
  \hat{\widetilde{IX}}
  \numberthis
  \label{eq:effective-hamiltonian-with-coefficients}
  \\
  \begin{aligned}
    &+
    \biggl(
      \frac{
        \epsilon_0
        + \delta\epsilon_0
        - \epsilon_1
        - \delta\epsilon_1
      }{2}
    \biggr)
    \hat{\widetilde{ZI}}
    +
    \biggl(
      \frac{\Delta_\drivelabel}{2}
      + \frac{\zeta_0' + \zeta_1'}{4}
    \biggr)
    \hat{\widetilde{IZ}}
    \\
    &+
    \biggl(
      \frac{\zeta_0' - \zeta_1'}{4}
    \biggr)
    \hat{\widetilde{ZZ}}
    \nonumber
    \,.
  \end{aligned}
\end{gather}
This is essentially the same effective Hamiltonian as
in \cref{eq:effective-hamiltonian-main}
when $\omega_\drivelabel$ is equal to the average target-qubit frequency
$\bar{\omega}_\targlabel$.
However, in the main text, we further subtract $\bar{\omega}_\contlabel$
from the control-qubit frequency,
leading to a shift in the $ZI$~coefficient.
This shift is trivial since a $Z$~rotation on the control qubit
commutes with the effective Hamiltonian.

We wish to emphasize that
\cref{eq:effective-hamiltonian-with-coefficients}
is an effective Hamiltonian in the sense that
the following holds (up to global phase) on the computational subspace,
\begin{align}
  &\mathcal{T}
  \exp
  \ab(
  -i\!
  \int_{t_0}^{t_1}
  \hat H_\CRlabel^\efflabel(J,\Omega(t))
  \,dt
  )
  \\
  &\hspace{7em}\propto
  \mathcal{T}
  \exp
  \ab(
  -i\!
  \int_{t_0}^{t_1}
  \hat H_\CRlabel(J,\Omega(t), t)
  \,dt
  )
  \,,
  \nonumber
\end{align}
where $\mathcal{T}$ indicates time ordering
and $\Omega(t_0) = \Omega(t_1) = 0$.

\section{Double well renormalization}
\label{sec:double-well-renormalization}

The potential energy for a fluxonium with
inductive energy $E_L$
and Josephson energy $E_J$
at half-flux bias is
\begin{equation}
  V_{E_L, E_J}(\varphi)
  \coloneqq
  \frac{1}{2}E_L \varphi^2
  + E_J\cos \varphi
  - E_J
  \,.
\end{equation}
Contrary to convention,
the extra $-E_J$ is included to zero the potential energy at $\varphi=0$.
Instead of providing the inductive and Josephson energies,
the double-well potential can alternatively be parameterized in terms of the interwell distance and barrier height.
Let $0 \leq a < \pi$ be the position of the first minima of the potential.
It is related to the inductive-to-Josephson energy ratio transcendentally through
\begin{equation}
  \sinc a
  =
  \frac{E_L}{E_J}
  \,,
  \label{eq:sinc-ratio-minima-position}
\end{equation}
where $\sinc x\coloneqq \sin x/x$.
For the existence of at least two minima, $E_L$ must be smaller than $E_J$.
Typical fluxonium $E_L/E_J$ ratios vary between $0.1$ and $0.5$
\cite{nguyenBlueprintHighPerformanceFluxonium2022}.
This corresponds to minimum positions in the range $\tfrac{1}{2}\pi < a < \tfrac{11}{12}\pi$.
The depth $b > 0$ of the two minima relative to zero is
\begin{equation}
  b
  =
  -V_{E_L, E_J}(a)
  =
  E_J
  \ab(
    1
    - \cos a
    - \frac{a}{2}\sin a
  )
  \,.
  \label{eq:double-well-depth}
\end{equation}
We can therefore reparameterize the fluxonium potential at half-flux bias
in terms of $a$ and $b$ as
\begin{equation}
  W_{a,b}(\varphi)
  \coloneqq
  b
  \ab[
    \frac{
      \frac{1}{2}
      (\sinc a)
      \varphi^2
      + \cos\varphi
      - 1
    }{
      1
      - \cos a
      - \frac{a}{2}\sin a
    }
  ]
  \,,
\end{equation}
which equals $V_{E_L, E_J}(\varphi)$
when
\cref{eq:sinc-ratio-minima-position,eq:double-well-depth}
hold.

\begin{figure}
  \begin{subcaptiongroup}
    \includegraphics{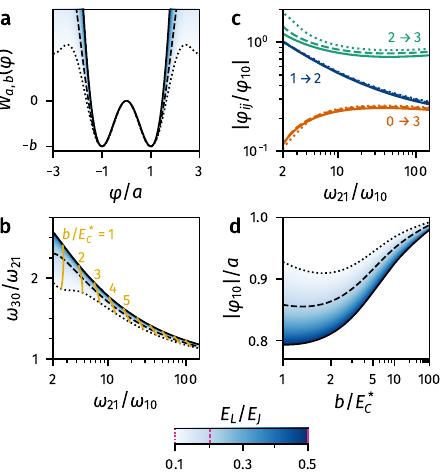}
    \phantomcaption\label{fig:double-well-renorm-potential}
    \phantomcaption\label{fig:double-well-renorm-anharmonicity}
    \phantomcaption\label{fig:double-well-renorm-matrix-elements}
    \phantomcaption\label{fig:double-well-renorm-phi-qubit}
  \end{subcaptiongroup}
  \caption{%
    (\subref{fig:double-well-renorm-potential})
    Comparison of the fluxonium double-well potential
    for varying $E_L/E_J$ ratios and constant barrier height $b$.
    The horizontal axis is normalized by $a$.
    \textbf{Solid},
    \textbf{dashed}
    and
    \textbf{dotted}
    traces
    correspond to $E_L/E_J=0.5$,
    $0.2$ and $0.1$, respectively.
    The \textbf{blue gradient} interpolates
    between the discrete traces
    for intermediate $E_L/E_J$ ratios
    with the corresponding scale given by the color bar at the bottom.
    (\subref{fig:double-well-renorm-anharmonicity})
    Ratio of the plasmon transition frequencies
    $\omega_{30}/\omega_{21}$
    as a function of
    $\omega_{21}/\omega_{10}$.
    \textbf{Yellow} contours denote lines of constant
    dimensionless barrier height
    $b/E_C^*$.
    (\subref{fig:double-well-renorm-matrix-elements})
    Normalized phase matrix elements $|\varphi_{ij}/\varphi_{10}|$
    as a function of
    $\omega_{21}/\omega_{10}$.
    (\subref{fig:double-well-renorm-phi-qubit})
    Qubit phase matrix element $|\varphi_{10}|$,
    normalized by half the interwell distance $a$,
    as a function of $b/E_C^*$.
  }
  \label{fig:double-well-renorm}
\end{figure}

For $a \ll \pi$,
the wavefunction is confined to a region where the cosine is quartic.
In this limit, the double-well potential is self-similar
under simultaneous rescaling of $a$ and $\varphi$
since
\begin{equation}
  W_{a,b}(\varphi)
  \approx
  b
  [
    (\varphi/a)^4
    - 2(\varphi/a)^2
  ]
  \,.
  \label{eq:fluxonium-quartic-approx}
\end{equation}
For the typical
inductive-to-Josephson energy ratios,
the quartic approximation is crude.
\Cref{fig:double-well-renorm-potential}
compares the potential landscape
for $E_L/E_J$ between $0.1$ and $0.5$
at a fixed barrier height $b$,
with the $\varphi$-coordinate normalized by $a$.
The solid, dashed and dotted traces represent the potential
for $E_L/E_J=0.5$, $0.2$ and $0.1$, respectively.
Both potential curves for $E_L/E_J=0.5$ and $0.2$
resemble a quartic double well
in the region of interest.
However,
$E_L/E_J=0.1$ is qualitatively different as it has additional local minima near
$\varphi/a=\pm3$.

\Cref{fig:double-well-renorm-anharmonicity}
shows the relationship between the ratio of the plasmon transition frequencies
$\omega_{30}/\omega_{21}$,
and the
ratio of the two lowest transition frequencies
$\omega_{21}/\omega_{10}$.
We see that
$\omega_{30}/\omega_{21}$
is well predicted by
$\omega_{21}/\omega_{10}$,
with increasing precision at larger values of
$\omega_{21}/\omega_{10}$.
The yellow contours denote lines of constant dimensionless barrier height
$b/E_C^*$,
where
$E_C^* \coloneqq (\pi/a)^2E_C$
is a renormalized charging energy,
motivated by the squeezing of $\varphi$
at larger $E_L/E_J$.
Varying $E_L/E_J$ 
at a fixed $b/E_C^*$
only slightly varies
$\omega_{21}/\omega_{10}$.
Since
the latter
is directly measurable,
it is a better defined
characterization of the fluxonium potential than
the dimensionless barrier height $b/E_C^*$.

\Cref{fig:double-well-renorm-matrix-elements}
shows the magnitude of the phase matrix elements
$\varphi_{ij} \coloneqq \langle i|\varphi|j\rangle$,
normalized by $\varphi_{10}$,
as a function of
$\omega_{21}/\omega_{10}$
for $E_L/E_J=0.1$, $0.2$ and $0.5$.
Both
$\varphi_{21}/\varphi_{10}$
and $\varphi_{30}/\varphi_{10}$
maintain very similar magnitudes
across the three traces,
indicating they are nearly independent of $E_L/E_J$.
There is greater variation in $\varphi_{32}/\varphi_{10}$,
as expected,
since both $2$- and $3$-state wavefunctions
are extended in $\varphi/a$ and
therefore sensitive to $E_L/E_J$.
\Cref{fig:double-well-renorm-phi-qubit}
shows $|\varphi_{10}|/a$
as a function of $b/E_C^*$.
We see that $\varphi_{10}$ scales roughly linearly with
the minima position $a$,
as expected.
This approximation is more accurate in the heavy-fluxonium regime,
where $b/E_C^* > 10$.

Our numerical study confirms the self-similarity predicted by
\cref{eq:fluxonium-quartic-approx}
holds well for typical fluxonium parameters.
That is,
the fluxonium eigenspectrum and relative phase matrix elements
(for the first four levels)
are only weakly dependent on $E_L/E_J$.
Meanwhile,
the absolute magnitude of the phase matrix elements scale
approximately linearly with $a$.

The choice of $E_L/E_J$
can be informed by the desire to minimize losses
due to flux noise or quasiparticle tunneling through the array
\cite{sunCharacterizationLossMechanisms2023,voolNonPoissonianQuantumJumps2014,wangHighcoherenceFluxoniumQubits2025}.
For a fixed qubit frequency
and dimensionless barrier height $b/E_C^*$,
\cref{eq:double-well-depth}
implies that $E_J$ scales approximately as $a^{-4}$,
and by \cref{eq:sinc-ratio-minima-position},
$E_L$ scales as $a^{-4}\sinc a$.
Decay via flux noise or quasiparticle tunneling
can be mitigated by decreasing $E_L$ which implies increasing $a$,
or equivalently, lowering $E_L/E_J$.
Dielectric loss, in contrast,
is largely independent of $E_L/E_J$
since its rate
is proportional to $\varphi_{10}^2/E_C$.
For a fixed qubit frequency and $b/E_C^*$,
the physical charging energy $E_C$ scales as $a^2$,
in the same way as $\varphi_{10}^2$.

\section{Time-glide-reflection symmetry}
\label{sec:time-glide-reflection-symmetry}

The fluxonium Hamiltonian,
\cref{eq:fluxonium-hamiltonian},
is invariant under parity inversion
$\hat\parity$,
which maps the coordinate $\varphi$ to $-\varphi$,
when biased at half-flux quantum,
$\varphi_\extlabel = \pi$.
We will only analyze the control fluxonium
in this appendix,
so we omit the `$\contlabel$' subscripts for simplicity.
The Hamiltonian with a time-periodic drive,
\cref{eq:control-hamiltonian},
breaks parity symmetry because
$\hat n$ is odd under parity.
However,
the Hamiltonian is invariant under the combined transformation
of a half-period translation in time followed by parity inversion:
$
  \hat\parity
  \hat H(t+\pi/\omega_\drivelabel)
  \hat\parity^\dagger
  = \hat H(t)
$.
This is also known as time-glide-reflection symmetry,
in analogy with spatial glide reflection symmetry
\cite{morimotoFloquetTopologicalPhases2017}.
In the extended Hilbert space,
time-glide-reflection is represented by the unitary transformation
$
  \hat\Mparity
  =
  \exp(\pi\hat\partial_\tau/\omega_\drivelabel)
  \hat\parity
$.
Likewise, the Floquet Hamiltonian,
\cref{eq:control-floquet-hamiltonian},
is invariant under
$\hat\Mparity$; that is,
$
\hat\Mparity\hat\floqham\hat\Mparity^\dagger
=
\hat\floqham
$.
Since time-glide-reflection is self-inverse,
$\hat\Mparity^2 = \hat\Id$,
its eigenvalues are either $+1$ or $-1$.
The two eigenspaces of $\hat\Mparity$
are invariant subspaces of
$\hat\floqham$,
so it suffices to restrict our attention to only
those eigenstates that are even under $\hat\Mparity$.

To numerically obtain Floquet modes,
one generally first computes the time-evolution operator
\begin{equation}
  \hat U(t_1, t_0)
  =
  \timeorder
  \exp
  \biggl(\!
    -i
    \int_{t_0}^{t_1}
    \hat H(t)
    \,dt
  \biggr),
\end{equation}
where $\timeorder$ indicates time ordering,
over one full period, $t_1 - t_0 = 2\pi/\omega_\drivelabel$.
Then, solving the eigenequation
\begin{equation}
  \hat U(t_0 + 2\pi/\omega_\drivelabel, t_0)
  |\Phi_j(t_0)\rangle
  =
  e^{-i\epsilon_j(2\pi/\omega_\drivelabel)}
  |\Phi_j(t_0)\rangle
\end{equation}
produces the Floquet modes $|\Phi_j(t)\rangle$ at $t=t_0$ and their quasienergies $\epsilon_j$.
The modes at later times can be reconstructed using
\begin{equation}
  |\Phi_j(t_0 + t)\rangle
  =
  e^{i\epsilon_j t}
  \hat U(t_0 + t, t_0)
  |\Phi_j(t_0)\rangle
  \,.
  \label{eq:floquet-mode-evolution}
\end{equation}
Time-glide-reflection symmetric modes further satisfy
\begin{equation}
  |\Phi_j(t_0)\rangle
  =
  \hat\parity|\Phi_j(t_0 + \pi/\omega_\drivelabel)\rangle
  \,.
  \label{eq:time-glide-reflection-symmetric-mode}
\end{equation}
Applying
\cref{eq:time-glide-reflection-symmetric-mode}
to
\cref{eq:floquet-mode-evolution}
when $t=\pi/\omega_\drivelabel$
gives
\begin{equation}
  \hat\parity
  \hat U(t_0 + \pi/\omega_\drivelabel, t_0)
  |\Phi_j(t_0)\rangle
  =
  e^{-i\epsilon_j(\pi/\omega_\drivelabel)}
  |\Phi_j(t_0)\rangle
  \,.
\end{equation}
Solutions of this eigenequation are time-glide-reflection symmetric Floquet modes.
Two benefits arise from this.
First,
the time-evolution operator only needs to be solved over half a period
which is more computationally efficient.
Second,
accidental degeneracies between opposite parity modes are prevented because
the quasienergies now wrap over $2\omega_\drivelabel$.
One drawback, however, is that
$|\Phi_j(t)\rangle$
can no longer connect to
$|j\rangle$ at zero drive amplitude since the odd
excited states change sign under time-glide-reflection.
In our calculations,
we use an alternative convention where
$\epsilon_j + \omega_\drivelabel$
for odd $j$
correspond to the bare energy at zero drive.

\section{Extra CNOT duration landscapes}
\label{sec:extra-cnot-duration-landscapes}

\begin{figure}
  \begin{subcaptiongroup}
    \includegraphics{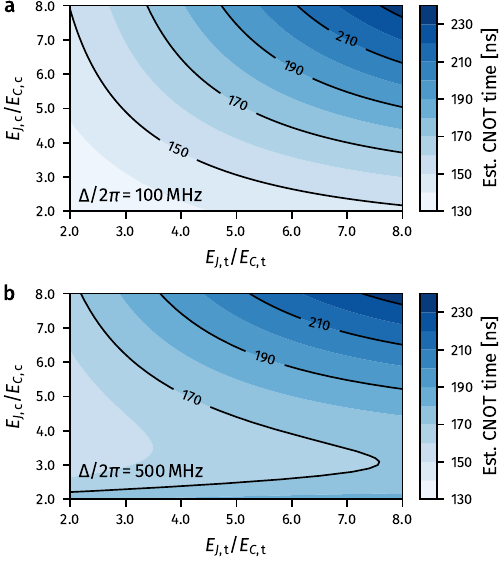}
    \phantomcaption\label{fig:cnot-duration-low-detuning}
    \phantomcaption\label{fig:cnot-duration-high-detuning}
  \end{subcaptiongroup}
  \caption{%
    Estimated CNOT time for two capacitively-coupled fluxonium qubits
    detuned by
    (%
      \subref{fig:cnot-duration-low-detuning},~%
      \subref{fig:cnot-duration-high-detuning}%
    )~%
    $
    \Delta/2\pi
    \coloneqq (\omega^{10}_\targlabel - \omega^{10}_\contlabel)/2\pi
    = (100, 500)\,\unit{\MHz}
    $,
    for varying Josephson-to-charging energy ratios
    $E_{J,q}/E_{C,q}$.
    The sum of the two qubit frequencies is fixed at
    $
    (\omega^{10}_\contlabel + \omega^{10}_\targlabel)/2\pi
    =
    \qty{1.3}{\GHz}
    $,
    the inductive-to-Josephson energy ratio for both qubits is
    $E_{L,q}/E_{J,q}=0.25$
    and
    the coupling strength $J$ is determined by
    $|\mu_{ZZ}|/{2\pi} = \qty{50}{\kHz}$.
  }
  \label{fig:cnot-duration-different-detuning}
\end{figure}

In addition to the parameters in the main text, we have estimated the CNOT time for
two fluxoniums
with $E_{L,q}/E_{J,q} = 0.25$,
detuned by
$(\omega^{10}_\targlabel - \omega^{10}_\contlabel)/2\pi = \qty{100}{\MHz}$
and \qty{500}{\MHz},
while fixing the sum of their frequencies at
$
(\omega^{10}_\contlabel + \omega^{10}_\targlabel)/2\pi
=
\qty{1.3}{\GHz}
$,
the same as in \cref{sec:cnot-duration-landscape}.
The CNOT duration landscape for the small detuning case,
\cref{fig:cnot-duration-low-detuning},
is very similar to both panels of
\cref{fig:cnot-duration}.
For the larger detuning case,
\cref{fig:cnot-duration-high-detuning},
CNOT gates generally take longer
but the overall trend remain similar for
$E_{J,\contlabel}/E_{C,\contlabel} > 3$.
For smaller $E_{J,\contlabel}/E_{C,\contlabel}$,
the control qubit's $1\!\to\!2$ frequency
lies closer to the target-qubit frequency,
strengthening the residual $ZZ$ interaction.
Maintaining $|\mu_{ZZ}|/2\pi$ at the \qty{50}{\kHz} threshold
requires a reduction in $J$,
which in turn increases the CNOT gate duration.

\section{CNOT fidelity computation}
\label{sec:cnot-fidelity-computation}

When
$\mu_{IZ}$ and $\mu_{ZZ}$
are both zero
in the effective CR Hamiltonian,
\cref{eq:effective-hamiltonian-main},
the resulting gate takes the form
of what we define to be an ideal CR unitary,
\begin{equation}
  \hat U_\CRlabel
  (\chi, \theta, \phi)
  \coloneqq
  \exp\ab[
    -i\ab(\!
      \frac{\chi}{2}
      \hat Z_\contlabel \hat X_\targlabel
      +
      \frac{\theta}{2}
      \hat X_\targlabel
      +
      \frac{\phi}{2}
      \hat Z_\contlabel
    \!)
  \!]
  \,.
\end{equation}
For the purpose of this appendix, we set aside the distinction between dressed and bare states.
The case $\chi=\tfrac{\pi}{2}$ is locally equivalent to the CNOT gate,
differing by only control-qubit $Z$ and target-qubit $X$ rotations.
We view this to be a trivial difference,
and therefore define a family of CNOT unitaries considered to be equivalent,
\begin{equation}
  \hat U_\CNOTlabel(\theta, \phi)
  \coloneqq
  \hat U_\CRlabel(
  \tfrac{\pi}{2},
  \theta-\tfrac{\pi}{2},
  \phi-\tfrac{\pi}{2}
  )
  \,,
\end{equation}
with the standard CNOT matrix given by $\theta=\phi=0$, up to an overall phase.
The average CNOT fidelity is then understood to mean the maximal average gate fidelity to
$\hat U_\CNOTlabel(\theta, \phi)$,
optimized over $\theta$ and $\phi$:
\begin{equation}
  F_\CNOTlabel(\hat M)
  \coloneqq
  \max_{(\theta,\phi)}
  F(\hat M, \hat U_\CNOTlabel(\theta, \phi))
  \,,
  \label{eq:cnot-fidelity-definition}
\end{equation}
where
\begin{equation}
  F(\hat M, \hat U_0)
  =
  \frac{1}{20}
  \bigl[
    \Tr(\hat M^\dagger\hat M)
    +
    |\Tr(\hat U_0^\dagger\hat M)|^2
  \bigr]
  \,,
  \label{eq:average-gate-fidelity-formula}
\end{equation}
is a formula for calculating the average gate fidelity
to an ideal two-qubit unitary $\hat U_0$
\cite{pedersenFidelityQuantumOperations2007}.
The imperfect gate $\hat M$ need not be unitary since it is usually obtained by
truncating a unitary acting on the full Hilbert space to the computational subspace.

It follows from \cref{eq:average-gate-fidelity-formula}
that the optimal angles $\theta^\star$ and $\phi^\star$
of \cref{eq:cnot-fidelity-definition}
satisfy
\begin{equation}
  (\theta^\star, \phi^\star)
  =
  \argmax_{(\theta,\phi)}
  \bigl|
    \Tr\bigl(
      \hat U_\CNOTlabel^\dagger(\theta, \phi)\hat M
    \bigr)
  \bigr|^2
  \label{eq:optimization-theta-phi}
  .
\end{equation}
By decomposing the trace as
\begin{equation}
  \bigl|
  \Tr\bigl(
    \hat U_\CNOTlabel^\dagger(\theta, \phi)\hat M
  \bigr)
  \bigr|
  =
  \bigl|
  z_+(\phi)
  e^{i\theta/2}
  + z_-(\phi)
  e^{-i\theta/2}
  \bigr|
\end{equation}
where
\begin{gather}
  z_\pm(\phi)
  \coloneqq
  r_{0\pm}
  e^{i(\gamma_{0\pm} + \phi/2)}
  \pm
  r_{1\pm}
  e^{i(\gamma_{1\pm} - \phi/2)}
  \,,\\
  r_{j\pm}
  e^{i\gamma_{j\pm}}
  \coloneqq
  \langle {j\pm} |
  \hat M
  | {j\pm} \rangle
  \,,
\end{gather}
we deduce the constraint
$
  \theta^\star
  =
  \angle
  z_-(\phi^\star)
  -
  \angle
  z_+(\phi^\star)
$,
which reduces
\cref{eq:optimization-theta-phi}
to a one-dimensional maximization,
\begin{equation}
  \phi^\star
  =
  \argmax_{\phi}
  \bigl\{
  |z_+(\phi)|
  + |z_-(\phi)|
  \bigr\}
  \,.
  \label{eq:phi-maximization}
\end{equation}
The magnitudes in
\cref{eq:phi-maximization}
in terms of $r_{j\pm}$ and $\gamma_{j\pm}$ are
\begin{equation}
  |z_\pm(\phi)|
  =
  \sqrt{
    r_{0\pm}^2
    + r_{1\pm}^2
    \pm 2r_{0\pm}r_{1\pm}
    \cos(
      \phi
      -
      \delta_\pm
    )
  }
  \,,
\end{equation}
where
$
  \delta_\pm
  \coloneqq
  \gamma_{1\pm}
  -
  \gamma_{0\pm}
$.
\Cref{eq:phi-maximization},
in general, has no closed-form solution.
However,
if
$|j\pm\rangle$
are approximate eigenvectors of $\hat M$,
we have $r_{j\pm}\approx1$
and this leads to
\begin{align}
  |z_+(\phi)|
  &\approx
  2
  \Bigl|
  \cos\Bigl(
    \frac{\phi - \delta_+}{2}
  \Bigr)
  \Bigr|
  \,,\\
  |z_-(\phi)|
  &\approx
  2
  \Bigl|
  \sin\Bigl(
    \frac{\phi - \delta_-}{2}
  \Bigr)
  \Bigr|
  \,,
\end{align}
whose sum is maximized by
\begin{equation}
  \phi^\star
  \approx
  \delta_-
  + \frac12
  \bmod\!(
  \delta_+ - \delta_-,
  2\pi
  )
  + \frac{\pi}{2}
  \,,
\end{equation}
where $\bmod(x, y)$ is the modulo operator that wraps $x$ within $[0, y)$.
This approximation can then be used as an initial guess
for solving \cref{eq:phi-maximization} iteratively.

In the presence of small $\mu_{IZ}$ or $\mu_{ZZ}$,
the final unitary will contain the Pauli products
$
\hat I_\contlabel
\hat Z_\targlabel
$,
$
\hat Z_\contlabel
\hat Z_\targlabel
$,
$
\hat I_\contlabel
\hat Y_\targlabel
$
and
$
\hat Z_\contlabel
\hat Y_\targlabel
$.
It is not possible to distinguish whether these errors originate from
$\mu_{IZ}$ or $\mu_{ZZ}$
by examining the final unitary alone,
as their effects are mixed by the dominant Hamiltonian terms.
Hence, we refer to both contributions collectively as ``phase error'' in 
\cref{fig:time-domain-amplitude-sweep-error}.

\section{Effective energy spectral density derivation}
\label{sec:effective-energy-spectral-density-derivation}

The effective Hamiltonian of
\cref{eq:effective-hamiltonian-main}
is valid when only the target-qubit transition is resonantly driven.
The adiabatic theorem implies that
off-resonant transitions do not induce population transfer
in the instantaneous eigenbasis.
Adiabaticity is violated when the
instantaneous energy gaps narrow
during the CR drive.
This violation can occur in two ways.
In the first case,
ramping the drive too quickly
leads to diabatic transitions between control-fluxonium eigenstates,
as shown in \cref{fig:control-qubit-collisions-peaks}.
In the second case,
the control fluxonium \emph{would} evolve adiabatically
in the absence of coupling ($J=0$),
but finite coupling permits
joint transitions between control-fluxonium Floquet eigenstates
and target-fluxonium eigenstates.
Diabatic errors in the second case
scale as $O(J^2)$
and can be analyzed
by building upon the semi-analytic framework developed in
\cref{sec:effective-cr-hamiltonian}.

\subsection{Drive-activated scattering}
\label{sec:drive-activated-scattering}

As defined in \cref{sec:semi-analytical-block-diagonalization},
the Floquet Hamiltonian
$\hat\floqham_\contlabel(\Omega)$
has eigenstates
$
\{|\Phi_{j,k}(\Omega)\rrangle\}_{j\in\NN_0,k\in\ZZ}
$
with eigenvalues
$\epsilon_j(\Omega) + k\omega_\drivelabel$.
Further, their phases are defined such that
$\Im\llangle\Phi_{j,k}(\Omega)|\partial_\Omega|\Phi_{j,k}(\Omega)\rrangle=0$.
If the drive envelope $\Omega(s)$ varies sufficiently slowly,
the evolution of the driven control qubit \emph{in isolation},
follows the adiabatic trajectory
\begin{align}
  \hat U_\contlabel(s_1, s_0)
  \coloneq
  \!
  \smash{\sum_{
    \substack{
      j\in\NN_{\mathrlap{0}}
      \\
      k\in\mathrlap{\ZZ}\phantom{\NN_{\mathrlap{0}}}
    }
  }}\,
  \biggl\{\!
  &
    \exp\biggl(
      {-i}\!\int_{s_0}^{s_1}\!\!
      \epsilon_j(\Omega(s))
      + k\omega_\drivelabel
      \,ds
    \biggr)
    \label{eq:isolated-control-qubit-adiabatic-evolution}
    \\
    &
    \times
    |\Phi_{j,k}(\Omega(s_1))\rrangle
    \llangle\Phi_{j,k}(\Omega(s_0))|
  \biggr\}
  \nonumber
  .
\end{align}
For the isolated target qubit, its evolution is simply
\begin{equation}
  \hat U_\targlabel(s_1, s_0)
  \coloneqq
  \sum_{j\in\NN_0} \exp\bigl(
    {-i}
    E_\targlabel^j
    \times(s_1 - s_0)
  \bigr)
  |j\rangle\langle j|
  \,.
  \label{eq:isolated-target-qubit-evolution}
\end{equation}
Our goal is to perturbatively study how the interaction
$J\hat n_\contlabel \hat n_\targlabel$
affects the joint evolution of the two qubits.
To facilitate this,
we move into an interaction picture
where the isolated evolutions,
\cref{%
eq:isolated-control-qubit-adiabatic-evolution,%
eq:isolated-target-qubit-evolution%
},
are rendered static by transforming states
from the Schr\"odinger picture by
the unitary
$
\hat A(s)
\coloneqq
\hat A_\contlabel(s)
\otimes
\hat A_\targlabel(s)
$
where
\begin{equation}
  \hat A_q(s) \coloneqq \hat U_q^\dagger(s, s_\reflabel)
  \label{eq:adiabatic-interaction-picture-transformation}
\end{equation}
and $s_\reflabel$ is an arbitrary reference time
when the drive is off, $\Omega(s_\reflabel) = 0$.
The full Hamiltonian,
\cref{eq:floquet-hamiltonian},
in this interaction picture is
\begin{align}
  \hat \floqham_\intlabel(s)
  \coloneqq{}&
  \hat A(s)
  \bigl[
    \hat \floqham(\Omega(s))
    -i\partial_s
  \bigr]
  \hat A^\dagger(s)
  \\
  \approx{}&
  J
  \bigl[
    \hat A_\contlabel(s)
    \hat n_\contlabel
    \hat A_\contlabel^\dagger(s)
  \bigr]
  \bigl[
    \hat A_\targlabel(s)
    \hat n_\targlabel
    \hat A_\targlabel^\dagger(s)
  \bigr]
  \,.
\end{align}
The approximation in the last line follows from the adiabatic theorem,
which states that
\begin{equation}
  i\partial_s
  \hat A^\dagger(s)
  \approx
  \hat\floqham_0(\Omega(s))
  \hat A^\dagger(s)
  \,,
  \label{eq:adiabatic-transformation-adiabatic-limit}
\end{equation}
where $\hat\floqham_0$ is the full Hamiltonian,
\cref{eq:floquet-hamiltonian},
with $J=0$.

The true adiabatic limit is generally unattainable
since small avoided crossings often appear in the quasienergy spectrum
(see, for example, \cref{fig:optimal-drive-amplitude-spectrum-small-ac})
\cite{weinbergAdiabaticPerturbationTheory2017,honeTimedependentFloquetTheory1997}.
Very narrow gaps
(e.g., less than $2\pi\times\qty{10}{\MHz}$),
are assumed to be traversed completely diabatically at the cost of
introducing
small discontinuities
in the unitary of
\cref{eq:isolated-control-qubit-adiabatic-evolution}.
However,
for ``intermediate'' gaps,
the adiabatic approximation breaks down.

Let the evolution unitary of the coupled system
in the interaction picture be
\begin{equation}
  \hat U_\intlabel(s_1, s_0)
  =
  \timeorder
  \exp\ab(\!
  -i
  \int_{s_0}^{s_1}
  \hat\floqham_\intlabel(\Omega(s))
  \,ds
  )
  \,,
  \label{eq:interaction-unitary-formal}
\end{equation}
where $\timeorder$ indicates time ordering.
Given two eigenstates
of the idle Hamiltonian
$\hat\floqham(\Omega=0)$,
say
$
|
\widetilde{\text{\i n}}
\rrangle
$
and
$
|
\rlap{\raisebox{-0.45ex}{$\widetilde{\phantom{out}}$}}
\outlabel
\rrangle
$,
the probability
$
|
\widetilde{\text{\i n}}
\rrangle
$
is driven to
$
|
\rlap{\raisebox{-0.45ex}{$\widetilde{\phantom{out}}$}}
\outlabel
\rrangle
$
is
\begin{equation}
  P^{\inlabel\to\outlabel}
  \coloneqq
  \bigl|
  \llangle
  \rlap{\raisebox{-0.45ex}{$\widetilde{\phantom{out}}$}}
  \outlabel
  |
  \hat U_\intlabel(c, -c)
  |
  \widetilde{\text{\i n}}
  \rrangle
  \bigr|^2
  ,
  \label{eq:transition-probability-dressed}
\end{equation}
where $c > 0$ is some arbitrary cutoff such that $\Omega(s)=0$ for $|s|>c$.
The transition probability does not depend on the value of $c$
because
$
|
\widetilde{\text{\i n}}
\rrangle
$
and
$
|
\rlap{\raisebox{-0.45ex}{$\widetilde{\phantom{out}}$}}
\outlabel
\rrangle
$
are eigenstates of
$
\hat\floqham(\Omega=0)
$.
To express
\cref{eq:transition-probability-dressed}
in terms of bare eigenstates,
we suppose the coupling strength $J$ is adiabatically switched on
by making the substitution $J\mapsto w_{b,c}(s) J$
where $w_{b,c}(s)$ is zero for $|s| > c$ and ramps to unity for $|s|$
smaller than another cutoff $b < c$.
The new cutoff is still large enough such that
$\Omega(s) = 0$ for $|s| > b$.
This ensures the fictitious tuning of $J$ does not influence the drive-activated scattering process.
The dressed states
in \cref{eq:transition-probability-dressed}
are then replaced by their bare counterparts,
$|\inlabel\rrangle$
and
$|\outlabel\rrangle$,
which are eigenstates of the \emph{uncoupled} idle Hamiltonian
$\hat\floqham_0(\Omega=0)$.
By expanding \cref{eq:interaction-unitary-formal}
as a Dyson series
and
assuming that
$|\inlabel\rrangle$
and
$|\outlabel\rrangle$
are orthogonal,
we can compute the scattering probability to leading order as
\begin{equation}
  P^{\inlabel\to\outlabel}
  \!
  =
  \!
  J^2
  \biggl|
  \int_{\mathrlap{-c}}^c
    \!
    w_{b,c}(s)
    \llangle \outlabel |
    \hat n_{\contlabel, \intlabel}(s)
    \hat n_{\targlabel, \intlabel}(s)
    | \inlabel \rrangle
    ds
  \biggr|^2
  \!\!\mathrlap{,}
  \label{eq:transition-probability-part-with-cutoff}
\end{equation}
where
$
\hat n_{q,\intlabel}
\coloneqq
\hat A_q
\hat n_{q}
\hat A_q^\dagger
$
are the interaction-picture charge operators.
(The equality in
\cref{eq:transition-probability-part-with-cutoff},
as well as in subsequent equations,
should be understood to hold only to leading order in $J$.)
In the limit $b,c\to\infty$,
the integral can be replaced by a Fourier transform:
\begin{equation}
  P^{\inlabel\to\outlabel}
  =
  J^2
  \bigl|
  \fourier[
    \llangle \outlabel|
    \hat n_{\contlabel, \intlabel}(s)
    \hat n_{\targlabel, \intlabel}(s)
    |\inlabel \rrangle
  ](0)
  \bigr|^2
  \!\!,
  \label{eq:transition-probability-in-out}
\end{equation}
where the Fourier transform is defined as
$
  \fourier[f(s)](\omega)
  =
  \int_{-\infty}^\infty
  f(s)
  e^{-i\omega s}
  ds
$
for convergent improper integrals
\cite{lighthillIntroductionFourierAnalysis1958}.

To utilize
\cref{eq:transition-probability-in-out}
semi-analytically,
suppose
$|\inlabel\rrangle$
and $|\outlabel\rrangle$
are simple product states of the form
$
  |ij^k\rrangle
  \coloneqq
  |e_k\rangle
  \otimes
  |i\rangle
  \otimes
  |j\rangle
$.
Since the Floquet Hamiltonian
is invariant up to a global energy shift
under the discrete translation
$|e_k\rangle \mapsto |e_{k+\ell}\rangle$,
it suffices to consider `in' states of the form
$|ij^0\rrangle$
and `out' states of the form
$|fg^k\rrangle$.
\Cref{eq:transition-probability-in-out}
then becomes
\begin{equation}
  P^{ij^0 \to fg^k}
  =
  J^2
  \bigl|
  n_\targlabel^{gj}
  \bigr|^2
  \bigl|
  \tilde{n}_{\contlabel,\intlabel}^{[k]fi}
  (-\omega_\targlabel^{gj})
  \bigr|^2
  \!\!,
  \label{eq:transition-probability-product-state}
\end{equation}
where
$
\tilde{n}_{\contlabel,\intlabel}^{[k]fi}(\omega)
\coloneqq
\fourier[n_{\contlabel,\intlabel}^{[k]fi}(s)](\omega)
$
is the Fourier transform of
$
  n_{\contlabel,\intlabel}^{[k]fi}
  \!\coloneqq
  \Tr(
  \hat n_{\contlabel,\intlabel}
  |e_0\rangle
  \langle e_k|
  \otimes
  |i\rangle
  \langle f|
  )
$.
This expression
can be simplified
by summing over the output indices:
\begin{align}
  P^{ij\to*}
  \coloneqq{}&
  \sum_{f,g \in \NN_0}
  \sum_{k \in \ZZ}
  P^{ij^0\to fg^k}
  \label{eq:transition-probability-product-state-summed}
  \\
  ={}&
  J^2
  \smash{\sum_{g\in\NN_0}}
  \bigl|n_\targlabel^{gj}\bigr|^2
  \ESD_i
  (-\omega_\targlabel^{gj})
  \,,
\end{align}
where
\begin{equation}
  \ESD_i(\omega)
  \coloneqq
  \sum_{k \in \ZZ}
  \sum_{f \in \NN_0}
  \bigl|
    \tilde{n}_{\contlabel,\intlabel}^{[k]fi}
    (\omega)
  \bigr|^2
  \label{eq:control-fluxonium-effective-esd-i-state}
\end{equation}
is the \emph{effective energy spectral density}
for the control-qubit charge operator,
when the qubit is
in the $i$~state.
The summand
in \cref{eq:transition-probability-product-state-summed}
is zero when
$i=f$,
$j=g$
and $k=0$,
so $P^{ij\to*}$ equals
the probability
for an initial $|ij^0\rrangle$
to scatter
to any \emph{orthogonal} final state.

Since the target is resonantly driven,
the non-separable states
$
  |i\pm^k\rrangle
  \coloneqq
  (
    |i0^k\rrangle
    \pm
    |i1^{k-1}\rrangle
  )/\sqrt2
$
are also good `in' and `out' eigenstates.
However, there is a minor subtlety in
the calculation of
$P^{i+\to*}$,
the scattering probability
for a target qubit in the `$+$' state.
If the final state is an orthogonal product state,
this probability is simply
\begin{equation}
  P^{i+^0\to fg^k}
  =
  \frac12
  \bigl(
    P^{i0^0\to fg^k}
    +
    P^{i1^0\to fg^k}
  \bigr)
  .
\end{equation}
However,
if the final state is $|f\pm^k\rrangle$,
the probability is
\begin{flalign}
  &P^{i+^0\to f\pm^k}
  =
  \frac14
  \bigl(
    P^{i0^0\to f1^{k-1}}
    +
    P^{i1^0\to f0^{k+1}}
  \bigr)
  \\
  &\quad\qquad\pm
  \frac{J^2|n_\targlabel^{10}|^2}{2}
  \Re\Bigl\{
    \overline{\tilde{n}_{\contlabel,\intlabel}^{[k+1]fi}(\omega_\targlabel^{10})}
    \tilde{n}_{\contlabel,\intlabel}^{[k-1]fi}(-\omega_\targlabel^{10})
  \Bigr\}
  \nonumber
  .
\end{flalign}
For the special case
$i=f$ and $k=0$,
we have
\begin{equation}
  P^{i+^0\to i\pm^0}
  =
  \frac12
  P^{i0^0\to i1^{-1}}
  \!\!
  \pm
  \frac{J^2|n_\targlabel^{10}|^2}{2}
  \Re\bigl\{
    \tilde{n}_{\contlabel,\intlabel}^{[-1]ii}(-\omega_\targlabel^{10})^2
  \bigr\}
  .
\end{equation}
This can be simplified by noting that the reference time
$s_\reflabel$
in
\cref{eq:adiabatic-interaction-picture-transformation}
is arbitrary.
We can therefore choose $s_\reflabel$ such that
the Fourier amplitude
$
\tilde{n}_{\contlabel,\intlabel}^{[-1]ii}(-\omega_\targlabel^{10})
$
is purely real,
leading to the simplification
$
P^{i+^0\to i+^0}
= P^{i0^0\to i1^{-1}}
$
and
$
P^{i+^0\to i-^0}
=
0
$.
We now have all the ingredients needed to sum
$P^{i+^0\to\outlabel}$
over orthogonal `out' states,
which yields
\begin{equation}
  P^{i+\to*}
  =
  \frac12
  \bigl(
    P^{i0\to*}
    +
    P^{i1\to*}
  \bigr)
  - P^{i0^0\to i1^{-1}}
  .
  \label{eq:scattering-probability-from-i-plus}
\end{equation}
If we assume
$\ESD_i(\omega)$
has only localized peaks
centered on the transition energies between Floquet eigenstates,
and that no accidental resonances exist,
we can use the approximation
$
\ESD_i(-\omega_\targlabel^{10})
\approx
|
  \tilde{n}_{\contlabel,\intlabel}^{[-1]ii}
  (-\omega_\targlabel^{10})
|^2
$
to express
\cref{eq:scattering-probability-from-i-plus}
also in terms of the effective ESD:
\begin{equation}
  P^{i+\to*}
  =
  \frac{J^2}{2}
  \sum_{g=2}^\infty
  \sum_{j=0}^1
  \bigl|n_\targlabel^{gj}\bigr|^2
  \ESD_i
  (-\omega_\targlabel^{gj})
  \,.
\end{equation}
This sum is equal to the \emph{leakage} probability
of the target qubit, averaged over its initial qubit states.
The formula
is unchanged
if the target qubit is initially in the `$-$' state,
that is,
$
P^{i-\to*}
=
P^{i+\to*}
$.

\subsection{Comparison with numerical simulations}
\label{sec:comparison-with-numerical-simulations-esd}

\begin{figure}
  \includegraphics{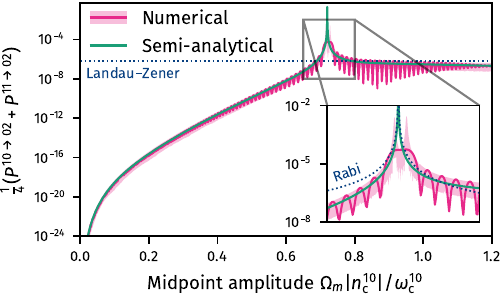}
  \caption{%
    Leakage error
    from \cref{fig:time-domain-amplitude-sweep-error},
    shown in \textbf{solid pink},
    for a calibrated CNOT gate 
    between fluxonium qubits with parameters given in
    \cref{table:time-domain-parameters}.
    The \textbf{filled light pink} area shows the \num{90}th percentile
    leakage error for a soft square CR pulse with
    plateau duration uniformly sampled up to \qty{1}{\ms},
    and the same ramp duration and carrier frequency as the calibrated CNOT pulse.
    The \textbf{solid green} line 
    uses the semi-analytic formula,
    \cref{eq:transition-probability-product-state},
    to compute the leakage probabilities.
    The \textbf{dotted} lines correspond to predictions from
    the Landau--Zener formula and,
    in the inset,
    Rabi's formula.
  }
  \label{fig:semi-analytic-leakage-comparison}
\end{figure}

In \cref{fig:semi-analytic-leakage-comparison},
we compare the semi-analytic prediction of
\cref{eq:transition-probability-product-state},
in solid green,
to the leakage error first shown in
\cref{fig:time-domain-amplitude-sweep-error},
reproduced here in solid pink.
Leakage induced by the CR pulse depends on the exact pulse duration
due to interference effects,
resulting in visible ripples in the solid pink curve.
To smooth over these transient features,
we also show in filled light pink,
the \num{90}th-percentile range of leakage errors
for a soft square CR pulse
with a plateau duration uniformly sampled up to \qty{1}{\ms},
and the same ramp duration and carrier frequency as the calibrated CNOT pulse.
In the semi-analytic calculation,
the drive envelope is assumed to have an infinitely long plateau,
which also eliminates transient effects.
Computing the spectrum of a function with an infinitely long plateau is discussed in
\cref{sec:ft-infinitely-long-square-pulse}.
Away from resonance,
the semi-analytical calculation agrees with the fully numerical result 
over many orders of magnitude,
confirming its high accuracy.
Near resonance (see inset)
it qualitatively deviates from the numerical result,
predicting only a single peak rather than the two observed numerically.
The two peaks originate from the Rabi splitting of
$|1+^k\rrangle$
and $|1-^k\rrangle$,
a feature not captured by leading-order perturbation theory.

Using
\cref{eq:transition-probability-product-state}
requires numerically computing the Fourier transform of interaction-picture matrix elements,
a procedure that is technically involved.
A simpler model treats the leakage process instead as
adiabatic traversal of an avoided crossing in the Floquet quasienergy spectrum.
Let the detuning between
$|11^4\rrangle$
and
$|02^0\rrangle$
be denoted by
\begin{equation}
  \Delta(\Omega)
  =
  [
    \epsilon_1(\Omega)
    - \epsilon_0(\Omega)
    + 4\omega_\drivelabel
  ]
  - \omega_\targlabel^{21}
  \,,
\end{equation}
and let
$\Omega_\aclabel$
be the amplitude at the avoided crossing,
which we define as minimizing
$|\Delta|$.
In this case,
$\Omega_\aclabel|n_\contlabel^{10}|/\omega_\contlabel^{10} = 0.72$
and
$|\Delta(\Omega_\aclabel)|/2\pi = \qty{50}{\kHz}$.
The diabatic traversal probability can be approximated using the
Landau--Zener formula 
\cite{shevchenkoLandauZenerStuckelberg2010}
\begin{equation}
  P_\LZlabel
  =
  \exp
  \ab(
  -2\pi
  \frac{
    |\Delta(\Omega_\aclabel)|^2
  }{
    4v
  }
  )
  \,,
\end{equation}
where
$
  v
  =
  |
  \dot\Omega
  \times
  (d\Delta/d\Omega)(\Omega_\aclabel)
  |
$
is the Landau--Zener velocity,
which we approximate by assuming the slope of the envelope at the avoided crossing is
$\dot\Omega\approx \Omega_\aclabel/t_\ramplabel$.
Since the avoided crossing is traversed twice, once during each ramp of the pulse,
the final leakage probability is
$P^{11\to02} \approx 2P_\LZlabel(1 - P_\LZlabel)$
\cite{shevchenkoLandauZenerStuckelberg2010}.
This approximation is shown as the
dotted horizontal line in
\cref{fig:semi-analytic-leakage-comparison}
and roughly matches the
numerical result when the avoided crossing is fully traversed
($\Omega_\midlabel>\Omega_\aclabel$).

The Landau--Zener formula is inaccurate if the system dwells near the avoided crossing
($\Omega_\midlabel\approx\Omega_\aclabel$).
In this case,
we turn to the standard Rabi oscillation formula,
\begin{equation}
  P^{11\to02}(\Omega_\midlabel)
  \approx
  \frac{
    |\Delta(\Omega_\aclabel)|^2
  }{
    4|\Delta(\Omega_\midlabel)|^2
    + |\Delta(\Omega_\aclabel)|^2
  }
  \label{eq:rabi-approximation-avoided-crossing}
  \,.
\end{equation}
We further simplify $\Delta(\Omega_\midlabel)$ by
Taylor-expanding $\Delta(\Omega)$
to linear order
around $\Omega=\Omega_\aclabel$,
making
\cref{eq:rabi-approximation-avoided-crossing}
a Lorentzian in $\Omega_\midlabel$.
The Rabi approximation is shown
as the dotted curve in the inset of
\cref{fig:semi-analytic-leakage-comparison}
and agrees well with the semi-analytical result
when very close to resonance,
but overestimates the error
for $\Omega<\Omega_\aclabel$
and underestimates
for $\Omega>\Omega_\aclabel$.

\section{Infinite soft square pulse spectrum}
\label{sec:ft-infinitely-long-square-pulse}

The definition of the effective ESD
contains the Fourier transform of terms like
$f(s) = a(s)e^{i\phi(s)}$,
where
\begin{equation}
  a(s)
  =
  \llangle
    \Phi_{j,k}(\Omega(s))
    |
    \hat n_\contlabel
    |
    \Phi_{i,0}(\Omega(s))
  \rrangle
\end{equation}
is a charge matrix element
in the instantaneous Floquet eigenbasis and
\begin{align}
  \phi(s)
  &=
  \int_{s_\reflabel}^s
  \Delta(\Omega(s'))
  \,
  ds'
  \,,\\
  \Delta(\Omega)
  &=
  \epsilon_j(\Omega)
  - \epsilon_i(\Omega)
  - k\omega_\drivelabel
\end{align}
is the dynamical phase accumulated by the
interaction-picture matrix element
and the Stark-shifted transition frequency,
respectively.
As we ultimately only care about the spectral density,
we set $s_\reflabel=0$,
since this only affects the overall phase of $f(s)$.
Let the drive envelope be a symmetric soft square pulse
with plateau duration $2\tau$ and midpoint amplitude $\Omega_\midlabel$,
we express this as
\begin{equation}
  \Omega(s)
  =
  \Omega_\midlabel
  \bigl[
    \rampfunc(s - \tau)
    + \rampfunc(-s - \tau)
    - 1
  \bigr]
  \,,
\end{equation}
where $\rampfunc(s)$ is a ramp function
that smoothly interpolates between
$\rampfunc(-\infty) = 1$
and $\rampfunc(+\infty) = 0$
over a characteristic ramp duration;
e.g., \cref{eq:planck-taper}.
We seek to compute the ESD
of $f(s)$
averaged over long $\tau$,
such that the $\sinc$ ripples in the spectrum are also averaged away.
This is not straightforward to directly calculate numerically,
so we first perform a series of analytic transformations to relate
the $\tau$-averaged spectrum to the spectrum of a compactly supported signal.

In the long $\tau$ limit, we have the pointwise convergence
\begin{equation}
  \lim_{\tau\to\infty}
  \Omega(s + \tau)
  =
  \Omega_\midlabel
  \rampfunc(s)
  \,,
  \label{eq:half-envelope-pointwise-convergence}
\end{equation}
which implies
\begin{align}
  b(s)
  &\coloneqq
  \lim_{\tau\to\infty}
  a(s+\tau)
  \,,\\
  \theta(s)
  &\coloneqq
  \lim_{\tau\to\infty}
  \bigl\{
    \phi(s+\tau)
    - \phi(\tau)
  \bigr\}
\end{align}
are both well defined and given by
\begin{align}
  b(s)
  &=
  \llangle
    \Phi_{j,k}(\Omega_\midlabel\rampfunc(s))
    |
    \hat n_\contlabel
    |
    \Phi_{i,0}(\Omega_\midlabel\rampfunc(s))
  \rrangle
  \,,\\
  \theta(s)
  &=
  \int_0^s
  \Delta(\Omega_\midlabel\rampfunc(s'))
  \,
  ds'
  \,.
\end{align}
Define
$g(s) \coloneqq b(s)e^{i\theta(s)}$.
For long $\tau$, we have
\begin{equation}
  f(s+\tau)
  \approx
  g(s)e^{i\phi(\tau)}
  \approx
  g(s)e^{i[\Delta(\Omega_\midlabel)\tau + \delta]}
  \,,
  \label{eq:half-signal-approx}
\end{equation}
where
$
\delta
\coloneqq
\lim_{\tau\to\infty}
\{
\phi(\tau)
-
\Delta(\Omega_\midlabel)\tau
\}
$.
At first sight,
it appears
this equation
establishes a relation between
$\fourier[f(s)](\omega)$
and
$\fourier[g(s)](\omega)$,
however, it only holds for $s$ greater than a finite lower bound
when $\tau$ is fixed.
To address this,
we perform the transformation
\begin{equation}
  f_1(s)
  \coloneqq
  \biggl(
  \Delta(\Omega_\midlabel)
  + i\frac{d}{ds}
  \biggr)
  f(s)
  \,,
\end{equation}
which corresponds to multiplication by
$\Delta(\Omega_\midlabel) - \omega$
in frequency space.
Define $g_1(s)$ analogously by replacing $f$ with $g$.
The benefit of doing this is that the following relation now holds for all $s$,
\begin{equation}
  f_1(s)
  \approx
  g_1(s-\tau)
  e^{i[\Delta(\Omega_\midlabel)\tau + \delta]}
  + g_1^*(-s-\tau)
  e^{-i[\Delta(\Omega_\midlabel)\tau + \delta]}
  \,.
  \label{eq:transformed-full-signal-approx}
\end{equation}
The underlying idea is that
$f_1(s)$ is constructed to vanish on the pulse interior
and $g_1(s)$ corresponds to the positive half of $f_1(s)$,
modulo the dynamical phase.
It follows from
\cref{eq:transformed-full-signal-approx}
that
\begin{equation}
  \bigl|
    \fourier[f_1(s)](\omega)
  \bigr|^2
  =
  2\bigl|
    \fourier[g_1(s)](\omega)
  \bigr|^2
  +
  \text{oscillatory}
  \,,
\end{equation}
where the omitted terms are oscillatory in $\tau$
and will be averaged away,
except when $\omega=\Delta(\Omega_\midlabel)$.
By using this equation and dropping the oscillatory terms,
the ESD of the original function $f(s)$ can be expressed as
\begin{equation}
  \bigl|
    \fourier[f(s)](\omega)
  \bigr|^2
  =
  \frac{
    2\bigl|
      \fourier[g_1(s)](\omega)
    \bigr|^2
  }{
    \bigl[
      \omega - \Delta(\Omega_\midlabel)
    \bigr]^2
  }
  \,.
\end{equation}
The contribution of the dropped terms
when $\omega=\Delta(\Omega_\midlabel)$
is irrelevant since
the spectrum
anyway diverges there.

We may factor $g_1(s)$ into $b_1(s)e^{i\theta(s)}$ where
\begin{align}
  b_1(s)
  &\coloneqq
  \bigl[
    \nu(s)
    - \Delta(\Omega_\midlabel)
  \bigr]
  b(s)
  + i\frac{db(s)}{ds}
  \label{eq:first-chirplet-amplitude-definition}
  \,,\\
  \nu(s)
  &\coloneqq
  \frac{d\theta(s)}{ds}
  =
  \Delta(\Omega_\midlabel\rampfunc(s))
  \,.
\end{align}
Notice that
$b_1(-\infty) = 0$
because
the first term of
\cref{eq:first-chirplet-amplitude-definition}
vanishes due to
$\nu(-\infty) = \Delta(\Omega_\midlabel)$
and the second term vanishes in either limit $s \to \pm\infty$.
This is consistent with the interpretation
that $g_1(s)$ corresponds to the positive half of $f_1(s)$.
Further,
if the matrix element is zero in the bare basis,
the second term also vanishes as $s\to\infty$
because $b(\infty) = 0$.
When this is the case,
$b_1(s)$ acts as a finite window for the chirp
$e^{i\theta(s)}$,
and we refer to $g_1(s)$ as the \emph{first ramp chirplet}.
Because the support of $g_1(s)$ is effectively finite,
its Fourier transform is readily computable numerically.

Otherwise,
if the matrix element is nonzero in the bare basis,
e.g. when $i=0$, $j=1$ and $k=0$,
then
$g_1(s)$ remains finite for positive $s$.
We can play the same trick again and define
\begin{equation}
  f_2(s)
  \coloneqq
  \biggl(
  \Delta(0)
  + i\frac{d}{ds}
  \biggr)
  f_1(s)
  \,,
\end{equation}
likewise for $g_2(s)$.
Factoring $g_2(s)$ into $b_2(s)e^{i\theta(s)}$
yields
\begin{align}
  b_2(s)
  \coloneqq{}&
  \bigl[
    \nu(s)
    - \Delta(0)
  \bigr]
  b_1(s)
  + i\frac{db_1(s)}{ds}
  \\
  ={}&
  C_0(s)
  b(s)
  +
  C_1(s)
  \frac{db(s)}{ds}
  - \frac{d^2b(s)}{ds^2}
  \,,
\end{align}
where
\begin{align}
  C_0(s)
  &\coloneqq
  \bigl[
    \nu(s) - \Delta(0)
  \bigr]
  \!
  \bigl[
    \nu(s) - \Delta(\Omega_\midlabel)
  \bigr]
  +
  i\frac{d\nu(s)}{ds}
  ,\\
  C_1(s)
  &\coloneqq
  i\bigl[
    2\nu(s)
    - \Delta(0)
    - \Delta(\Omega_\midlabel)
  \bigr]
  \,.
\end{align}
The window $b_2(s)$ is effectively finite since
$C_0(\pm\infty) = 0$,
so we call $g_2(s)$ the \emph{second ramp chirplet}.
The ESD of $f(s)$ is then recovered as
\begin{equation}
  \bigl|
    \fourier[f(s)](\omega)
  \bigr|^2
  =
  \frac{
    2\bigl|
      \fourier[g_2(s)](\omega)
    \bigr|^2
  }{
    \bigl[
      \omega - \Delta(0)
    \bigr]^2
    \bigl[
      \omega - \Delta(\Omega_\midlabel)
    \bigr]^2
  }
  \,.
\end{equation}
As expected,
if the matrix element does not asymptotically vanish,
its ESD has poles at both the bare and Stark-shifted transition frequencies.

\section{Extra control--spectator transitions}
\label{sec:additional-frequency-collision-analysis}

\begin{figure}
  \begin{subcaptiongroup}
    \includegraphics{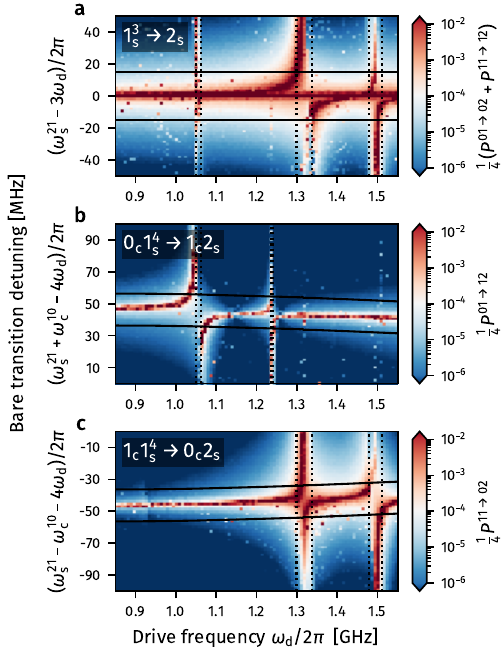}
    \phantomcaption\label{fig:control-spectator-leakage-third-harmonic}
    \phantomcaption\label{fig:control-spectator-leakage-fourth-harmonic-plus}
    \phantomcaption\label{fig:control-spectator-leakage-fourth-harmonic-minus}
  \end{subcaptiongroup}
  \caption{%
    Spectator-fluxonium $1\!\to\!2$ error
    induced by driving
    (\subref{fig:control-spectator-leakage-third-harmonic})
    near the third subharmonic of its transition frequency,
    (\subref{fig:control-spectator-leakage-fourth-harmonic-plus})
    near the fourth subharmonic and assisted by
    a control-qubit \emph{excitation}
    and
    (\subref{fig:control-spectator-leakage-fourth-harmonic-minus})
    near the fourth subharmonic and assisted by
    a control-qubit \emph{decay}.
    The vertical axis is the detuning between
    the bare transition frequency and
    a drive harmonic.
    \textbf{Solid} black lines define a \qty{\pm15}{\MHz} detuning window
    in (\subref{fig:control-spectator-leakage-third-harmonic})
    and a \qty{\pm10}{\MHz} window centered on the Stark-shifted transition frequency in
    (%
      \subref{fig:control-spectator-leakage-fourth-harmonic-plus},
      \subref{fig:control-spectator-leakage-fourth-harmonic-minus}%
    ).
    \textbf{Dotted} black lines reproduce the collision windows from
    \cref{table:control-qubit-collision-windows}
    for $\Delta p = 0.8$.
  }
  \label{fig:control-spectator-leakage}
\end{figure}

\begin{figure}
  \begin{subcaptiongroup}
    \includegraphics{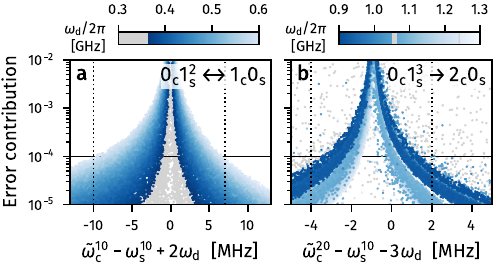}
    \phantomcaption\label{fig:control-and-spectator-iswap}
    \phantomcaption\label{fig:control-and-spectator-leakage}
  \end{subcaptiongroup}
  \caption{%
    (\subref{fig:control-and-spectator-iswap})~%
    iSWAP error
    and
    (\subref{fig:control-and-spectator-leakage})~%
    control-qubit leakage error
    assisted by a spectator decay
    when the detuning
    (horizontal axis)
    between the Stark-shifted transition frequency
    and
    (\subref{fig:control-and-spectator-iswap})~%
    the second
    or
    (\subref{fig:control-and-spectator-leakage})~%
    third drive harmonic
    is small.
    The vertical axis
    in
    (\subref{fig:control-and-spectator-iswap})
    is
    $
    \frac15(
      P^{01\to10}
      + P^{10\to01}
    )
    $
    and in
    (\subref{fig:control-and-spectator-leakage})
    $
    \frac14P^{01\to20}
    $.
    The \textbf{blue shading} of the scatter points corresponds
    to the color bar above.
    The extracted collision bounds
    (\textbf{dotted vertical}) 
    are:
    (\subref{fig:control-and-spectator-iswap})~%
    ($-10$ to $+7$)~\unit{\MHz}
    and
    (\subref{fig:control-and-spectator-leakage})~%
    ($-4$ to $+2$)~\unit{\MHz}.
  }
  \label{fig:control-and-spectator-secondary}
\end{figure}

In this appendix, we only consider the case where
the CR drive amplitude induces $\Delta p = 0.8$
and the control and spectator qubits are coupled with a strength of $J/2\pi = \qty{100}{\MHz}$,
consistent with \cref{sec:control-spectator-transitions}.

Beyond the errors discussed in the main text,
we present here the leakage error induced when driving near the third subharmonic of
a spectator's $1\!\to\!2$ transition frequency, see 
\cref{fig:control-spectator-leakage-third-harmonic}.
These errors are almost identical to
\cref{fig:control-target-third-harmonic-strong},
where a \emph{target's} $1\!\to\!2$ transition is driven.
From this similarity,
we assert that even though the spectator's qubit transition
is not resonantly driven by the CR drive,
its leakage probability is the same as if it were a resonantly driven target qubit.
\Cref{%
  fig:control-spectator-leakage-fourth-harmonic-plus,%
  fig:control-spectator-leakage-fourth-harmonic-minus%
}
shows the $01\!\to\!12$
and $11\!\to\!02$
leakage errors,
respectively,
induced when driving near the fourth subharmonic of the transition frequencies.
These plots are similar to
\cref{%
  fig:control-target-fourth-harmonic-plus,%
  fig:control-target-fourth-harmonic-minus%
},
except the figures in the main text
correspond to a larger drive amplitude inducing $\Delta p = 1.0$.
As expected,
a smaller amplitude leads to narrower error peaks.
Therefore,
the \qty{\pm10}{\MHz} collision bounds
centered on the Stark-shifted transition frequency
are the worst-case values,
applicable when the drive amplitude must be increased
beyond inducing $\Delta p = 0.8$
to detune the resonance from occurring on the pulse plateau.

Driving near the second subharmonic of the $01\!\to\!10$ frequency
can induce iSWAP-like errors,
see
\cref{fig:control-and-spectator-iswap}.
However, its collision bounds,
($-10$ to $+7$)~\unit{\MHz},
are not very large
and the spectator frequency needs to be
larger than around \qty{1}{\GHz} to satisfy this resonance condition.
In the same vein,
driving near the third subharmonic of the $01\!\to\!20$ transition frequency can induce leakage,
see
\cref{fig:control-and-spectator-leakage},
but this resonance is narrow,
($-4$ to $+2$)~\unit{\MHz},
and requires a drive over \qty{0.9}{\GHz}.
Because of the narrow bounds
and high qubit frequencies involved,
these two collisions were excluded from the final collision model.

\section{Target-qubit frequency allocation}
\label{sec:target-qubit-frequency-allocation}

The target-qubit parameters of
\cref{table:monte-carlo-parameters}
are determined by minimizing the collision probability
in one vertex neighborhood of a control qubit.
We fix the nominal control-qubit parameters at
$E_{J,\contlabel}/2\pi = \qty{4.0}{\GHz}$,
$E_{C,\contlabel}/2\pi = \qty{1.2}{\GHz}$
and $E_{L,\contlabel}/2\pi = \qty{0.4}{\GHz}$,
and the nominal target-qubit charging and inductive energies
at $E_{C,\targlabel}/2\pi = E_{L,\targlabel}/2\pi = \qty{1}{\GHz}$.
Hence,
we optimize over only the
nominal target-qubit Josephson energies.
The charging energies of the target qubits are chosen to be lower than that of the control qubit
since we expect the capacitive loading on the targets to be greater
in the heavy-hex lattice due to their higher neighbor count.
The nominal inductive energy of the target qubits is set to ensure
their qubit frequencies span
\qty{0.4}{\GHz} to \qty{1.0}{\GHz}
as $E_{J,\targlabel}/2\pi$
varies from \qty{3}{\GHz} to \qty{5}{\GHz}.

When calculating the collision probability of one vertex neighborhood,
we consider only collisions involving the qubit frequencies,
corresponding to
types 1, 8 and 9 of
\cref{table:frequency-collision-model}.
Additionally,
any realized qubit frequency over \qty{1}{\GHz} is also considered unacceptable.
In this simplified model,
we represent
the frequency detunings of the collisions
as correlated Gaussian random variables,
with their standard deviations calculated from first-order perturbation theory.
Accordingly, the probability of \emph{simultaneous} collisions
is given by the integral of a multivariate Gaussian over a hyperrectangle.
This can be computed with high numerical accuracy
using the Genz algorithm, implemented in SciPy as
\verb|scipy.stats.multivariate_normal.cdf|
(available since version 1.0.0)
\cite{genzNumericalComputationMultivariate2016,virtanenSciPy10Fundamental2020}.
The probability of \emph{one or more} collisions is then computed using the
inclusion--exclusion principle,
which is tractable for up to roughly 10 random variables
\cite{grinsteadCombinatorics2003}.
Compared to \cref{sec:zero-collision-yield},
we use a more precise method here
to calculate the zero-collision yield
such that the objective function
remains smooth even in the low-disorder regime.

\begin{figure}
  \begin{subcaptiongroup}
    \includegraphics{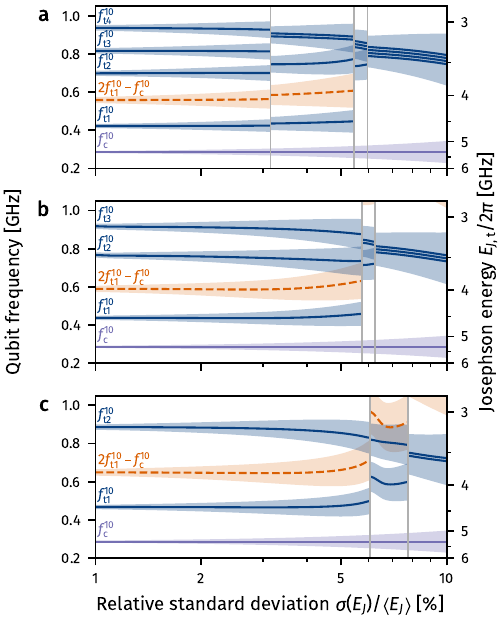}
    \phantomcaption\label{fig:ej-allocation-square}
    \phantomcaption\label{fig:ej-allocation-hex}
    \phantomcaption\label{fig:ej-allocation-heavy-hex}
  \end{subcaptiongroup}
  \caption{%
    Optimal nominal qubit frequencies
    (\textbf{solid blue} for the targets
    and \textbf{solid violet} for the fixed control)
    as a function of the Josephson-energy RSD
    that minimize the collision probability within a single vertex neighborhood
    of a control qubit in the
    (\subref{fig:ej-allocation-square})
    square lattice,
    (\subref{fig:ej-allocation-hex})
    hexagonal lattice
    and
    (\subref{fig:ej-allocation-heavy-hex})
    heavy-hexagonal lattice.
    Target-qubit frequencies should also avoid collision type~9
    (\textbf{dashed orange})
    from
    \cref{table:frequency-collision-model}.
    The \textbf{shaded} area around each line covers the region within one standard deviation
    of the mean.
    Discontinuities in the optimal allocation
    (\textbf{vertical lines})
    occur when multiple frequencies overlap and merge with increasing RSD.
    When they merge,
    the nominal frequencies are artificially offset to visualize their multiplicity.
    A conversion to the corresponding target-qubit Josephson energy
    is shown on the right spine.
    The other qubit parameters are the same as in
    \cref{table:monte-carlo-parameters}.
  }
  \label{fig:ej-allocation}
\end{figure}

In the low-disorder regime,
each target qubit has a unique frequency assignment,
see \cref{fig:ej-allocation}.
However, as the disorder increases,
the distinct frequencies spread and eventually coalesce,
collapsing into a single nominal frequency when the RSD is at \qty{10}{\percent}.
Since the optimal allocation in the low-disorder regime
remains valid up to modest RSD values
(around \qty{3}{\percent}),
we take these to define the ideal target-qubit parameters in
\cref{table:monte-carlo-parameters}
 
\bibliographystyle{apsrev4-2-trunc10}
\bibliography{references}

\end{document}